\documentclass[12pt]{article}

\usepackage[utf8]{inputenc}
\usepackage[english]{babel}
\usepackage{graphicx}
\usepackage{bm}
\usepackage{dcolumn}
\usepackage{amsmath}
\usepackage{epsfig,amssymb,euscript,mathrsfs}
\usepackage{array,calc,epsfig}

\usepackage[normalem]{ulem}
\usepackage{amsfonts}
\usepackage{slashed}
\usepackage{caption}
\usepackage{pdfpages}
\usepackage{braket}
\usepackage{rotating}
\usepackage{bbm}

\usepackage{tikz-cd}
\usepackage[noadjust]{cite}
\usepackage{chngcntr}
\usepackage[pdfpagelabels]{hyperref}
\usepackage{color}
\usepackage{booktabs}
\usepackage{comment}
\usepackage{cases} 
\usepackage{empheq}
\usepackage{cancel}
\usepackage{xfrac} 
\usepackage{float} 
\usepackage{subfig}
\usepackage{tabulary}   
\usepackage{makecell}
\usepackage{diagbox}

\newcommand{\numberset}{\mathbb}

\newcommand{\R}{\numberset{R}}
\newcommand{\C}{\numberset{C}}
\newcommand{\Z}{\numberset{Z}}

\newcommand{\p}{\partial}

\newcommand{\T}{\mathcal{T}}
\newcommand{\A}{\mathcal{A}}

\renewcommand{\d}{\mathrm{d}}
\let\oldcup\cup
\renewcommand{\cup}{\mathbin{\,\oldcup\,}}
\renewcommand{\Im}{\text{Im}}
\renewcommand{\ker}{\text{Ker}}
\newcommand{\Tr}{\text{Tr}}
\newcommand{\1}{\mathbf{1}}

\newcommand{\TR}{\mathsf{T}}

\newcommand{\Ag}{\mathsf{A}}
\newcommand{\Gg}{\mathsf{G}}
\newcommand{\spin}{\text{Spin}}
\newcommand{\SO}{\text{SO}}
\newcommand{\U}{\text{U}}
\newcommand{\SU}{\text{SU}}

\renewcommand{\hom}{\text{Hom}}

\newcommand{\tor}{\text{Tor}}

\newcommand{\link}{\text{Link}}
\newcommand{\inters}{\text{Int}}
\newcommand{\PD}{\text{PD}}

\newcommand{\CS}{\text{CS}}
\newcommand{\spinc}{\text{spin}_c}

\def\bal{\begin{align}}
\def\eal{\end{align}}
\def\be{\begin{equation}}
\def\ee{\end{equation}}

\def\dag{\dagger} 
\def\a{\alpha} 
\def\b{\beta} 
\def\g{\gamma} 
\def\G{\Gamma} 
\def\D{\Delta} 
\def\de{\delta} 
\def\e{\epsilon}

\def\l{\lambda}

\def\r{\rho} 
\def\s{\sigma} 
\def\S{\Sigma}

\def\w{\omega} 
\def\W{\Omega} 
 
\def\th{\theta}
\def\wt{\widetilde} 
\def\wh{\widehat}

\numberwithin{equation}{section}

\let\oldlgraf\{ 
\renewcommand{\{}{\left \oldlgraf}

\let\oldrgraf\}
\renewcommand{\}}{\right \oldrgraf}

\begin{document}

\begin{titlepage} 
\begin{center} 
\vspace{1cm}
{\LARGE Gauging in superconductors}

\smallskip
 {\LARGE and other electronic systems}
\vspace{1cm}

Marcus Berg${}^{a,b}$, Andrea Cappelli${}^c$,  and Riccardo Villa${}^c$

\vspace{1cm}
{\em ${}^{a}$ Department of Engineering and Physics, Karlstad University,\\
Karlstad, Sweden}
\\
{\em ${}^{b}$ Nordita, KTH Royal Institute of Technology and
Stockholm University, \\ Stockholm, Sweden}
\\
{\em ${}^{c}$ INFN, Sezione di Firenze,\\
  Via G. Sansone 1, 50019 Sesto Fiorentino - Firenze, Italy}\\
\end{center} 

\vspace{1cm}

\begin{abstract}
  Ordinary, s-wave superconductors have been recognized as
  being topological phases of matter, in which the dynamical gauge field
  implies less understood global features. Using the 
  tools of topological field theories and generalized symmetries, we provide
  an updated description of these systems.
  At very low energies, the Higgs model reduces to
  the BF theory, which exhibits topological
  order. Furthermore, the gauge field must be a $spin_c$ connection,
  to describe the spin of fermions forming Cooper
  pairs. Gauging implies that superconductors are inherently bosonic
  systems, yet they are endowed with a gravito-magnetic anomaly
  that is the remnant of their fermionic origin. We recognize that
  this anomaly is related to the Gaiotto-Kapustin-Thorngren
  bosonization, achieved via gauging fermion parity $(-1)^F$,
  now included in the gauge dynamics. This
  anomaly characterizes gauged electronic matter in great generality in
  three and four spacetime dimensions, forbidding
  trivial massive phases at low energy.  It holds beyond
  the validity of the Higgs model, and in other kinds of
  superconductors as well. It also appears in the nontrivial
  massless phase of three-dimensional electrodynamics, recently
  understood. 
  
\end{abstract} 
 
\vfill 
\end{titlepage} 
\pagenumbering{arabic} 
\newpage

\tableofcontents
\newpage
\section{Introduction}
\label{intro}

Superconductors (SC) are usually described as the Higgs phase of a
$\U(1)$ gauge theory with a complex scalar field $\Phi$ representing
the condensate.  Let us write the simpler relativistic Lagrangian for
concreteness and parameterize $\Phi=\frac{1}{\sqrt{2}}\r e^{i\phi}$,
where the real part acquires a vacuum expectation value
$\langle \r^2 \rangle =v^2$ and the phase $\phi \sim \phi+ 2\pi$
defines a compact dimensionless scalar\footnote{ In Eq. \eqref{hi-2}
we use the differential forms for all fields,
  e.g. $a=a_\mu \d x^\mu$, $f=\d a=(f_{\mu\nu}/2)\d x^\mu\wedge \d x^\nu$.
   We also consider $3+1$ Euclidean spacetime $X$ and set
 $\hbar=c=e=1$.}
\begin{align}
  \label{hi-1}
S&=\int_X \left\vert (\p_\mu -iq a_\mu)\Phi \right\vert^2 +
   V\left(\Phi^\dag \Phi \right) + \frac{1}{4} f_{\mu\nu}^2
  \\
  &= \int_X \frac{\r^2}{2} (\d \phi- qa) \wedge *(\d\phi-qa) +
    \frac{1}{2} \d a\wedge *\d a + \frac{1}{2}
    \d \rho \wedge *\d \rho + V\left(\r\right)\, .
  \label{hi-2}
\end{align}
Note that the field $\Phi$
has charge $q\in \Z$ (so $\phi\to \phi+q\alpha$, $a \to a+\d \a$
under the gauge transformation $\a$), with $q=2$ for
the Cooper pair.

\paragraph{Low-energy limit and gauging.}
Let us consider the action \eqref{hi-2} in the low-energy limit, which
corresponds to large values of $v$.
Both photon and Higgs masses are $O(v)$ and quench the dynamics of
the respective fields: one is left with the first term in action,
\be
\label{s-wei}
S\to S[\d\phi-qa]=\int_X \frac{v^2}{2}
(\d \phi- qa) \wedge *(\d\phi-qa),
\ee
where $\r=v$. The corresponding equations of motion,
\be
\label{flux-q}
a=\frac{1}{q} \d\phi\,,
\ee
are actually a constraint for $a$, which looks like a pure
gauge in the superconductor bulk (Meissner effect). This field
could be disregarded, were it not for its global features.
As a matter of fact, microscopic analyses of superconductivity,
such as the BCS theory and the BdG equation do not generically
take it into account \cite{AltlandSimons,FradkinBook}.
These approaches are consistent
because they aim at describing local physics.

In modern studies of topological phases of matter, and in this work,
one is instead interested in the global features and global degrees of freedom
that are present at energies below the gap $O(v)$. Their existence
distinguishes a topological phase from a trivially gapped one.
These aspects are described by the action $S[\d\phi-qa]$,
in which the dynamical gauge field, i.e. ``gauging'', is essential.

This point of view was considered long ago by Weinberg
in a beautiful paper \cite{weinbergSC}, where he showed that some 
defining features  of SC can be obtained by studying $S[\d\phi-qa]$.
He stresses the fact that these are exact properties, stemming from
topological and geometrical properties of the gauge field, being
independent of the specific form of dynamical terms into
the action (usually more involved than those considered in \eqref{hi-1}).
These properties are, beside the Meissner effect:
\begin{itemize}
\item flux quantization in units of $\Phi_o/q$, as follows
  by integrating \eqref{flux-q}
  on a closed contour inside the SC ($\Phi_o=2\pi$ is the flux quantum,
  thus half-flux in the standard case);
\item current flowing without applied potentials (zero resistivity).
\item Josephson effect, which arises at the junction of two SCs
  separated by a thin insulating barrier: an alternating current is found
  when a potential is applied across the junction and a continuous one occurs
  without potential.
\end{itemize}

\paragraph{Topological order and BF theory.}
Superconductors were clearly identified as being topological phases of
matter after another important feature was remarked in Ref.
\cite{hansson2004superconductors}.
This is the topological order, the ground-state degeneracy
which depends on the spatial topology of the system and
is associated to a {\it non-local} order parameter
\cite{mcgreevy2023generalized}.
This degeneracy can be understood as a consequence of half-flux
quantization. It can be seen at the classical level by evaluating the action
$S[\d\phi-qa]$ on its solitonic ground-state solutions
(setting $q=2$ hereafter).

Consider a 3d spatial geometry containing $n=1,2,3$ independent
non-contractible loops, 
(resp. a thick cylinder, a thick two-torus, a three-torus).
Non-trivial solutions of \eqref{flux-q} are labeled by
$\oint_{\G_i} a =0,\pi, 2\pi,\dots$
in any loop $\G_i$, $i\le n$, whose values are determined by
magnetic flux lines which are linked with them.
At the quantum level, these solutions
correspond to degenerate ground states
which are characterized by the expectation value of Wilson loops
$W(\G_i)=\exp i\oint_{\G_i} a=\pm 1$. It follows that the topological-order
degeneracy is $2^n$ ($q^n$ in general).

The action $S[\d\phi-qa]$ can be recast into a more familiar expression
by introducing the two-form field $b$ dual to $\phi$.
We rewrite the action as follows,
\be
\label{b-term}
S=\int_X \frac{v^2}{2} (u- qa) \wedge *(u-qa) +\frac{i}{2\pi} b\wedge \d u \,.
\ee
Integrating $b$ gives $\d u=0$ with $2\pi$-quantized holonomies,
thereby recovering the original phase $u=\d \phi$
with correct $2\pi$ periodicity.
Conversely, eliminating  $u$ by substituting its equation
of motion $u-q a = (i/2\pi v^2) *\d b$ in the action, leads to
\be
S[a,b]=\int_X \frac{i q}{2\pi}  b\wedge\d a +
\frac{1}{8\pi^2 v^2} \d b\wedge * \d b \,.
\ee
Upon neglecting the second subleading term for $v\to\infty$, we obtain
the level-$q$ topological BF theory
\be
\label{BF-act}
S_{BF}[a,b]=\frac{iq}{2\pi}\int_X b\wedge\d a\,.
\ee
In this theory, correlation functions of 
Wilson loops $W(\G)=\exp i\oint_{\G} a$ and surface
operators $U(\S)=\exp i\int_\S b$  express
the Aharonov-Bohm phases $\Theta=2\pi/q$ of particles
encircling flux lines, as  semiclassically described before.
We conclude that in the deep infrared, the Abelian Higgs model reduces to
the level-$q$ BF field theory, which describes the 
$\Z_q$ topological gauge sector of the Higgs phase.\footnote{
  This result can be summarized as ``higgsing the U(1) theory
  down to $\Z_q$.'' The explicit discrete form of BF theory
  will be given later.}

The BF theory also describes other systems at low energy,
such as the topological insulators (TI) \cite{cappellianomaliescondmat}.
However, there is an
important difference: in that case the fields $a,b$ represent
``hydrodynamic'' matter fluctuations, and parameterize the associated
currents. For example, $J =\frac{1}{2\pi} *\d b$ is the $\U(1)$
current which then couples to the electromagnetic field
$A$, treated as a classical background,\footnote{We adopt the convention
  of representing dynamical fields lowercase $(a)$ and backgrounds uppercase
  $(A)$.   Furthermore, letters $a,b,c,\dots$ usually denote
  $1-,2-,3-,\dots$ form fields.}
namely this is not gauged. 
It follows that the global $\U(1)$ symmetry together with time reversal
invariance characterize these systems, through their anomalies
and associated responses. 

In the case of superconductors, 
the back-reaction of the matter degrees of freedom on $a$ is clearly shown
by the Meissner effect, thus gauging is essential.
This implies that the $\U(1)$ symmetry  becomes
a redundancy in the description of the system.
It follows that SC and other phases
of electrodynamics should be characterized by further
global symmetries and classical backgrounds, involving the
generalized symmetries \cite{gaiotto2015generalized,mcgreevy2023generalized}.
This is a central theme of current investigations
of gauge theories. In the following section, we 
give a detailed description of these symmetries in the Higgs model.

\paragraph{Higgs model of boson and fermions: $spin_c$ connection.}  The
Higgs model does not discriminate whether the $\Phi$
field is fundamental or a composite field and, in the latter case,
whether it is a bound state of fermionic particles, as it happens for
superconductors. The question to be addressed is whether
there is a remnant in the low-energy topological theory when the
Higgs field represents Cooper pairs.

As seen before, the observables of the BF theory are
the Wilson loops of $a$ and surfaces of $b$.
The loop of $a$ describes
a spin-zero quasiparticle, because it is not affected by
rotations at any point of the line. However, it is known that 
superconductors have fermionic quasiparticle excitations, which are
ultimately a superposition of an electron and a hole state
\cite{AltlandSimons,CasalbuoniSC}. The earlier description is
therefore not completely correct.  This issue was already noticed for
example in \cite{higgsSPTII2023,vonKeyserlingk2014WWModelBF}
and also in \cite{thorngren2014framed}, where a solution was proposed.

The basic observation, stressed in
\cite{seibergwitten2016gappedTI}, is that excitations in condensed matter
systems obey the $\Z_2$ parity rule between charge and spin, $Q=2S$ mod $2$,
because nuclei are neglected.
It follows that the correct, faithful symmetry is
$\spin_c(d)$, which is defined by the following quotient
\begin{equation}
\label{spinc grp}
    \spin_c(d) \simeq \frac{\spin(d)\times \U(1)}{\Z_2}\,.
\end{equation}
This identifies fermion number $(-1)^F$ symmetry, that we call $\Z_2^f$,
with the $\Z_2$ subgroup of $\U(1)$.
This fact has some geometrical and topological consequences.

We recall that spin manifolds, where fermions can live, are characterized by the
trivialization of the second Stiefel-Whitney class
$w_2(TX) \in H^2(X;\Z_2)$, i.e. $w_2=\d\eta$, for some
$\Z_2$-valued one-form $\eta$. Actually, a non-vanishing flux
$\int_\S w_2=1$, with $\S\subset X$, would be an obstruction to consistently
define sign choices for spinors on any patch \cite{nakahara}. 
In $\spinc$ manifolds, one can use the Abelian gauge field to
compensate the $w_2$ flux and then define spinors. This implies
the modified flux quantization \cite{seibergwitten2016gappedTI}
\begin{equation}
    \label{spinc connection}
    \int_\S \frac{\d a}{2\pi} + \frac{1}{2}w_2(TX) =n \in \Z.
\end{equation}
In this case, the field $a$ is called a spin$_c$ connection.

It follows that theories with spin$_c$ symmetry can be defined on
manifolds which are not necessarily spin ($\C \mathbb{P}^2$,
for example).\footnote{
  All oriented $2+1d$ manifolds are spin; those 
  in $3+1d$ are all $\spinc$, but some are not spin.
  Spin$_c$ symmetry is further reviewed in appendix A
  of \cite{CappelliVillaBosDual2025}, where non-orientable
  manifolds are also considered.}
This in turn puts stronger conditions on the consistency of the theory, stemming
from the generalized flux quantization \eqref{spinc connection}.
In physical terms, the general idea is to probe the theory with most
general gravitational backgrounds for revealing characteristic
properties, such as anomalies and responses
\cite{witten2016fermion,cappellianomaliescondmat}. 

Even on spin manifolds, the spin$_c$ symmetry has an interesting
consequence that helps resolve the earlier problem with
Wilson loops in the Higgs model.  On non-spin manifolds, line operators
must be associated to surfaces  containing  $w_2$ flux\footnote{
Two extensions $\oint_\G a= \int_\S \d a$ and
$\oint_\G a=\int_{\S'} \d a$, $\p \S =\p \S'=\G$, without $w_2$ flux
can differ $\int_{\S\cup \S'}\d a=\pi n$ by  
  \eqref{spinc connection} and be inconsistent.}  
\begin{equation}
    e^{i\oint_\G a}e^{i\pi\int_\S w_2}, \qquad \p \S= \G \,.
  \end{equation}
They are not ``genuine'' line operators.
However, on spin manifolds, where $w_2 =\d \eta$, they do become
line operators of $a'=a+\pi\eta$, which is an ordinary gauge
field obeying standard quantization $\int_\S \d a'=2\pi n$.
We can write
\begin{equation}
\label{spinc wilson line}
W(\G)= e^{i \oint_\G a}=
e^{i \oint_\G a'}e^{i\pi\oint_\G\eta}\,.
\end{equation}
The additional neutral loop operator is a
sign factor, $\exp( i\pi\oint_\G\eta)=\pm 1$, which flips under
$2\pi$ rotations of the framing of $\G$ \cite{thorngren2014framed}.
Therefore, the Wilson line of the $\spinc$ connection $a$ carries a
$\SO(d)$ projective representation and describes a fermionic
particle (with charge one). Clearly,
this argument applies to odd powers of \eqref{spinc wilson line},
while even powers are independent of $\eta$: particles with even
charge are bosons. 
We remark that $\exp( i\pi\oint_{\g_i}\eta)=\pm 1$
evaluated on non-contractible loops $\g_i$ of $X$, also determines the choice
of spin structure \cite{gaiottokapustinspinTQFT1,gaiottokapustinspinTQFT2,
  thorngren2014framed}.
For this reason, the one-form $\eta $ is called the $\Z_2^f$ background field,
or simply ``spin structure''.
 
In conclusion, the Higgs model with spin$_c$ gauge field can describe
superconductors. Note that the $\spinc$ symmetry is motivated
a-priori, but is also a necessity. 

The global properties of the $\spinc$ connection can be
included in the (bosonic) BF theory \eqref{BF-act} as follows ($q=2$)
\begin{equation}
\label{BF-w2}
S_{BF}  =i\int \frac{2}{2\pi}b\wedge \left(\d a+ \pi w_2 \right) +
\frac{1}{2\pi}\d a \wedge B_m\,,
  \end{equation}
  because the $b$ equation of motion enforces the flux condition
  \eqref{spinc connection}.
  This improved topological theory effectively describes fermionic
  quasiparticles in the SC at low energy.

  In the action \eqref{BF-w2} we also included the 2-form background $B_m$
  for the ``magnetic'' 1-form global symmetry $\U(1)^{(1)}_m$,
  which acts by $b\to b+\g, B_m\to B_m-2\g$
as better explained in the next section.
This symmetry of the Higgs model, inherited from the Maxwell theory, is
actually broken by the term involving $w_2$, leading to a
mixed magnetic-gravitational anomaly. It can be canceled using
anomaly inflow from a $5d$ manifold $Y$, with $X=\p Y$, by the topological
term\footnote{Hereafter, imaginary actions are clearly identified mod $2\pi i$.}
\begin{equation}
  \label{an-b}
    \A =i\pi \int_Y w_2 \wedge \frac{\d B_m}{2\pi}. 
\end{equation}
This is the anomaly mentioned in \cite{higgsSPTII2023}.
It is a characteristic feature of superconductors, i.e. of the
Higgs mechanism made by Cooper pairs.

The presence of this anomaly puts a constraint on effective low-energy
theories of more general superconductors, beyond the simple Higgs
model \eqref{hi-1}.
It requires the presence of topological or massless excitations at low
energy that reproduce it. In particular, it forbids a trivial massive
phase, which is possible for a fundamental Higgs particle.
It could be relevant, for example, in superconductivity by
condensation of an even number of fermions
\cite{BergFradkinKivelson2009SC4e,Vishwanath2026TSC4e,GaoWangZhangYang2026SC4ePrimary,SC2e4eQPT2026,SC3D4e6e2026,GaoWangYangWu2025TSC2ne}, corresponding to a
Higgs field with charge $q=2n$. In this case, $\U(1)$ is
higgsed to the subgroup $\Z_{2n}$, which clearly contains $\Z_2^f$
(see section 4).

\paragraph{Bosonization by gauging $\Z_2^f$.} 
A general map between fermionic and bosonic theories in any dimension
has been formulated by Gaiotto, Kapustin, Thorngren and others, starting from
their basic topological properties
\cite{gaiottokapustinspinTQFT1,gaiottokapustinspinTQFT2,
  thorngren2020anomalies,kapustinthorngren2017}.
Let us briefly review it.\footnote{ For an extensive introduction see
  \cite{CappelliVillaBosDual2025}, section 2 and appendix
  A.  } The map involves summing the fermionic partition function
$Z_f[\eta]$ over the spin structures $\eta$,
\be
\label{Zb}
Z_b[C]=\sum_\eta Z_f[\eta] (-1)^{\int_X \eta\cup C}\,.
\ee
The resulting bosonic partition function\footnote{In this equation, we use
the notation of cochains and cup product, which
is convenient for fields taking values in the group $\Z_n$
\cite{kapustinseiberg2014}. Cochains can be mostly represented by
continuous flat U(1) forms, with exceptions to be discussed.
The correspondence is $ C\equiv (2\pi/n) \wt C$, where the
flat $p$-form $C$ obeys the quantization $\int_{\S_p} C =(2\pi/n) \Z$,
such that $\exp ( i\int_{\S_p} C) \in \Z_n $; the 
cochain $\wt C$ is integer valued, $\int_{\S_p} \wt C =0,1,\dots,n-1 $.
Formulas involving cochains will be recognizable by the explicit 
cup product, while the wedge will be mostly omitted in
continuum notation. The twiddle over cochains
will also be omitted when clear from the context.}
$Z_b[C]$ depends on the $\Z_2$
flat background gauge field $C\in H^{d-1}(X;\Z_2)$ which couples to $\eta$.
The sum in \eqref{Zb} can be thought of as gauging the fermion parity symmetry
$\Z_2^f$.
The map \eqref{Zb} can be inverted by gauging $C$ in $Z_b[C]$, as follows
\be
\label{Zf}
Z_f[\eta]=\sum_{c\in H^{d-1}(X;\Z_2)} Z_b[c] (-1)^{\int_X \eta\cup c}\,,
\ee
being a kind of $\Z_2$ Fourier transform.

Equation \eqref{Zf} has the following interpretation. The bosonic theory
$Z_b[C]$ no longer knows of the existence of a spin structure and can be
defined without mentioning it. Instead, it depends on the twisted sectors
specified by the values of the $(d-1)$-form $C$ background.
This means that it is endowed with the higher-form symmetry $\Z^{(d-2)}_2$, dual
of $\Z^f_2$. However, notice that it
is not invariant with respect to it, because
\be
\label{Z-an}
Z_b[C+\d\l]=Z_b[C](-1)^{\int_X \eta\cup \d \l}=
Z_b[C](-1)^{\int_X w_2\cup \l}\,,
\ee
owing to $w_2=\d\eta$.
Forgetting momentarily the mid expression in this formula,
we can say that the bosonic
theory has an anomaly with respect to the $\Z^{(d-2)}_2$ symmetry.
This can be expressed by inflow from $d+1$ dimensions of the following
topological action, with $X=\p Y$,
\be
\label{an-c}
{\cal A} =i\pi \int_Y w_2\cup C\,.
\ee

We have found that any fermionic theory $Z_f[\eta]$ can
be bosonized by means of \eqref{Zb} (at topological level).
The opposite fermionization map is obtained by using \eqref{Zf},
subjected to the following conditions and specifications.
The bosonic theory $Z_b[C]$ should have a $\Z^{(d-2)}$ symmetry with the
characteristic anomaly \eqref{an-c}.
The fermionic $Z_f[\eta]$ is defined on a spin manifold $X$, 
over which the $\Z_2$ field $\eta$ is defined, obeying
$w_2=\d\eta$. In such a case, the Stokes theorem
can be applied to the quantity ${\cal A}$ in \eqref{an-c}, reducing it to 
the counterterm $i\pi\int_X \eta \cup C$. This is indeed added to the sum
\eqref{Zf} for trivializing the anomaly of $Z_b[C]$.

In conclusion, the bosonic theories suitable for fermionization should have
the anomaly \eqref{an-c}, which is then trivialized by introducing the
spin structure. We finally remark that in two dimensions
this approach corresponds to the familiar bosonization
\cite{ginspargCFT,thorngren2020anomalies,tongArf,backfiringbosonization2024}:
actually, the anomaly \eqref{an-c} vanishes in $2d$ and there are
no restrictions to fermionization.

\paragraph{Gauging a $\spinc$ symmetry and bosonization.}
In the previous discussion, we established that superconductors and other
condensed matter systems involve
the gauging of the $\spinc$ symmetry. Given that the fermion parity
symmetry is included in it, $\Z_2^f\subset \U(1)$, owing to \eqref{spinc grp},
the bosonization map is automatically done by gauging.

It follows that gauged systems are intrinsically bosonic: yet they are 
characterized by the anomaly \eqref{an-c} due to their
fermionic origin.
The relation between gauging and bosonization proves that
the characteristic anomaly \eqref{an-c} has a more fundamental origin,
and it is valid in all possible phases, beyond the Higgs one.
The only assumptions is that the microscopic theory of
condensed matter systems is electrodynamics of electrons (and possibly
other matter obeying the spin-charge parity rule).
This is the main message of this work. In particular,
\begin{itemize}
  \item
The anomaly \eqref{an-b} earlier derived within the analysis
of the Higgs phase should match the one of bosonization \eqref{an-c}.
The precise relation between the two expressions is described in section 3,
by performing the gauging in two steps, first over $\Z^f_2$ and then
over the remaining ``bosonic'' symmetry $\U(1)_b=\U(1)/\Z^f_2$. This 
discussion is rather technical and is drawn by the general analysis of
group extensions in Ref. \cite{Villa2026Extensions}.
Section 3.4 further analyzes the gravito-magnetic anomaly, pointing
  out the need for time-reversal symmetry in three dimensions.
\item
  The bosonization anomaly \eqref{an-c} should characterize all
  low-energy limits of three and four dimensional electrodynamics of
  fermions. For example,
the strongly interacting superconductors and the topological superconductors
not described by the Higgs (mean-field) model.
Another instance is the nontrivial massless phase of $N_f=2$
electrodynamics in $3d$ recently discussed in
Ref.\cite{DumitrescuNiroThorngren2024QED3}.
The present understanding of these systems will be  summarized in
section 4. We shall also discuss the minimal topological theory leading
to the anomaly, which comes in two forms, the $\spinc$ BF theory \eqref{BF-w2}
and the twisted BF theory. 
\end{itemize}

Appendices A and B describe several properties of the two BF topological
theories involved in this paper, the $\Z^f_2$ gauge theory and the twisted
BF theory and compute their Wilson loop observables.
Appendix C evaluates the number of $\spinc$ structures.
Finally the Conclusion and Outlook discusses some open questions on
the phases of gauged electronic systems.

\section{Abelian Higgs model}

In this section, we summarize the modern analysis of the Abelian
Higgs model, describing the generalized global symmetries, anomalies
and low-energy limit
\cite{hansson2004superconductors,weinbergSC,witten2007SC,WeinbergQFTII}.
This will set the stage for the discussion in the sections 3 and 4.
The Higgs field will be assumed to be a charge $q$ scalar field
and its compositeness will be considered in the next sections.

\subsection{Generalized symmetries and mixed anomaly}

We rewrite the Abelian Higgs action \eqref{hi-2} as
 \begin{equation}
\label{Abelian Higgs}
  S[B_e,B_m] = \int_X  \frac{\rho^2}{2} \vert\d \phi-qa\vert^2+
  \frac{1}{2}\vert \d \rho \vert^2 + V(\rho) +
 \frac{1}{2}\vert \d a-B_e\vert^2 + \frac{i}{2\pi} B_m \d a\,,
\end{equation}
using the notation $\vert v\vert^2=v\wedge *v$. In this expression, we also
added the backgrounds $B_e,B_m$ for the generalized symmetries,
but, as a first step, let us momentarily put them to zero.
As already said, the U(1) symmetry is
gauged and becomes a redundancy. At the same time,
the dynamical field $a$ defines further currents.

The generalized electric symmetry acts on Wilson loops
$W(\G)=\exp i\oint_\G a$ by a phase, which is obtained by $a\to a+\b_e$,
where $\b_e$ is a closed form. The corresponding Noether
current is the 2-form current $J_e= f =\d a $,
which obeys the equations of motion $\d *\! f= *j$, where $j$ is the 
Higgs field current.
$J_e$ would be conserved in absence of matter, i.e. in free
Maxwell theory, leading to the global one-form symmetry $\U(1)^{(1)}_e$.
Nonetheless, if matter has a minimal charge $q\in\Z$, 
the total charge $Q(\S_{d-2})=\int_{\S_{d-2}} *f $ is well defined\footnote{
  Using $\int_\S *f -\int_{\S'}  *f =
  \int_{V} \d *f =\int_{V}*j=0$  mod $q$  with $\S \cup \S'=\p V$.}
(i.e. topological)
modulo $q$.
Then, the finite transformations $U_k=\exp(i2\pi k/q\,Q(\S_{d-2}))$ are also
well defined and generate the $\Z_q$ group.
Therefore, the  Higgs model with $q$-charge condensate
has the global $\Z_{q,e}^{(1)}$ symmetry, being the remaining of
Maxwell $\U(1)^{(1)}_e$.

The corresponding invariance of the action \eqref{Abelian Higgs} is
verified by transforming
$a \to a+ \beta_e$ and $\d \phi \to \d \phi + q \beta_e$ (for $B_e=B_m=0$).
The shift of the compact scalar is
consistent with the quantization $\oint_\gamma \d \phi =2\pi \Z$ if
$\oint_\gamma \beta_e = (2\pi /q)\Z$. So $\beta_e$ represents a $\Z_q$
valued cocycle,\footnote{See footnote 7.} checking the $\Z_{q,e}^{(1)}$
symmetry. Notice that for $q=1$, $\beta_e$ is closed
  and $\oint \beta_e =2\pi\Z$. Thus, it parameterizes the
  $\U(1)^{(0)}$ gauge transformation and there is no electric one-form
  symmetry left.

The magnetic generalized symmetry is the dual symmetry acting on `t Hooft
loops, which are $(d-3)$ defects in general dimension. The
$(d-2)$-form current is $J_m=* f/2\pi$, and
is conserved by the Bianchi identity $\d f=0$. This is a ``topological''
current\footnote{Like the kink current of the $2d$ compact boson.}
and it would be the Noether current in the dual theory.
Therefore, the
Higgs model, as well as Maxwell, possesses the $\U(1)^{(d-3)}_m$
magnetic symmetry. This is actually present in any Abelian
gauge theory in absence of magnetic charges.
In conclusion, the Higgs model  has the
$\Z_{q,e}^{(1)}\times \U(1)_{m}^{(d-3)}$ generalized symmetry
\cite{gaiotto2015generalized}.

In the action \eqref{Abelian Higgs}, the backgrounds $B_e, B_m$ are
coupled to the respective currents $J_e= f $ and $J_m=* f/2\pi$
(the local term $ B_e^2$ is added for convenience).
$B_m$ is a U(1) $(d-2)$-form gauge field and $B_e$ a 2-form
flat field, with holonomies $e^{i\oint B_e}= e^{i\frac{2\pi}{q}n}$,
$n\in\Z$, for representing a discrete
$\Z_q$ field $\wt B_e\in Z^{2} (X;\Z_q)$,
by $B_e = \frac{2\pi}{q} \wt B_e$. In presence of backgrounds, the symmetries
become local, acting by
\be
\label{local-symm}
B_m \to B_m +\d \b_m,\qquad\quad
B_e \to B_e +\d \b_e, \ \ \ a\to a+\b_e,\ \ \ \d \phi \to \d \phi+q\b_e\,,
\ee
where now $\b_m$ and $\b_e$ are not required to be flat. 

While the action is invariant under each individual symmetry, there is a
mixed electromagnetic anomaly $\Z_{q,e}^{(1)}\times\U(1)_{m}^{(d-3)}$
\cite{gaiotto2015generalized}, which can be represented using inflow by the
topological action in $d+1$ dimensions
\begin{equation}
\label{AHM mixed anomaly}
\A = i\int_Y  B_e \wedge \frac{\d B_m}{2\pi} =
i \int_Y \frac{2\pi}{q} \wt B_e \cup \frac{\d B_m}{2\pi},
\qquad\qquad {\rm for}\ \  \Z_{q,e}^{(1)}\times\U(1)_{m}^{(d-3)}.
\end{equation}
In this equation, we wrote both the continuum and cochain expressions for
clarity.

\subsection{The Higgs phase}

We consider the action \eqref{Abelian Higgs} with the standard potential
\begin{equation}
\label{higgs potential}
    V(\rho) = m^2 \rho^2 + \lambda \rho^4, \qquad\quad \lambda >0.
\end{equation}
The Coulomb phase for $m^2 >0$ contains a massive scalar field coupled to the
electromagnetic field. At energies below the gap,
the scalar field can be integrated out and only the electromagnetic
field remains, resulting in the free Maxwell theory.
There is a massless photon, which is the Nambu-Goldstone
boson for the spontaneous symmetry breaking of $\U(1)_{m}^{(d-3)}$
\cite{gaiotto2015generalized,lake2018}. It follows that also
$\Z_{q,e}^{(1)}$ is spontaneously broken, being in fact enhanced
to the $\U(1)^{(1)}_e$ symmetry in IR Maxwell theory. Their mixed
anomaly matches the UV anomaly \eqref{AHM mixed anomaly}.

In the Higgs phase $m^2 <0$, the scalar field condenses, 
$\braket{\rho^2}= v^2$. This phase is fully gapped
but not trivial, since it must match the anomaly \eqref{AHM mixed
  anomaly}. Below the gap $O(v)$,  the $\U(1)$ gauge group is higgsed to
$\Z_q$, as described in the Introduction.
The action \eqref{Abelian Higgs} reduces to \eqref{s-wei},
now including the backgrounds,\footnote{
  The subleading Maxwell term is momentarily kept
  for showing the $B_e$ coupling.}
\begin{equation}
\label{higgs phase bkgd}
S[B_e,B_m] = \int \frac{v^2}{2} \vert \d \phi-qa \vert^2 +
\frac{1}{2}\vert \d a -B_e\vert^2 + \frac{i}{2\pi} B_m \d a.
\end{equation}

Notice that the vacuum expectation value $v$ of the Higgs scalar allows
the phase $\phi$ to be a well-defined field.
There is an emergent winding symmetry $\Z_{q,w}^{(d-2)}$
associated with it,  generated by
\begin{equation}
\label{Winding sym defects}
U(\gamma) = \exp\left(i\theta \oint_\gamma \frac{\d \phi}{2\pi}\right),
\qquad\quad \theta = \frac{2\pi}{q}n, \ n \in \Z_q\,.
\end{equation}
The defects $\U(\g)$ are topological since $\d^2 \phi =0$ and the
condition $\theta \in \Z_q$ ensures that they are gauge invariant
under $\phi \to \phi + q \alpha$ with
$\oint \d \alpha = 2\pi \Z$.

The next step is to dualize $\phi$ into the $(d-2)$-form $b$. Following
the same steps as in the Introduction (see \eqref{b-term}), but
in presence of backgrounds, we write
\begin{equation}
  \label{u-higgs}
  S[B_e,B_m] = \int \frac{v^2}{2} \vert u-qa\vert^2 +
  \frac{1}{2}\vert \d a -B_e\vert^2 + \frac{i}{2\pi} B_m \d a
  +\frac{i}{2\pi} b\d u - \frac{iq}{2\pi} bB_e.
\end{equation}
This action including the $b$ field is still
gauge invariant (provided that $B_e$ is a
$\Z_q$ gauge field, as indeed it is): $u$  transforms as
$\d \phi$ under the gauge and generalized symmetries (see \eqref{local-symm})
and there is an extra gauge symmetry associated to the $b$ field,
$b\to b + \d \lambda$. For the global symmetries,
$\U(1)_{m}^{(d-3)}$ works as before, while the symmetry
$\Z_{q,e}^{(1)}$ requires the introduction of the coupling $bB_e$ in
\eqref{u-higgs}, to compensate for the transformation of the new term $b\d u$.
The anomaly is still \eqref{AHM mixed anomaly}.

Integrating back $b$ in \eqref{u-higgs} now gives the condition
$\d u -q B_e =0$, involving the background $B_e$.
We observe that this continuous flat field is a proxy for the
$\Z_q^{(2)}$ field $\wt B_e$, and that $qB_e$  gives trivial
holonomies $e^{i\int q B_e}=1$.
Thus, we can solve the above equation within the ``gauge condition''
$q B_e=0$, leading to the solution\footnote{
  Alternatively, we can write $q B_e=\d \wt u$,
  with $\int \d\wt u=2\pi n$ and solve $u+\wt u=\d\wt\phi$,
  but $\wt\phi$ is just a redefinition of $\phi$.}
 $u=\d\phi$ and the original action
\eqref{higgs phase bkgd}.

Since $b$ is the
dual of $\phi$, it is naturally charged under the winding (magnetic) symmetry
of $\phi$. Thus, it also transforms under this emergent symmetry.
Note that the Wilson surfaces $\exp(i\int_{\Sigma_{d-2}}b)$
are the 't Hooft lines of $\phi$, the objects charged under the
winding symmetry.

Integrating out $u$ gives $u - qa = \frac{i(-1)^d}{2\pi v^2} * \d b$, and
substituting into the action yields
\begin{equation}
\label{higgs phase with b}
  S[B_e,B_m] = \int_X \frac{iq}{2\pi} b \d a - \frac{iq}{2\pi} bB_e +
  \frac{i}{2\pi} B_m \d a +\frac{1}{2}\vert \d a-B_e\vert^2 +
  \frac{1}{8\pi^2 v^2} \vert\d b\vert^2.
\end{equation}    
    
\paragraph{Topological BF theory with backgrounds.}
In the low-energy limit
$v \to \infty$, we can neglect the dynamical term of $b$,
as done in the Introduction. Also integration over $b$ makes $a$ flat and
we can consistently ignore the Maxwell term. Thus, we obtain
\begin{equation}
    \label{IR Abelian higgs phase}
    S_{IR} = \int_X \frac{iq}{2\pi} b \d a - \frac{iq}{2\pi} bB_e+
    \frac{i}{2\pi} B_m \d a.
\end{equation}
This is the BF theory \eqref{BF-act} derived earlier, with in addition
the backgrounds $B_e,B_m$ of the original Higgs model. These are useful
to match symmetries and anomalies, as discussed in the next paragraph.

We first recall general aspects of BF theories \cite{kapustinseiberg2014}
and then compare with the result \eqref{IR Abelian higgs phase}. 
The theory has a $\Z_{q}^{(1)}\times\Z_{q}^{(d-2)}$ symmetry
generated by the Wilson operators of $b$ and $a$. They are
also the charged objects under each other symmetry, as shown by the
linking action
\begin{equation}
    \label{BF link action}
    \braket{e^{in\oint_\g a} e^{im\oint_\S b}} = e^{2\pi \frac{mn}{q}\link(\g,\S)}.
  \end{equation}
  The expectation values of one kind of operators (or the other) characterize 
  the degenerate ground states, say on the torus spatial geometry
  $S^1\times S^1\times S^1$,  which lead to the topological order $q^3$
  \cite{hansson2004superconductors}.
  
The theory has a mixed anomaly because the symmetry defects are
charged with respect to the other symmetry, as seen in \eqref{BF link action}.
This anomaly can be computed by turning on
the following flat background gauge fields $B$ (for $\Z_q^{(1)}$) and $C$ (for
$\Z_q^{(d-2)}$), as follows\footnote{Thus $C$ is a 3-form in $4d$.}
 \begin{equation}
   \label{BF theory with backgrounds}
   S_{BF} = \frac{iq}{2\pi}\int_X  b \d a - bB - a C.
\end{equation}
The generalized symmetry transformations with backgrounds are
\be
\label{g-symm-bkgd}
b \to b + \g, \qquad C \to C+ \d \gamma, \quad {\rm and}\quad
a\to a + \b, \qquad B \to B + \d \b\,,
\ee
where $\g$ and $\b$ are the corresponding parameters.
The non-invariance of the action \eqref{BF theory with backgrounds}
can be canceled by the following
$d+1$ topological term, which expresses the mixed anomaly,
\begin{equation}
\label{BF mixed anomaly}
\A = \frac{iq}{2\pi}\int_Y B \wedge C  =
\frac{2\pi i}{q} \int_Y  \wt B\cup \wt C, \qquad\quad {\rm for}\quad
\Z_q^{(1)}\times\Z_q^{(d-2)}.
\end{equation}
The second, discrete, expression involves the cochains
$\wt C \in Z^{d-1}(Y;\Z_q)$ and $ \wt B \in Z^{2} (Y;\Z_q)$.

\paragraph{Symmetries and anomaly matching.}
It is now possible to understand the fate of UV symmetries 
in the low-energy Higgs and Coulomb phases
(they are summarized in Table \ref{fig table AHM}).
The UV Higgs model \eqref{Abelian Higgs} has a
$\Z_{q,e}^{(1)}\times\U(1)_{m}^{(d-3)}$
symmetry. The magnetic symmetry $\U(1)_{m}^{(d-3)}$ confines and
it is not present in the low energy theory
\eqref{BF theory with backgrounds} (apparently, see later).
It is not spontaneously broken because the Higgs phase is gapped. Actually, 
the transition from Higgs to Coulomb phase is characterized by 
the spontaneous symmetry breaking of $\U(1)_m^{(d-3)}$.

The electric symmetry is instead spontaneously
broken in the Higgs phase, since its charged objects, the Wilson lines of $a$,
are present in the IR spectrum of \eqref{IR Abelian higgs phase}: this
implies that the Higgs phase has $\Z_q$ topological order and the
IR symmetry $\Z_q^{(1)}$ is identified with $\Z_{q,e}^{(1)}$. The
Wilson lines of $a$ become topological in the IR, generating the
$\Z_q^{(d-2)}$ symmetry of \eqref{IR Abelian higgs phase}, which is the
already noticed winding symmetry $\Z^{(d-2)}_{q,w}$ \eqref{Winding sym
  defects}. Indeed, $b$ transforms
under it as $b \to b+\g$ and the $\Z_q^{(d-2)}$ symmetry defects act
on the Wilson surfaces of $b$ according to \eqref{BF link action}.

Comparing the backgrounds \eqref{BF theory with backgrounds} and
\eqref{IR Abelian higgs phase} yields the identification
\begin{equation}
    \label{C=dB magnetic sym frac in IR SC}
    B=B_e\,, \qquad\quad  C = - \frac{1}{q} \d B_m \,, \qquad\quad
    \left( \wt C = -\frac{\d B_m}{2\pi} \mod q\right),
\end{equation}  
which allows us to match the UV \eqref{AHM mixed
  anomaly} and IR anomalies \eqref{BF mixed anomaly}
This correspondence was obtained by explicit computation
in the simple Higgs model. In
strongly-coupled systems, where this model would not be valid, the
Eqs. \eqref{C=dB magnetic sym frac in IR SC} could still be argued by
imposing anomaly matching.

While $B =B_e$ clearly follows from the
discussion above, the relation involving $C$ and $B_m$ is a new piece
of information: it says that there is a {\it symmetry fractionalization} \cite{barkeshli2014symfrac,DumitrescuCordovaBrennanSymFrac2022,brennanJacobsonRoumpedakis2025SymFrac,KomargodskiHsinSymFrac2022} in
the IR between the $\Z_{q,w}^{(d-2)}$ and $\U(1)_m^{(d-3)}$ symmetries, which is
forced by anomaly matching
\cite{Hsin2025GenSymPhTr,HsinTurzillo2019SETQuantumSpinLiquids}.
This phenomenon can be described as follows (having in mind $d=4$).
The operators charged under $\U(1)_m^{(d-3)}$ are the `t Hooft lines describing
the monopoles of $a$, which are confined in the IR. The charged operators
of $\Z_{q,w}^{(d-2)}$ are the Wilson surfaces of $b$: inserting
$e^{i\int_\S b}$ in the action \eqref{BF theory with backgrounds}
yields
\begin{equation}
\label{ch2 b source of 1/q da}
\frac{\d a}{2\pi}= -\frac{1}{q} \de_\S, \qquad\quad \de_\S =
\text{Poincar\'e dual of $\S$}.
\end{equation}
Therefore, the Wilson surfaces of $b$ actually are spacetime
representations of monopoles lines of $a$ (magnetic flux lines) with
fractional charge $1/q$.  While genuine monopoles are confined ($d-3$
objects, i.e. spacetime lines), ``wrongly quantized'' monopole lines
($d-2$ surfaces) survive in the IR.  We see that the $b$-Wilson
surface inserts a codimension-2 magnetic defect carrying the minimal
flux $2\pi/q$. In $q=2$ superconductors, these are the
Abrikosov-Nielsen-Olesen vortex lines, carrying flux $\pi$.

\begin{table}
    \centering
\begin{tabulary}{\textwidth}{|L|L|L|L|L|}
\hline
\diagbox[width=3.5cm,height=2.2em]{\makecell{Phases}}{\makecell{Symmetry}} &
\makecell{$\U(1)_m^{(1)}$} &
\makecell{$\Z_{2,e}^{(1)}$} &
\makecell{IR emergent} &
\makecell{Notes} \\
\hline
\makecell{Coulomb phase} &
\makecell{SSB} &
\makecell{SSB} &
\makecell{$\Z_{2,e}^{(1)}$ enhanced \\to $\U(1)_e^{(1)}$} &
\makecell{Massless\\ NGB: photon} \\
\hline
\makecell{Superconductor\\ (Higgs phase)} &
\makecell{Unbroken + \\ Fractionalized} &
\makecell{SSB} &
\makecell{Winding \\ symmetry $\Z_2^{(2)}$} &
\makecell{Fractionalization:\\ $\wt C =\frac{\d B_m}{2\pi} \mod 2$} \\
\hline
\end{tabulary}
\caption{The table summarizes the relevant symmetries of the Abelian
  Higgs model and their realization in the two phases. Here the
  spacetime dimension is $d=4$ and the Higgs field has charge $q=2$
  (generalizations are given in the text).}
\label{fig table AHM}
\end{table}

\paragraph{Remark on $q=1$.}
Formally, the previous analysis can also be done in the $q=1$ case
of fundamental Higgs scalar, leading to
\eqref{IR Abelian higgs phase}. However, there is no mixed anomaly
\eqref{AHM mixed anomaly}, since the electric one-form symmetry
is completely broken, and the compact scalar can be removed by a gauge
transformation. One is left with a massive vector field
(unitary gauge) in a trivially gapped phase. Indeed, $q=1$ BF
theory  has no topological order and it is, in fact, the
trivial theory \cite{Witten2003SL2Z} (i.e. $Z=1$ on any closed
manifold). However, if some other global symmetry is present this
discussion could be enriched, in the sense that the $q=1$ Higgs phase
can support a SPT response. This point
of view has been developed in \cite{higgsSPTII2023,higgsSPTI2022}.

\paragraph{Historical note.} The fact that 
a $\U(1)$ gauge theory with charge $q$ condensate
reduces to a $\Z_q$ gauge theory in the IR was already noted in
\cite{KraussWilczek1989DiscreteGaugeSym}, where the authors also
commented on the presence of vortices with minimum flux $2\pi/q$ and
the consequent non-trivial Aharonov-Bohm (AB) phases for particles
that go around them. Ref.
\cite{ReznikAharonov1989ScSB}
is also interesting, where it is argued that the SC is
transparent to a `modular' part of the electromagnetic field (a SC
does not expel the EM field completely). This is because condensates
of charge $q$ do not screen a $\Z_q$ charge, which is an equivalent
way to say that inside the SC there is a $\Z_q$ gauge theory.\footnote{
  The argument is the following. Take a
  type II SC with a vortex line. A particle going around it gets an AB
  phase. Now we go to another reference frame in which the
  vortex goes around the particle. The result should be the same,
  with the AB phase produced by a dipole with magnetic moment that
  goes around in electric field. This `dual' AB
  effect is called the Aharonov-Casher effect. The presence of the electric
  field is necessary for consistency, so it cannot be completely
  screened and a $\Z_q$ part survives inside the SC.} The
low-energy quasiparticles do  carry a  $\Z_q$
charge, that is the remnant after screening.


\section{Gauging $\spinc$ symmetry and bosonization}
\label{Grp ext sec}

In the Introduction we said that electrodynamics of fermions obey
the spin-charge parity rule and
thus possess (at least) the $\spinc$ symmetry
\cite{seibergwitten2016gappedTI,wittenwebofduality},
\begin{equation}
\label{spinc group def}
    \spin_c(d) \simeq \frac{\spin(d)\times \U(1)}{\Z_2}.
\end{equation}  
In this case, the fermion parity symmetry $\Z_2^f \subset Z(\spin(d))$ is 
a subgroup of $\U(1)$. Therefore, a system with dynamical $\spinc$
symmetry automatically realizes the $\Z_2^f$ gauging, which
corresponds to the
Gaiotto-Kapustin-Thorngren bosonization in any dimension
\cite{gaiottokapustinspinTQFT1,gaiottokapustinspinTQFT2,thorngren2020anomalies,
  kapustinthorngren2017,CappelliVillaBosDual2025,tongArf}.
It follows that gauged fermionic systems are bosonic.

In this section, we discuss the general relation between dynamical
spin$_c$ symmetry and bosonization, and show that its characteristic
gravito-magnetic anomaly \eqref{an-b}, already observed in superconductors,
is actually present in all phases of gauged fermionic matter.

\subsection{Group extensions and symmetry fractionalization}

We start with an introduction of two phenomena which
occur in systems with generalized symmetries and are relevant
for the discussion in this section.

Consider a theory $\T$ having two symmetries, say $A$ and $G$,
such that the total symmetry is not simply a direct product
$A \times G$, but they fit together into a bigger group $E$ with
$E/A\simeq G$. In this case $A$ and $G$ are not really independent
symmetries (the background gauge transformations mix the two
groups). This situation is described by the following exact sequence
\begin{equation}
\label{grp ext}
A \to E \to G \simeq E/A \,.
\end{equation}
 A typical case is for $A$ Abelian in the center of $E$: this is
called a \textit{central extension} of $G$ by $A$. There is a
bijective map between such central extensions and $H^2(BG;A)$, with
$BG$ the classifying space of $G$.\footnote{$H^2(BG;A)$, the standard
  simplicial cohomology of $BG$, is called the group cohomology of
  $G$, $H^2_{grp}(G;A)$ \cite{dijkgraafwitten}.}

\begin{figure}
\begin{center}
\includegraphics[width=0.3\textwidth]{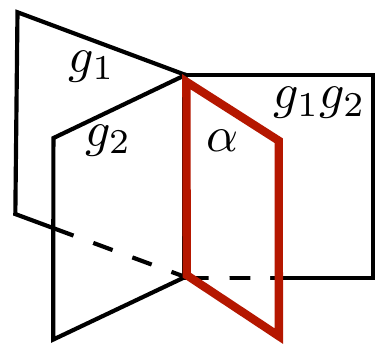}
\caption{In the fusion of two $G$ defects,
an $A$ defect is also emitted, as in \eqref{prodrule}.}
\label{defectfig}
\end{center}
\end{figure}

Concretely, a central extension like \eqref{grp ext}, given by
an element $[\a]\in H^2(BG;A)$, can be thought of as a twisted product
of $A$ and $G$: every element of $E$ can be written as a pair $(a,g)$,
$a\in A$, $g \in G$, with product rule
\begin{equation} \label{prodrule}
     (a_1,g_1) \cdot (a_2,g_2) = (a_1+a_2+\a(g_1,g_2),g_1\cdot g_2).
\end{equation}
As depicted in fig.\ \ref{defectfig}, when two symmetry defects of $G$
fuse, at their junction an $A$ symmetry defect $\a(g_1,g_2)$
is also emitted. Take $A$ a
finite group. The dual statement in terms of the background gauge
fields of $A$ (call it $\Ag$)\footnote{Note that the symbol A is used
  for the group in this context, whereas in the previous sections it
  was the gauge field, which is now denoted $\Ag$.} and $G$ (call it
$\Gg$) is \cite{tachikawaGaugeFiniteGroups,bhardwajTachikawa2017}
\begin{equation}
    \label{gauge field grp ext}
    \d \Ag = \Gg^*\a.
\end{equation}
It is convenient here to think about $\Gg$ as defining a principal $G$
bundle over $X$, which is specified by a map, which we call $\Gg$
again, $\Gg:X\to BG$ \cite{MilnorStasheff,FomenkoFuchs2016}, so that
$\Gg^*\a \in Z^2(X;A)$ is the pullback of $\a$. $\Gg$ is a source for
$\Ag$: the symmetry defects of $A$ can end on $G$-defects and thus the
gauge field of $A$ is no longer closed when $\Gg \neq 0$.

Another phenomenon that could arise is called \textit{symmetry
  fractionalization}
\cite{barkeshli2014symfrac,KomargodskiHsinSymFrac2022,DumitrescuCordovaBrennanSymFrac2022,brennanJacobsonRoumpedakis2025SymFrac,HsinShao2020}. Generally,
this refers to the fact that loop operators may carry a
projective representation of the (zero-form) symmetry group $G$,
classified by $H^2(BG;\U(1))$ (typically this is given by fractional
quantum numbers with respect to the ordinary representations of $G$,
hence the name). If there is a one-form symmetry $A^{(1)}$, the
junction of two symmetry defects of $G$ can be decorated by a symmetry
defect of $A^{(1)}$. This effect can be captured by an element in
$[\th]\in H^2(BG;A)$, which, similar to the discussion above leading
to \eqref{gauge field grp ext}, implies the following relation
between the background gauge fields
\begin{equation}
\label{sym frac gauge field}
    \Ag = \Gg^*\th.  
\end{equation}
If $W(\g)$ is charged under $A^{(1)}$, this shifts the action of $G$
on $W(\g)$ by an additional phase given by the $A^{(1)}$-symmetry
defect that appears in the junction, modifying its projective
representation.
An example of symmetry fractionalization was already encountered in the
previous discussion of the Higgs model \eqref{C=dB magnetic sym frac in IR SC}.\footnote{The fact that \eqref{C=dB magnetic sym frac in IR SC} is \eqref{sym frac gauge field} perhaps requires an explanation. In this case fractionalizations are classified by $\th \in H^{d-1}(B^{d-2}\U(1);\Z_q)$. Since $B\U(1)\simeq K(\Z,2)$, $B^{d-2}\U(1)\simeq K(\Z,d-1)$, whose universal cohomology class corresponds to the first Chern class of $B_m$, i.e. $\d B_m/2\pi$. Therefore $[\th] \in H^{d-1}(K(\Z,d-1),\Z_q) \simeq \hom(\Z,\Z_q)$, which is just the mod $q$ reduction of $\d B_m/2\pi$. This is indeed \eqref{C=dB magnetic sym frac in IR SC}.} 

The other interesting case is when $W(\g)$ is itself topological, so
it is a generator of some $(d-2)$-form symmetry $H^{(d-2)}$. The fact
that the $H^{(d-2)}$ symmetry defects carry a projective
representation of $G$ implies a mixed anomaly between the two
symmetries
\cite{tachikawaGaugeFiniteGroups,brennanJacobsonRoumpedakis2025SymFrac}. This
will be clear in the following, see \eqref{group extension
  anomaly}.

Consider the case $d=4$, where
also the magnetic symmetry is one-form and the 't Hooft operators
are lines. As discussed in the Introduction, we are interested in
extended defects carrying a projective representation
of $\SO(d)$. Usually the 
Wilson loop $W(\g)=\exp i\oint_\g a$ represents the worldline of a
charged bosonic particle (same with the 't Hooft operator $H(\g)$
for a monopole). However, this representation can be twisted by 
choosing a non-trivial fractionalization class \eqref{sym frac
  gauge field}, i.e. $[\th] \in H^2(B\SO;\Z_2)$,\footnote{Notice that
  for \eqref{sym frac gauge field}, one should consider
  $H^2(B\SO;\U(1))$, but this is actually induced by
  $H^2(B\SO;\Z_2)$. Accordingly, only a $\Z_2$ subgroup of either
  one-form symmetries plays a role here and this discussion applies
  always to bosonic theories (spacetime symmetry $\SO(d)$) with
  $\U(1)$ gauge group where at least a $\Z_2$ subgroup of the electric
  or magnetic symmetry survives.} such that
\begin{equation}
\label{sym frac Maxwell}
    B_e= \pi \Gg^* \th= \pi w_2(TX),\qquad B_m =\pi \Gg^*\th = \pi w_2(TX),
\end{equation}
where $\Gg: X \to B\SO$ is a background for the spacetime $\SO(d)$
symmetry and $w_2(TX)=w_2$ is the second Stiefel-Whitney class.
The first choice shifts the projective
representation class of the Wilson line $W$, so that now it represents
a fermionic particle, while the second choice makes the 't
Hooft line $H$ fermionic \cite{DumitrescuCordovaBrennanSymFrac2022,
  wittenNewSU2Anomaly2018, thorngren2014framed}. 

The fractionalization of the Wilson lines amounts to turning on
$B_e = \pi w_2$. This does not modify the local dynamics of $a$, but
the new gauge invariant connection $a_f$, such that
$\d a_f= \d a +\pi w_2$, has the fractional flux quantization
\eqref{spinc connection},
which we repeat here,
\begin{equation}
    \label{spinc connection 2}
    \int_\S \frac{\d a_f}{2\pi} + \frac{1}{2}w_2(TX) =n \in \Z,
\end{equation}
namely $a_f$ is a spin$_c$ connection. We conclude that
\eqref{spinc connection 2}
 is equivalent to the fractionalization choice for $B_e$
in \eqref{sym frac Maxwell}, i.e. $B_e = \pi w_2$, and
it implies a mixed gravitational anomaly
\begin{equation}
    \label{EM mixed grav anom spinc}
    \A_c =i\pi  \int_Y w_2 \cup \frac{\d B_m}{2\pi} .
\end{equation}
This is the specialization of the (bosonic) Higgs model anomaly \eqref{AHM mixed anomaly}
of the previous section, when $B_e$ is given the above spacetime property.

Another anomaly comes from
the group extension \eqref{grp ext}
\cite{tachikawaGaugeFiniteGroups,bhardwajTachikawa2017}.
Consider a theory $\T$ with symmetry group $E$ which is given by the
central extension \eqref{grp ext}, specified by $[\a] \in
H^2(BG;A)$. Take $A$ a finite Abelian group and $E$ as a whole is not
anomalous.
If $Z[\Ag]$ denotes the partition function of the original theory
$\T$ in presence of its background, gauging $A$ leads to
the partition function of $\T/A$ 
\begin{equation}
\label{general gauged theory discrete sym}
    \wh Z[\wh \Ag] = \sum_{[a]\in H^{p+1}(X;G)} Z[a] e^{2\pi i\int_X a \cup \wh A}.
\end{equation}
This theory possesses the dual symmetry
$\wh A^{(d-2)}$, so that its total symmetry is
$\wh A^{(d-2)}\times (E/A\simeq G)$. However, the Wilson lines of $a$,
i.e. the symmetry defects of $\wh A^{(d-2)}$, are not really
topological operators when $\Gg$, the background gauge field of $G$, is turned
on, because of \eqref{gauge field grp ext}. This implies a mixed
anomaly between $\wh A^{(d-2)}$ and $G$ which can be seen directly at
the level of partition function \eqref{general gauged theory discrete
  sym}. After a background gauge transformation
$\wh \Ag\to \wh \Ag + \d \l$, $\wh Z$ is not invariant since
$\d a = \Gg^* \a \neq 0$, so
\begin{equation}
  \wh Z[\wh \Ag + \d \l,\Gg]= e^{2\pi i \int_X \Gg^*\a \cup \l}
  \wh Z[\wh \Ag,\Gg].
\end{equation}
By inflow, this is described by the following $d+1$ dimensional anomaly term
\begin{equation}
\label{group extension anomaly}
 {\cal A}=   2\pi i\int_Y \Gg^*\a \cup \wh \Ag \,.
\end{equation}
This prevents gauging
$\wh A^{(d-2)}$ in general, but it actually trivializes when $G$ is
extended by the dual symmetry $A$. In this case, there
exists $\Ag$ such that $\d \Ag = \Gg^*\a$ and thus the anomaly can be
removed by a local counterterm $\Ag \cup \wh \Ag$, which allows to
gauge back $\wh A^{(d-2)}$ and $\T/A/\wh A^{(d-2)} \simeq
\T$. Basically \eqref{group extension anomaly} tells that when
$\Gg \neq 0$, it is not possible to gauge $\wh A^{(d-2)}$ without
turning on a background gauge field for the dual symmetry $A$,
according to \eqref{gauge field grp ext}.

The anomaly \eqref{group extension anomaly} is the one mentioned in
the discussion about symmetry fractionalization of a topological line,
below \eqref{sym frac gauge field}. For finite Abelian $A$,
$\wh A \simeq A$, so the dual symmetry is really $A^{(d-2)}$. The
condition \eqref{gauge field grp ext} implies that, after gauging, the
topological lines of $A^{(d-2)}$ should be attached to a surface
operator $ e^{i\int_\S \Gg^*\a}$, which is an anomaly inflow for the
theory on the line, where the element $[\a] \in H^2(BG;A)$ can be
thought of as specifying a projective representation of $G$ (with some abuse of language \cite{brennanJacobsonRoumpedakis2025SymFrac, Villa2026Extensions}). This says
that the topological lines of $A^{(d-2)}$ carry a projective
representation of $G$, given by $\a$, and the mixed anomaly
\eqref{group extension anomaly} reflects this fact.

\subsection{Gauging fermion parity $\Z_2^f=(-1)^F$}
\label{bosonization sec}

The fermion parity symmetry $\Z_2^f$ is a symmetry of all
(relativistic) fermionic theories $\T_f$. It arises from the
requirement of extending the local Lorentz (Euclidean) group $\SO(d)$
to $\spin(d)$ in order to define fermions on a
manifold $X$. As a result, we have the following sequence
\begin{equation}
\label{fermion parity extension}
\Z_2^f \to \spin(d)\to\SO(d) \simeq \frac{\spin(d)}{\Z^f_2}\,,
\qquad\qquad [\a]\in H^2(B\SO; \Z_2),
\end{equation} 
which shows that the basic  fermionic symmetry can
be decomposed in a triple $(\SO(d),\Z_2^f, \a)$, where $\SO(d)$ is the
bosonic part. According to \eqref{grp ext}, the
background gauge field $\eta$ for $\Z_2^f$ satisfies
\begin{equation}
\label{gauge field of Z2f is spin str}
    \d \eta = \omega^* \alpha = w_2(TX), 
\end{equation}
where $\omega$ is a background gauge field for $\SO(d)$, i.e. the spin
connection. From \eqref{gauge field of Z2f is spin str} we see that
the gauge field $\eta$ for $\Z_2^f$ is the spin structure of $X$ itself
\cite{gaiottokapustinspinTQFT1,thorngren2014framed}.

Gauging $\Z_2^f$ means summing over the spin structures in
$\T_f$.\footnote{See also
  \cite{backfiringbosonization2024,PBSmithDavighi202BosCoho} for a
  slightly different perspective.}
As a result, the final theory $\T_b = \T_f/\Z_2^f$ is bosonic.
This is the  bosonization map\footnote{For a review of this
subject and details on the following expressions,
see section 2 and appendix A of \cite{CappelliVillaBosDual2025}.
Further aspects will be discussed in section 4.}
\cite{gaiottokapustinspinTQFT1,gaiottokapustinspinTQFT2,thorngren2020anomalies,
  kapustinthorngren2017} introduced in section 1. The bosonic theory
$\T_b$ is characterized by the anomaly \eqref{group extension
  anomaly}, which in this case is a mixed gravitational anomaly with
the symmetry $\Z_2^{(d-2)}$, dual of $\Z^f_2$, with background\footnote{
  Here $C$ has the role of $\wh A$
  in \eqref{group extension anomaly},} $[C]\in H^{d-1}(X; \Z_2)$,
\begin{equation}
\label{standard bos anom}
\A_b = i\pi \int_Y w_2(TY) \cup C =i\pi \int_Y Sq^2C =
i \pi \int_Y C \cup_{d-3} C\,.
\end{equation}
We remark that the $\Z_2^{(d-2)}$ symmetry for which $C$ is the background
 is not a symmetry of the original fermionic theory,
 but appears only after gauging fermion parity.
 
The anomaly \eqref{standard bos anom}
has two forms: the first was already presented in
\eqref{an-c}, while the second (middle formula) involves the Steenrod
square $Sq^2$ of $C$ \cite{Steenrod1947}, then explicitly written in
terms of higher cup products\footnote{Note that $\cup_0\equiv \cup$,
  the algebraic rules for the needed $\Z_2$ case are given in appendix A.}
\cite{Steenrod1947,kapustinseiberg2014}. The equality between the two anomaly
expressions follows from the Wu formula
\cite{MilnorStasheff,kapustinthorngren2017}. This anomaly
trivializes when $\Z_2^f$ is part of a group extension satisfying
\eqref{gauge field of Z2f is spin str}, which means that $X$ (and $Y$)
is a spin manifold. In this case, it is possible to introduce a local
counterterm $\eta \cup C$ that allows to gauge $C$ and recover
$\T_f \simeq \T_b/\Z^{(d-2)}$, as discussed in the
Introduction.\footnote{Still, \eqref{standard bos anom} is a true
  anomaly for a bosonic theory, which does not need, and it is not
  given, a dependence on the spin structure.}
Note that, in general this bosonization anomaly is distinct from the 
$\Z^{(1)}_q\times\Z^{(d-2)}_q$ mixed anomaly in the Higgs model
\eqref{AHM mixed anomaly}.

The anomaly \eqref{standard bos anom} can also be understood
from the defect perspective mentioned before. After gauging $\Z_2^f$,
its Wilson line carries a surface dependence given by
$w_2(TX)=\w^*\a$, where $[\a]\in H^2(B\SO;\Z_2)$ specifies a
projective representation of $\SO(d)$. Therefore, the topological line
generating the dual symmetry $\Z_2^{(d-2)}$ is a fermion.
The anomaly \eqref{standard bos anom} reflects
the fact that its generator is a fermionic line.

This discussion can be generalized in presence of other
symmetries. The fermionic symmetry data are given by a triple
$(G_b,\Z_2^f,\a)$ characterizing the extension
\begin{equation}
\label{generic fermionic ext}
    \Z_2^f \to G_f\to G_b, \qquad [\a]\in H^2(BG_b;\Z_2),
\end{equation}
where $G_b$ is a generic bosonic symmetry that mixes the spacetime
Lorentz part with other internal symmetries. This is analogous to
\eqref{fermion parity extension}. When the bosonic group has the form
$\SO(d)\times G_b$, i.e. there is no mixing between internal and
spacetime symmetries and we neglect time reversal, there is the
decomposition (using $H^1(B\SO;\Z_2)=0$)
\begin{equation}
    H^2(B(\SO \times G_b);\Z_2)\simeq H^2(B\SO;\Z_2)\oplus H^2(BG_b;\Z_2),
\end{equation}
which implies that $\a = w_2(B\SO)+w_2(BG_b)$ and thus
\begin{equation}
\label{spinG structure}
    \d \eta = w_2(TX) + \G^*w_2(BG_b),
\end{equation}
where $\eta$ is the $\Z_2^f$ gauge field and $\G$ is the 
gauge field for $G_b$. In this case, $\eta$ corresponds to a
spin$_{G}$ structure on $X$
\cite{avisisham1980spinG,TDBrennan2023SpinG,TDBrennanIntriligator2023SpinG,
  wittenNewSU2Anomaly2018,thorngren2020anomalies}
(see also appendix A.4 of \cite{CappelliVillaBosDual2025}). After
gauging $\Z_2^f$, the bosonic theory has the anomaly
\begin{equation}
\label{bos anom with G_b}
    \A_b = i\pi \int_Y (w_2(TY) + \G^*w_2(G_b)) \cup C,
\end{equation}
which generalizes \eqref{standard bos anom}.

\subsection{Gauging the spin$_c$ symmetry}
\label{U(1) extensions sec}

In this section we show that all fermionic systems
with a dynamical $\spinc$ connection in $d \ge 3$ are characterized by  a specific gravito-magnetic anomaly.
We already found this anomaly in the Higgs phase,
see Eq. \eqref{an-b}, using the argument on fermionic Wilson lines.
The same argument has been reformulated
in the context of fractionalization in section 3.1, see
Eq. \eqref{EM mixed grav anom spinc}.

In the following, we provide a general proof of its validity and
clarify its origin from bosonization.
This is a rather relevant result because it forbids the existence
of a trivially gapped phase in gauged electronic systems.\footnote{ See also
\cite{TDBrennan2023SpinG,TDBrennanIntriligator2023SpinG} for similar
discussions.}

\paragraph{Direct argument.} A simple general argument \cite{higgsSPTII2023}
can be formulated by considering
the magnetic symmetry $\U(1)^{(d-3)}_m$, which is present in any Abelian gauge
theory (without magnetic charges).
We consider the coupling to the magnetic background 
$\int_X \d a \,B_m/2\pi$ and test whether it
is invariant under large gauge transformations. To this effect, we
extend it to $d+1$ dimensions
\begin{equation}
  \label{ext magnetic coupling}
  \frac{i}{2\pi} \int_X \d a\,  B_m =
  2\pi i \int_Y  \frac{\d a}{2\pi}   \frac{\d B_m}{2\pi},
\qquad\qquad \p Y = X.
\end{equation}
The difference between two such extensions is given by the r.h.s.
expression evaluated for closed $Y$. If this is a trivial phase,
the extensions are equivalent,
thus making \eqref{ext magnetic coupling} globally well-defined.
Actually, this is not true for a $\spinc$ connection $a$,
owing to the fractional fluxes \eqref{spinc connection 2}.
The problem can can be cured by adding the following $d+1$ term
\begin{equation}
  \label{spinc magnetic coupling}
2\pi i \int_Y  \frac{\d a}{2\pi}   \frac{\d B_m}{2\pi} \ \ \to \ \ 
  2\pi i \int_Y\left(  \frac{\d a}{2\pi} +\frac{1}{2}w_2\right)
  \frac{\d B_m}{2\pi} \,.
\end{equation}
We conclude that if $Y$ is a spin$_c$ manifold, on which $a$ is
extended as a spin$_c$ connection,\footnote{ It is not always the case
  that such $Y$ exists in 4d. A proper way to define the coupling can
  be achieved using differential cohomology
  \cite{tachikawa2020pformanom}. This also gives the anomaly
  \eqref{anom for gauge spinc} \cite{Villa2026Extensions}.}  the extra
$d+1$ term restores gauge invariance of the original theory on $X$.
Using inflow from $Y$ to $X$, this term represents an anomaly given by
the SPT phase
\begin{equation}
\label{anom for gauge spinc}
\A_c = i\pi \int_Y w_2(TY) \cup \frac{\d B_m}{2\pi} =
i\pi \int_Y \frac{\d B_m}{2\pi} \cup_{d-3} \frac{\d B_m}{2\pi} .
\end{equation}
Therefore, this argument shows that the anomaly \eqref{an-b} is
present in any theory with a dynamical spin$_c$ connection and
magnetic symmetry $\U(1)^{(d-3)}_m$.

\paragraph{Two-step gauging.} We now prove that the anomaly comes
from the bosonization procedure underlying
dynamical $\spinc$ symmetry. We rewrite the symmetry \eqref{spinc group def}
as an extension \eqref{generic fermionic ext} for the bosonic group
$\SO(d) \times \U(1)_b$:
\begin{equation}
\label{spinc group ext}
\qquad \Z_2^f \to \spin_c(d)  \to \SO(d) \times \U(1)_b,
\qquad\quad \spin_c(d) \simeq \frac{\spin(d)\times\U(1)_f}{\Z_2} \,.
\end{equation}
We add the subscript to distinguish between the two
$\U(1)$ groups, whose only difference is the charge unit.
The content of \eqref{spinc group ext} is
that also $\U(1)_b$ is non-trivially extended by $\Z_2^f$,
\begin{equation}
\label{spinc u1 extension}
    \Z_2^f\to \U(1)_f \to \U(1)_b \,.
  \end{equation}
In this case, the gauge field $\eta$ for $\Z_2^f$ satisfies, as in
\eqref{spinG structure},
\begin{equation}
\label{spinc structure}
    \d \eta= w_2(TX) + c_1(A_b)|_2 = w_2(TX) + 2 c_1(A_f)|_2,
\end{equation}
and it is called a $\spinc$ structure on $X$. This says that $A_f$
(later simply $A$) is a spin$_c$ connection with
flux quantization \eqref{spinc connection 2}.
In Eq.\eqref{spinc structure}, the first Chern class  $c_1(A)= \d A/2\pi$
is taken modulo $2$. Note the factor of two between the two
expression for $A_b$ and $A_f$.

As shown in \cite{Villa2026Extensions}, gauging $\U(1)_f$ in
\eqref{spinc u1 extension} can be understood as a two steps process in
which the $\Z_2^f$ symmetry is first gauged producing the bosonic theory 
$\T_b\simeq \T_f/\Z_2^f$, and then the remaining $\U(1)_b$
group follows, such that $\T_f/\U(1)_f\simeq \T_f/\Z_2^f/\U(1)_b$. The theory $\T_b$ has 
symmetry $\SO(d)\times \U(1)_b\times \Z_2^{(d-2)}$ and mixed
anomaly \eqref{bos anom with G_b}
\begin{equation}
\label{bos spinc anomaly}
    \A_b = i\pi \int_Y (w_2(TY) + c_1(A_b)|_2) \cup C\,.
  \end{equation}
The anomaly term for $A_b$ does not prevent us from gauging $\U(1)_b$ as well
(and indeed the total group \eqref{spinc u1 extension} is anomaly free
by assumption), but, analogous to \eqref{group extension anomaly} in
the opposite direction, it implies that, after gauging $\U(1)_b$, the
dual symmetry is part of a non-trivial extension. More precisely, if
we denote by $\U(1)_{m,b}^{(d-3)}$ the dual magnetic symmetry of
$\U(1)_b$, the total symmetry after also gauging $\U(1)_b$ is
\cite{Villa2026Extensions}
\begin{equation}
\label{magnetic extension}
    \U(1)_{m,b}^{(d-3)}\to \U(1)_{m}^{(d-3)}\to \Z_2^{(d-2)},
\end{equation}
reconstructing the full magnetic symmetry $\U(1)_m^{(d-3)}$ of
$\T/\U(1)_f$. Concretely, the content of \eqref{magnetic extension}
can be encoded in the following relation between the backgrounds:
\begin{equation}
\label{Bm = Bmtilde + C}
    c_1(B_m) = 2c_1(B_{m,b}) - \overline C,
\end{equation}
where $B_{m,b}$ is the background for $\U(1)_{m,b}^{(d-3)}$,
$c_1(B_m)=\d B_m/2\pi$ is the generalized Chern class for the higher form
$B_m$ and $\overline C$ is an integral lift $C$.

The interpretation of \eqref{Bm = Bmtilde + C} is as follows.
The two backgrounds
are locally proportional, $B_m=2 B_{m,b}$, but the $B_m$ fluxes
are generically not even.
The $\Z_2$ background $C$ takes care of its additional fluxes.
As a consequence, 
\begin{equation}
\label{dBm=Cmod2}
c_1(B_m)  = \frac{\d B_m}{2\pi}=C \mod 2 \,.
  \end{equation}
Therefore, the bosonic symmetry $\Z_2^{(d-2)}$ is part of the
topological data of the magnetic symmetry $\U(1)_m^{(d-3)}$ of $\U(1)_f$.
As discussed, the mixed anomaly term involving $A_b$ in \eqref{bos spinc
  anomaly} is what gives the extension \eqref{magnetic extension} in
the final theory, while the gravitational term $w_2$ 
remains in all steps of the process.
Finally, the general form of the gravito-magnetic anomaly is
given by the expression \eqref{anom for gauge spinc}, which is the
bosonization form \eqref{standard bos anom}
with the identification \eqref{dBm=Cmod2} between $\Z_2^{(d-2)}$
and $\U(1)_m^{(d-3)}$ backgrounds.
The direct and two-step gaugings are summarized
in Figure \ref{fig spinc gauging}.

In conclusion, we have shown that the anomaly \eqref{anom for gauge spinc}
of gauged fermionic systems in $d\ge 4$ is valid
in great generality, and follows from the bosonization
included in gauging $\spinc$ symmetry. The $d=3$ case is more subtle
and it is discussed in the next section.

\begin{figure}
    \centering
    \begin{tikzpicture}[
      >=Latex,
    ]
    
    \node (tl) at (0,7) {$\begin{aligned}
        &\left(\T_f,\, \spin_c(d)\simeq
          \frac{\spin(d)\times\U(1)_f}{\Z_2}\right)\qquad\\
        &\qquad\Z_2^f \to \U(1)_f \to \U(1)_b
      \end{aligned}$};
    \node (tr) at (9,7) {$\begin{aligned}
        (\T_f/\U(1)_f,\, & \SO(d)\times \U(1)_m^{(d-3)})\\
        \qquad\U(1)_{m,b}^{(d-3)} \to & \U(1)_{m}^{(d-3)}\to \Z_2^{(d-2)}\\
        \A_c = i\pi& \int_Y w_2\cup \frac{\d B_m}{2\pi}
      \end{aligned}$};

    \node (c)  at (4.5,4) {$\begin{aligned}
        (\T_b\simeq \T_f/\Z_2^f,\, &\SO(d)\times \U(1)_b\times\Z_2^{(d-2)})\\
        \A_b = i\pi \int_Y& (w_2+2c_1(A_f)) \cup C
      \end{aligned}$};
    
    \draw[->, thick] (tl) -- node[above] {$/U(1)_f$} (tr);

    \draw[->, thick] (tl) -- node[midway, sloped, above] {$/\Z_2^f$} (c);
    \draw[->, thick] (c) -- node[midway, sloped, above] {$/\U(1)_b$} (tr);

    \end{tikzpicture}
    \caption{Gauging $\U(1)_f$ in \eqref{spinc group ext} and
      \eqref{spinc u1 extension} in one and two steps.  Direct gauging
      is the top arrow, the two-step gauging is the two-arrow lower
      path, where one first gauges $\Z_2^f$ (bosonization) and then
      gauges the residual bosonic $\U(1)_b$, recovering the same final
      theory and gravito-magnetic anomaly.}
    \label{fig spinc gauging}
\end{figure}

\subsection{Further aspects of the gravito-magnetic
    anomaly}

In this section we discuss the anomaly \eqref{anom for gauge spinc}, which we rewrite here for convenience, using the notation $\d B_m/2\pi =c_1(B_m)\coloneqq c_1$ and $c_1|_2$ for its mod 2 reduction:
\begin{equation}
     \A_c = i\pi \int_Y w_2 \cup c_1|_2 = i\pi \int_Y c_1|_2 \cup_{d-3} c_1|_2.
\end{equation}
Generically, in $d$-dimensions, this is a bosonic mixed global anomaly between spacetime and a $\U(1)^{(d-3)}$ symmetry, classified by the \textit{torsion} part of the oriented bordism group $\W_{d+1}^{\SO}(K(\Z,d-1))$, with $K(\Z,d-1) \simeq B^{d-2}\U(1)$ the classifying space for $\U(1)^{(d-3)}$, $K(\Z,n)$ being an Eilenberg-MacLane space of type $n$,\footnote{$K(\Z,n)$ is a space with $\pi_n=\Z$ and $\pi_i=0$ for $i\neq n$.} according to the cobordism conjecture \cite{freed2014AnomAndInvFTs,kapustinBordism,kapustinbordismTI,kapustinFermionicBordism,yonekura2018CobordismSPT}.  As we will see, this is indeed an anomaly for a bosonic theory in $d=4$, but this is not the case in $d=3$ unless time reversal is considered.

In $d=3$, the $\U(1)$ symmetry involved is a standard zero-form symmetry and the relevant bordism group is $\W_{4}^{\SO}(B\U(1)) \simeq \Z \oplus \Z$. It is easy to understand the two $\Z$ factors: they represent the familiar purely gravitational and $\U(1)$ perturbative anomalies of a two-dimensional theory with anomaly polynomials 
\begin{equation}
    \frac{1}{3}p_1 \sim \Tr(R\wedge R), \qquad c_1 \cup c_1 \sim F\wedge F,
\end{equation}
which detect the two generators of $\W_{4}^{\SO}(B\U(1))$. $p_1$ is the first Pontryagin class and $R$ the curvature form \cite{nakahara,MilnorStasheff}. Notice that there is no torsion part and indeed $c_1|_2\cup c_1|_2$ is just the mod 2 reduction of the familiar perturbative anomaly. This says that \eqref{anom for gauge spinc} is not a global anomaly for the three-dimensional theory. Concretely, this can be seen from the fact that it can be removed by a local Chern-Simons (CS) counterterm for $B_m$. Indeed, the coupling 
\begin{equation}
\label{CS ct B}
    \int_X \frac{1}{2\pi}a \d B_m  + \frac{1}{4\pi} B_m\d B_m
\end{equation}
is well-defined on $X$ \cite{wittenwebofduality,seibergwitten2016gappedTI}, allowing to remove the apparently anomalous variation. One could wander that the level 1 CS term for $B_m$ should not be allowed in a bosonic theory; however, the composition of the two terms as \eqref{CS ct B}, for a $\U(1)$ gauge field $B$ and a spin$_c$ connection $A$, requires only a spin$_c$ structure to be defined \cite{seibergwitten2016gappedTI}. Since $A\to a$ is gauged in \eqref{CS ct B}, the resulting coupling is bosonic.

The counterterm \eqref{CS ct B} manifestly breaks time reversal symmetry. This is fine, since we never assumed it in the first place and we were in fact considering the oriented bordism group. We can however wonder what happens if we enforce also a time reversal symmetry $\TR$, that rules out $\TR$-breaking terms like \eqref{CS ct B}. In this latter case the global anomalies are classified by $\W_{4}^{\rm O}(B\U(1))$. We can guess the possible classes detecting its generators, namely
\begin{equation}
    w_1^4, \qquad w_2^2, \qquad w_1^2\cup c_1|_2, \qquad w_2 \cup c_1|_2,
\end{equation}
so that $\W_{4}^{\rm O}(B\U(1)) \simeq \Z_2^4$. The first two classes are the usual purely gravitational anomalies of a bosonic theory in $3d$, given $\W_{4}^{\rm O}(pt)$ (notice that $w_2^2 =p_1$ mod 2, so that $w_2^2$ is the remnant of $p_1$ restricted to unoriented manifolds). The last one is our anomaly \eqref{anom for gauge spinc}, showing that, if we rule out $\TR$-breaking terms, \eqref{anom for gauge spinc} is indeed an anomaly. Because $w_1\neq 0$, 
\begin{equation}
    w_2\cup c_1|_2 = c_1|_2 \cup c_1|_2 + w_1^2 \cup c_1|_2.
\end{equation}
The extra term with $w_1$ is expected if we reformulate our previous discussion leading us to \eqref{anom for gauge spinc} starting from Pin$_c$ manifolds \cite{CappelliVillaBosDual2025,thorngren2020anomalies} (taking O$(d)$ instead of $\SO(d)$ in \eqref{spinc group ext}). If we restrict to orientable manifolds, $w_1=0$, we recover \eqref{anom for gauge spinc}. The resulting anomaly can be indeed represented by inflow by a $\th =\pi$ angle for the $B_m$ field, 
\begin{equation}
    \pi \frac{\d B_m}{2\pi} \wedge \frac{\d B_m}{2\pi},
\end{equation}
which is the familiar three-dimensional mixed $\TR-\U(1)$ anomaly \cite{redlich1984gauge,redlich1984parity,witten2016fermion} suitably normalized for a bosonic theory. 

Consider now the case $d = 4$, with the relevant bordism group $\W_{5}^{\SO}(B^2\U(1))$. It is easy to see that it is not possible to construct integer invariants using $R$ and $c_1$ (of degree 3) by dimensional reason. We can use instead the $\Z_2$ invariants  
\begin{equation}
    w_2w_3, \qquad w_2\cup c_1|_2,
\end{equation}
showing that $\Omega_5^{\SO}(B^2\U(1)) \simeq \mathbb{Z}_2^2$ (at least). The first term is the famous gravitational anomaly of all-fermion electrodynamics \cite{kapustinBordism,thorngren2014framed,DumitrescuCordovaBrennanSymFrac2022,wangwenwitten2017}, while the second one is our anomaly \eqref{anom for gauge spinc}, which is therefore a fully-fledged anomaly in $d=4$. These anomalies are still there also when time reversal is considered, together with the other possible invariant $c_1|_2 \cup w_1^2$, showing that $\Omega_5^{\rm O}(B^2\U(1)) \simeq \mathbb{Z}_2^3$ (at least). A similar discussion applies to $d>4$, where $w_2\cup c_1|_2$ cannot be encoded in a integer invariant by dimensional reasons.\footnote{Notice the following interesting fact. In principle, our anomaly \eqref{anom for gauge spinc} is always equivalent to $c_1\cup_{d-3}c_1 $ mod 2, which can then be always derived by the integer valued object $c_1\cup_{d-3}c_1$, as we in fact do in $d=3$. However, using the properties of the cup products \cite{kapustinseiberg2014,kapustinthorngren2017}, one can show that $\d (c_1\cup_{d-3}c_1 )=(-1)^{d-1} 2(c_1\cup_{d-4}c_1)$. So only for $d=3$ we obtain an integer cohomology class $c_1\cup c_1$ out of $c_1$. For $d\geq 4$, $c_1\cup_{d-3}c_1$ is always closed only mod 2, defining a mod 2 cohomology class out of $c_1$. This is our $w_2\cup c_1$.}

The bottom line of this discussion is that the anomaly \eqref{anom for gauge spinc} is always there for $d\geq 4$. For $d=3$, it is actually a mixed anomaly with time reversal. In the following section, we use anomaly matching to constrain gapped phases of electronic systems in both $d=4$ and $d=3$. In this latter case, an unbroken $\TR$ symmetry of the UV microscopic theory must be assumed to rely on \eqref{anom for gauge spinc}.\\

\noindent\textit{Observation.} As reviewed in the introduction and in section \ref{bosonization sec}, in a modern bosonization approach, after gauging $\Z_2^f$ the bosonic theory has a particular anomaly for the dual symmetry $\Z_2^{(d-2)}$, namely \eqref{standard bos anom}, that trivializes on spin manifolds. On such manifolds the anomaly can be removed by a local counterterm involving the spin structure, allowing to gauge back $\Z_2^{(d-2)}$ to recover the fermionic theory. Let us make this explicit for the spin$_c$ case.

For a fermionic theory with a spin$_c$ structure \eqref{spinc group def}, we can gauge the $\U(1)$ symmetry to get a bosonic theory with anomaly \eqref{anom for gauge spinc}, i.e. $w_2\cup c_1|_2$, which detects a $\Z_2$ part of $\W_{d+1}^{\rm SO}(B^{d-2}\U(1))$. This trivializes obviously on spin manifolds, where $w_2 =\d \eta$, but also on spin$_c$ manifolds: in such cases, by definition, $w_2$ is the reduction mod 2 of an integer class $\l\in H^2(X;\Z)$ and therefore $w_2\cup c_1|_2$ is actually the reduction mod 2 of the integer invariant $\l\cup c_1$. As discussed above, this implies that the torsion part $\Z_2\subset \W_{d+1}^{\rm SO}(B^{d-2}\U(1))$ actually comes from a free part $\Z \subset \W_{d+1}^{\spin_c}(B^{d-2}\U(1))$, under the map $\W_{d+1}^{\spin_c}(B^{d-2}\U(1))\to \W_{d+1}^{\SO}(B^{d-2}\U(1))$ obtained by simply forgetting the spin$_c$ structure. This implies that the anomaly trivializes and, in fact, it can be removed by a local counterterm involving a background spin$_c$ structure $A_0$ (such that $2c_1(A_0)=\l$). This counterterm is simply $A_0\d B_m/2\pi$. Adding this term allows to gauge back $\U(1)_m^{(d-3)}$ to recover the original fermionic theory.

\section{Properties of systems with
  dynamical spin$_c$ connection}

In this section, we discuss some technical and physical aspects of
matter with gauged $\spinc$ symmetry, in particular superconductors.
We shall find how to reconstruct the topological theory in $d\ge 3$
from the knowledge of the bosonization anomaly \eqref{anom for gauge
  spinc} without relying on the existence of the Abelian Higgs model.
From this perspective, the coupling $\int b\cup w_2$ will naturally
emerge by anomaly matching.  This will allow us to infer some general
consequences for more general gapped superconductors as, for example,
superconductors with $q>2$ pairing and topological superconductors in
$2+1$ dimensions.

In $d=3$ it is assumed also a microscopic time reversal symmetry to
ensure that \eqref{anom for gauge spinc} is not trivial. However, we
do not claim that time reversal symmetry is essential for
superconductors \cite{TRbreakingOnSC2025}, so it can be broken in the
IR.  We will also make a brief comment on three-dimensional QED$_3$.
The anomaly \eqref{anom for gauge spinc} is not there for odd number
$N_f$ of flavours, since time reversal is broken, but it appears in
QED$_3$ with $N_f$ even, as the $N_f=2$ case discussed in
\cite{DumitrescuNiroThorngren2024QED3}.
 
\subsection{Minimal topological theory in the gapped phase}

In the first two sections of this paper we analyzed the low-energy
limit of s-wave superconductors, obtained the BF topological theory and
its gravito-magnetic anomaly \eqref{an-b}.  In the previous section,
we showed that this is a general consequence of gauging the $\spinc$
symmetry and it originates from bosonization (cf. \eqref{standard
  bos anom}, \eqref{anom for gauge spinc}).  We now consider general
gapped phases, find the minimal topological theory compatible with the
anomaly and compare with the earlier Higgs case.

\subsubsection{$\Z_2^f$ gauge theory}

Consider a system with a dynamical spin$_c$ connection, i.e. with
symmetry group \eqref{spinc group def} and gauged $\U(1)_f$.
If a suitable Higgs field is present, the system can be
driven to a Higgs phase at low enough energies. This phase
should match the anomaly \eqref{anom for gauge
  spinc}. In general, owing to the spin-charge relation implied by
\eqref{spinc group def}, the bosonic Higgs field
has even charge $q=2n$. It follows that $\U(1)_f$ can be
higgsed  to even cyclic subgroups $\Z_{2n}$. The minimal option is
$\Z_2 =\Z_2^f$ and in general the $\Z_2^f$ gauge theory,
$\Z_2^f\subset \Z_{2n}$,
is always present at low energy.\footnote{
  In this work, superconductors are always 
  intended as the Higgs phase of an $\U(1)$ gauge theory. Depending on
  the matter sector, there could be also massless excitations. What we
  say here applies only to gapped SCs, but one could argue that
  gapless SCs will still have the same topological part.
  However, in that case the massless degrees of freedom could realize
  the anomaly \eqref{anom for gauge spinc} by themselves, so anomaly
  matching alone does not constrain the topological sector.}

In the simplest case of an s-wave superconductor,
$\U(1)_f$ is higgsed to $\Z_2^f$, and the minimal low-energy
theory with anomaly \eqref{anom for
  gauge spinc} was already guessed in \eqref{BF-w2}.
We rewrite it here in discrete form,
\begin{equation}
\label{SC TQFT with w2 and Bm}
S[B_m] =i\pi\int_X b \cup (\d a+ w_2(TX)) +
\left[\frac{\d B_m}{2\pi}\right]_2\cup a,
  \end{equation}
  knowing that U(1) forms $a,b$ map into $\Z_2$ cochains\footnote{
    Omitting tildes in the following expressions involving cup
    products.}  by $a = \pi \wt a$, $b = \pi \wt b$.  Eq. \eqref{SC
    TQFT with w2 and Bm} is clearly the level-2 BF theory \eqref{BF
    theory with backgrounds} with fractionalization choice $B=w_2$ and
  $C=\d B_m/2\pi$.  In particular, the choice $B=w_2(TX)$ is what
  makes the Wilson line of $a$ fermionic. 

Let us analyze this action for $B_m=0$
\begin{equation}
\label{SC TQFT with w2}
    S_{\mathbb Z_2^f}=i\pi\int_X b \cup (\d a+ w_2(TX)).
  \end{equation}
This can be viewed as the minimal $\Z_2^f$ gauge theory,
  namely a $\Z_2$ gauge theory for the spin structures (see appendix A for
  further discussion).
It is just the ordinary $\Z_2$ BF theory with the insertion
of a $b$ defect,
\be
\label{}
\int_X b\cup w_2 =\int_\S b, \qquad\quad \S={\rm PD}(w_2)\,, \ee which
is supported on $\S$, the Poincar\'e dual of $w_2$.  A typical example
of non-spin $3+1d$ manifold is $\mathbb{CP}^2$, in which $w_2$ is
non-trivial on submanifolds $\mathbb{CP}^1\subset \mathbb{CP}^2$,
$\int_{\mathbb{CP}^1} w_2=1$. It follows that superconductors on
$\mathbb{CP}^2$ have a persistent $\pi$-flux line in the ground state,
in absence of external magnetic fields
\begin{equation}
  \label{persistent}
  \left\langle\exp\left(i\pi\int_{\S} b\right)
  \right\rangle \neq 0. \qquad 
  \quad \S={\rm PD}(w_2) 
\end{equation}
Since $\mathbb{CP}^1={\rm PD}(b)$, 
the non-trivial value $1=\int_X b\cup w_2 = {\rm Int}(\S,\mathbb{CP}^1) $
tells that the surface $\S$ intersects $\mathbb{CP}^1$ once.

The direct observation of this ground-state flux is clearly not possible
in a laboratory: a more mundane possibility would be to consider
a tabletop geometry which realizes the same topology with the help
of suitable boundary conditions and circuits, in the same spirit of
Ref.\cite{Kitaev:flux}. 
We remark that the existence of this defect implies that
the Higgs field vanishes on it, i.e. the Higgs phase does not
extend everywhere in the ground state (more on this later).

While the ordinary BF theory has the $\Z_2^{(1)}\times\Z_2^{(d-2)}$ symmetry,
the $\Z_2^{(d-2)}$ part is broken in the action \eqref{SC TQFT with w2} by
the presence of non-trivial $w_2$ (or defect) on 
non-spin manifolds. Indeed, the variation
of the action under $b \to b+\g$ gives $\D S =i\pi\int_X \g w_2 $,
which is the bosonization anomaly \eqref{standard bos anom}.

Finally, we can explicitly show the relation
between anomalous bosonic theory and fermionic theory under the
bosonization map (see also appendix A).
Adding the $C$ background for the $\Z_2^{(1)}$ symmetry 
with the term $i\pi\int C \cup a$ in \eqref{SC TQFT with w2} instead breaks
gravitational invariance, since this is not invariant under
$w_2 \to w_2 + \d f$, $a\to a+f$. This can be cured by
introducing a coupling to a background spin structure $\eta$ as
$i\pi\int C \cup \eta$ \cite{thorngren2020anomalies,CappelliVillaBosDual2025},
transforming \eqref{SC TQFT with w2} into a fermionic theory:
\begin{equation}
    S'=i\pi \int_X b \cup (\d a+w_2) + C \cup a + C \cup \eta.
\end{equation}
This is now both gravitational and gauge invariant under $b\to b+\g$
and $C\to C+\d\g$. Gauging $C$ fixes $a =\eta$ and gives the
partition function
\begin{equation}
Z[\eta]=  \sum_{b,a,c} (-1)^{\int_X b \cup (\d a+w_2) + c \cup a + c \cup \eta} =
  \sum_a \delta (a-\eta) \sim \1_{\eta},
\end{equation}
where $\1_\eta$ is the trivial theory in the space of fermionic
theories.  This is precisely the inverse bosonization step discussed
after \eqref{standard bos anom}, where the anomalous bosonic theory is
converted back into a fermionic one by adding the counterterm
$\eta \cup C$ and gauging $C$.

\paragraph{Derivation of the $\Z_2^f$ theory.} 
Consider the gapped phase where $\U(1)_f$ is higgsed to $\Z_2^f$. In the IR,
the minimal option is just a theory for the $\Z_2$ remnant of the
spin$_c$ connection $a$. To describe this, we notice that an ordinary
flat $\U(1)$ gauge field is an element of $H^1(X;\R/\Z)$,\footnote{In
  this notation the Wilson lines are suitably normalized as
  $e^{2\pi i\oint a}$.} while a flat spin$_c$ connection can be
regarded as cochain $a\in C^1(X;\R/\Z)$ such that $\d a = 1/2 \,w_2$
(mod 1), where $1/2 \cdot$ is a map from $\Z_2$ to $\R/\Z$. This is
the discrete analogue of \eqref{spinc connection} and ensures that Wilson
lines of $a$ live on the boundary of a $w_2$ surface (i.e. they are
fermions). Notice that $a_b=2a $ satisfies $\d a_b = w_2=0 \mod 1 $,
so $a_b \in Z^1(X;\R/\Z)$ correctly. If we further higgs
the spin$_c$ connection to $\Z_2$ values, then, by suitably
rewriting $a \to 2a$,\footnote{If $a$ is $\Z_2$ valued, it means that
  $\oint a\in\{0,1/2\}\subset \R/\Z$. However, it is more common to
  represent $\Z_2$ gauge fields with values in $\{0,1\}$
  ($\cdot 2: \R/\Z\to \Z_2$ is the inverse of
  $\cdot 1/2:\Z_2\to \R/\Z$).} we obtain $\d a =w_2$ (mod 2). Together
with the magnetic coupling, the resulting low energy theory
is\footnote{Rigorously, the magnetic coupling of \eqref{flat spinc theory} can be obtained from the ordinary one refined in differential cohomology \cite{tachikawa2020pformanom} and restricting to zero curvature. To restrict to $\Z_2$ values, we are also rescaling $a \to 2a$, as said before (so $a$ in \eqref{flat spinc theory} or \eqref{SC TQFT with w2 and Bm} is actually $2a$ after higgsing).}
\begin{equation}
\label{flat spinc theory}
    Z[B_m] = \sum_{a\in C^1(X;\Z_2)\,|\, \d a =w_2} e^{\pi i\int_X a \cup c_1(B_m)}.
\end{equation}
Using a Lagrange multiplier $b$ to enforce $\d a= w_2$, we recover the
action \eqref{SC TQFT with w2 and Bm}, which is thus the minimal
$\Z_2$ gauge theory with anomaly \eqref{anom for gauge spinc}.
Let us add some comments.

From the perspective of \eqref{flat spinc theory}, it is clear that
the emergent $\Z_2^{(d-2)}$ symmetry is a non-confined part of the
magnetic symmetry surviving in the low energy limit. More precisely,
if we view the gauge group $\U(1)$ as in \eqref{spinc u1 extension},
the magnetic symmetry is \eqref{magnetic extension}: the surviving
symmetry is the dual symmetry of $\Z_2^f$, while $\U(1)_{m,b}^{(d-3)}$
is confined (indeed, $a_b=2a$ is completely higgsed, so its associated
magnetic symmetry $\U(1)_{m,b}^{(d-3)}$ completely confines; only
$\eta$ survives). Therefore, the
fractionalization mentioned before, namely $C= c_1(B_m)|_2$, is
actually a condition already present in the UV \eqref{Bm = Bmtilde +   C} 

In the IR, the condition $\d a =w_2$ implies that $a$ is
effectively an emergent spin structure.  Actually, the theory
\eqref{SC TQFT with w2} has zero partition function on non-spin
manifolds. This can be understood directly from the symmetry
structure \eqref{spinc group def}: if $\U(1)$ is higgsed to $\Z_2$,
the low energy symmetry is\footnote{This is noted also in
  \cite{seibergwitten2016gappedTI,wittenNewSU2Anomaly2018}, albeit for
  a background spin$_c$ connection.}
\begin{equation}
    \frac{\spin(d)\times \Z_2^f}{\Z_2}\simeq \spin(d),
\end{equation}
with the condition that $\Z_2^f \subset \spin(d)$ is actually
gauged.\footnote{Consistently, the real global symmetry is always
  $\SO(d)$, both in the UV \eqref{spinc group def} (with $\U(1)$
  gauged) and in the IR.} Physically, this is consistent with the fact
that the low energy quasiparticle excitations 
look like neutral fermions at long distances, they carry just the $(-1)^F$
charge.

One can understand the emergence of the spin structure as follows.
The condensing Higgs field is a section for the associated
vector bundle of the $\U(1)_b$ principal bundle with connection
$2a=a_b$. If $\U(1)_f$ is higgsed everywhere to $\Z_2^f$, this means
that the Higgs field is non-zero everywhere, i.e. it is a non-zero
global section in mathematical terms. But a principal bundle with a
non-zero global section is trivial \cite{nakahara}, therefore
$c_1(a_b)=0$ and the spin$_c$ condition \eqref{spinc structure}
reduces to a spin structure ($a$ in \eqref{SC TQFT with w2} can be
thought of as $\eta$ in \eqref{spinc structure} with trivial
$c_1(a_b)$). So a spin$_c$ connection can be higgsed everywhere only
on spin manifolds. On more general $\spinc$ manifolds, since
$c_1(a_b)$ is necessarily nontrivial, it follows that the Higgs field
must vanish somewhere. Indeed it vanishes at the ground-state
vortex line \eqref{persistent} found earlier.

This argument will be extended to the case of $q=2n>2$
superconductors in section 4.2.1.

\subsubsection{Twisted BF theory and uniqueness}

Consider \eqref{SC TQFT with w2} and integrate $a$ to make $b$ flat.
The resulting expression can be transformed using the Wu formula
\begin{equation}
\label{Wu-gen}
S=i\pi\int_X b\cup w_2 = i\pi \int_X b\cup_{d-4} b,
\qquad\quad ({\rm flat} \ b),
\end{equation}
on $d$-dimensional closed orientable $X$, where $b$ is a $d-2$ form.
Note that the relation introduces a dependence on $d$, so the action
should be independently analyzed in each case (see appendix A for the
complete discussion).  In particular, it vanishes for $d=2,3$ (cup
with negative index is not defined), thus the first non-trivial case
is $d=4$ for the action and $d=3$ for the anomaly in $d+1$, as in
\eqref{standard bos anom}. It fits the fact that all orientable
manifolds are spin in $d\le 3$. In $d=4$ it also vanishes on spin
manifolds, where $w_2=\d\eta$.

One can add again a Lagrange multiplier $\wh a$ enforcing $\d b=0$ to
extend the r.h.s of \eqref{Wu-gen} to non-flat $b$. We thus obtain
the following actions
\be
\label{SC TQFT bb}
S=i\pi \int_X b \cup \d \wh a + b\cup_{d-4} b + b \cup_{d-3} \d b\,,
\ee
which defines the so-called twisted BF theory.
In this expression, additional terms $g(b)$ which vanish for $\d b=0$
could be added: the choice $g(b)=b \cup_{d-3} \d b$ is fixed by
requiring gauge invariance under $b\to b +\d \l$
when $b$ is not closed, owing to the noncommutativity of the cup
product.\footnote{Calculations need
  the algebraic rules of higher cup products, which are described
  in \cite{kapustinseiberg2014}: they greatly simplify in the $\Z_2$ case and
are given in appendix A.}

A nice feature of the action \eqref{SC TQFT bb} is that it is written
entirely in terms of the bosonic field $b$. However, it still
depends on the background geometry, as does \eqref{SC TQFT with w2}, but
in implicit form. Actually, it is also not $\Z_2^{(d-2)}$ invariant on
  non-spin manifolds: under $b\to b +\d \l$, the variation 
  $\D S=i\pi\int_X \g \cup_{d-4} \g$ is the same as that of  $\Z^f_2$
  theory \eqref{SC TQFT with w2}, owing to the Wu formula again.
  Note that the field $\wh a$ is a-priori unrelated to $a$, the
  $\Z_2$ remnant of electromagnetism.

\paragraph{$3+1d$ theory.}
The torsion term $b\cup b$ looks like a topological remnant of an
interaction between vortices.\footnote{Interactions between vortices
  are present in superconductors, although they are not described at
  topological level
  \cite{jacobsRebbi79,VVintSC2011,MeasureVortexIntSC}. The force is
  attractive for type I SC, in which vortices do not form a stable
  configuration, and repulsive for type II, so that a vortex of
  $n\pi$-flux decays in $n$ $\pi$-flux vortices.}  However, it seems
that it does not produce any new relevant correlator in the
topological theory \cite{putrovwang2017LinkNum}, besides making the
Wilson line of $\wh a$ fermionic as $a$ in \eqref{SC TQFT with w2}
(this is explained in appendix \ref{app twisted BF}).  This theory
appeared in \cite{kapustinthorngren2017} as the seed bosonic theory
used to classify fermionic SPTs by fermionization.

The theory \eqref{SC TQFT bb} has a
counterpart formulated with continuous $\U(1)$ gauge fields, which is
properly the twisted BF theory studied in the literature
\cite{kapustinseiberg2014,gaiotto2015generalized,HsinLamSeiberg20181formSym3d4d,BrennanHong2023LecturesGenSym}.
This reads (dropping the hat on $a$ in the following)
 
\begin{equation}
\label{twisted BF continuum}
S = \int_X \frac{q}{2\pi} b \wedge\d a + \frac{pq}{4\pi} b \wedge b,
\qquad\quad q\in \Z, \;p\in \Z_{2q}.
\end{equation}
The relation with \eqref{SC TQFT bb} is
for $q=p=2$, by sending $b\to b/\pi$, $ a\to a/\pi$
(generically $qp \in 2\Z$). One can check that for these $p,q$ values 
the two theories have the same symmetries and anomalies.

The theory \eqref{twisted BF continuum} has the symmetry
$\Z_{\gcd(q,p)}^{(1)}\times \Z_{\gcd(q,p)}^{(2)}$
\cite{kapustinseiberg2014,gaiotto2015generalized}.  Writing $q=rL$ and
$p=sL$ with $L\coloneqq\gcd(q,p)$, the $\Z_2\subset \Z_L^{(2)}$ subgroup has
the bosonization anomaly precisely when $rs$ is odd; equivalently, the
generating line is fermionic (see appendix \ref{app twisted BF} for a
detailed discussion). This theory is also related to oblique
confinement \cite{fradkin2022ObliqueTI} and its connection to
superconductors was mentioned in
\cite{vonKeyserlingk2014WWModelBF}. For $q=p=2$ it is sometimes called
``fermionic'' toric code (although clearly bosonic).  Note that in the
continuum theory \eqref{twisted BF continuum}, the specific
properties on non-spin manifolds, relevant here, are sometimes
disregarded in the literature.
Correlation functions of Wilson loop and surface operators are extensively
discussed in appendix B, both for the discrete \eqref{SC TQFT bb} and continuum
theory \eqref{twisted BF continuum}.

\paragraph{Uniqueness of $3+1d$ twisted BF theory.}
In the previous section we argued that the topological theory, required
for anomaly matching, for $q=2$ is \eqref{SC TQFT with w2 and Bm},
which is a natural choice given \eqref{flat spinc theory}. Using
\eqref{SC TQFT bb}, we can show that this result is unique. The
assumption is that the gapped Higgs phase has an unscreened $\Z_2$
gauge field and this is the basic degree of freedom in the
IR. Therefore, the Higgs phase is described by a $\Z_2$ gauge theory
and we know that possible actions for a discrete $G$ gauge theory are
classified by Dijkgraaf-Witten terms $H^4(BG;\R/\Z)$
\cite{dijkgraafwitten}.\footnote{This kind of reasoning is useful to
  avoid seemingly non-trivial actions that instead are trivial. For
  example, for $a$ valued in $\Z_2$, one could consider in principle
  also $a^4= a\cup a \cup a \cup a$. However, this is zero:
  $a^4=Sq^1 (a^3)=w_1\cup a^3=0$ on oriented manifolds with
  $[w_1]=0$.} We can also be slightly more general as follows. First,
plain $G$ gauge theory in $d=4$ has a dual formulation in terms of a
$\wh G^{(1)}$ gauge theory (namely, \eqref{SC TQFT with w2} is viewed
as a theory for the Lagrange multiplier $b$). We can thus consider
also generalized DW actions classified by $H^4(B^2\wh
G;\R/\Z)$. Second, we can be more general by using the cobordism
approach to classify such actions
\cite{kapustinBordism,kapustinbordismTI,kapustinFermionicBordism,
  OhmoriPutrovWangAl2018fTQFTs}. However,
for $d=4$, $\hom(\tor (\W_4^{\SO}(X)),\R/\Z) \simeq H^4(X;\R/\Z)$,
thus the two classifications coincide.\footnote{Moreover,
  $\W_4^{\SO}(X)\simeq \W_4^{\SO}(pt)\oplus \wt \W_4^{\SO}(X)$, where
  $\W_4^{\SO}(pt)\simeq\Z$ and $\wt \W_4^{\SO}(X)$ is the reduced
  bordism group, and $\wt \W_4^{\SO}(X)\simeq H_4(X;\Z)$. This means
  that the DW classification takes into account also mixed
  gauge-gravitational terms, as shown by the equivalence of \eqref{SC
    TQFT bb} and \eqref{SC TQFT with w2}.} For $G=\Z_2$,
$H^4(B\Z_2;\R/\Z)=0$ and $H^4(B^2\Z_2;\R/\Z)\simeq \Z_4$
\cite{EilenbergMacLaneOriginal}. The only possible actions are
therefore
\begin{equation}
  \label{Poin-act}
    2\pi i \frac p 4 \int \mathcal{P}(b), \qquad p\in \Z_4,
\end{equation}
where $\mathcal{P}: H^2(X;\Z_2)\to H^4(X;\Z_4)$ is the Pontryagin square \cite{kapustinthorngren2013TQFTlattice,CordovaBeniniHsin2group,whitehead1949,whitehead1950}, defined as 
\begin{equation}
\label{Pontr square def text}
   \mathcal{P}(b)= \wt b\cup \wt b- \wt b\cup_1 \d \wt b,
\end{equation}
for an integer lift $\wt b$ of $b$ (see appendix A and B for more details).
Eq.  \eqref{Poin-act} matches precisely the twisted BF
theory \eqref{SC TQFT bb}, and in the continuum
\eqref{twisted BF continuum} (where $\mathcal{P}(b) \sim b\wedge b$).

The conclusion is that bosonic $\Z_2$ gauge theories in $d=4$ are all
described by \eqref{twisted BF continuum} for $q=2$ and there are four
options: $p=0,1,2,3$, with symmetry
$\Z_{\gcd(2,p)}^{(1)}\times \Z_{\gcd(2,p)}^{(2)}$. For $p=1,3$ there
are no higher-form symmetries (and thus no topological order) and the
anomaly \eqref{anom for gauge spinc} cannot be matched. $p=2$, namely
\eqref{SC TQFT bb} and \eqref{SC TQFT with w2}, is the only
possibility.\footnote{$p=0$ is just level 2 BF theory that matches
  \eqref{anom for gauge spinc} with the fractionalization choice
  $B=w_2$. However, in this case, the theory is actually $p=2$ in
  disguise.}

Furthermore, the possible $\Z_2$ topological orders in
four dimensions have been classified in \cite{JohnsonFreyd2020Z2TO4d}. This
is different from our approach above, where we searched
$\Z_2$ gauge theories (and indeed we found some, $p=1,3$, that
have no topological order).\footnote{The result for the topological
  order in \cite{JohnsonFreyd2020Z2TO4d} is also much more general.}
However, the result in \cite{JohnsonFreyd2020Z2TO4d} shows that there
are three possible $\Z_2$ topological orders: level 2 BF, the theory
\eqref{SC TQFT with w2} and again \eqref{SC TQFT with w2} with an
additional coupling to $w_3$, which has a pure gravitational anomaly
\cite{thorngren2014framed,wangwenwitten2017}. This is a $\Z_2$ gauge
theory, but it did not appear in our argument above for two reasons:
first, in $d=4$ $w_3$ is always trivial; second, related to the first
one, the relevance of this case is that it has a pure gravitational
anomaly and thus should be regarded as a relative theory on the
boundary of a five-dimensional gravitational SPT phase $\int w_2 w_3$.

\paragraph{$2+1d$ theory.} 
In the last paragraph of section three, we said that the bosonization
anomaly \eqref{anom for gauge spinc} is preserved in $ 2+1$ dimensions
only in presence of time-reversal symmetry.  Ordinary s-wave superconductors
are not expected to break the time-reversal symmetry which is
naturally present in condensed matter systems, thus the discussion of anomaly matching can be applied. In three dimensions, the theory \eqref{SC TQFT with w2} is equivalent to plain
BF theory, i.e. the standard toric code \cite{HsinShao2020}. Since
every manifold is spin, $w_2=\d \eta$, the action can be rewritten as
$b\d c$, with $c= a+\eta$.  The reduction to ordinary toric code is
possible because level-2 BF theory already contains a fermionic anyon.
There are three anyons, $e$, $m$ and $f=em$, and $f$ is the fermion,
as further described in the appendix A.2.  The diagonal symmetry
$\Z_2^{(1)}\subset \Z_2^{(1)}\times \Z_2^{(1)}$, generated by this
fermion, is anomalous, with the required expression \eqref{standard
  bos anom}.  With the choice of fractionalization $C=c_1(B_m)$, it
also gives \eqref{anom for gauge spinc}. The toric code is indeed
obtained by gauging fermion parity in the trivial fermionic theory
\cite{gaiottokapustinspinTQFT1,gaiottokapustinspinTQFT2}.  In
comparison, the $3d$ theory \eqref{SC TQFT with w2} uses a different
basis: $m$ and $f$ are associated to the fields in the action,
i.e. $m$ is the Wilson line of $b$ and $f$ is the Wilson line of $a$.

\paragraph{Uniqueness of the $2+1d$ theory.} Similar to our previous
discussion, we could think of twisting the $\Z_2$ gauge theory with DW
actions $H^3(B\Z_2;\R/\Z)\simeq \Z_2$. The only non-trivial option is
to consider a Chern-Simons action that in the continuum reads
\cite{seibergwitten2016gappedTI}
\begin{equation}
    \int \frac{2}{2\pi} b\d a + \frac{2}{4\pi} a\d a.
\end{equation}
However, one can check that this has an extra anomaly part that does
not match \eqref{anom for gauge spinc}. In fact, there is no fermionic
line in the spectrum of the theory with $p=2$
\cite{seibergwitten2016gappedTI}. We are thus left with \eqref{SC TQFT
  with w2} only.

\subsection{Other superconductors}

As already discussed, the anomaly \eqref{anom for gauge spinc} characterizes all
electronic systems coupled to a dynamical spin$_c$ 
symmetry, forbidding a trivially gapped phase. It follows that this
anomaly must be matched also in
topological superconductors and charge $q=2n$ superconductors.

We start with a general observation involving the anomaly \eqref{anom
  for gauge spinc}. A simple way to match this anomaly with a low energy
theory is to assume that the magnetic symmetry is still there, acting
on some operator surviving in the IR. However, this implies that
$\U(1)_m^{(d-3)}$ is spontaneously broken, thus the theory is
massless. If we instead want to characterize a gapped phase, the
magnetic symmetry should confine and there should be an emergent
symmetry with anomaly \eqref{anom for gauge spinc}, after a suitable
fractionalization choice. By matching the degree in \eqref{anom for
  gauge spinc}, this symmetry should have a $(d-1)$-form background
$C$ and it is therefore a discrete $(d-2)$-form symmetry, generated by
a topological line operator. Since the anomaly is $\Z_2$ valued, the
general case involves a $\Z_{2n}^{(d-2)}$ emergent symmetry in the IR
with a self-anomaly, such that $C|_2 = c_1(B_m)|_2$ reproduces
\eqref{anom for gauge spinc}.

A consequence of this discussion is that every theory with a
$\Z_{2n}^{(d-2)}$ symmetry with anomalous $\Z_2$ subgroup and
bosonization anomaly \eqref{standard bos anom} (its
generator is a fermionic line), is a possible candidate IR phase by
anomaly matching. This is the class of bosonic theories that arise
from the bosonization of fermionic theories.\footnote{See the
  discussion on the emergent spin structure
  below \eqref{SC TQFT with w2 and Bm}.} The generating line $\psi$ of
$\Z_2^{(d-2)}$ can be cut open with local fermions on its ends,
non-genuine local operators in the twisted sector of the bosonic
theory. Therefore, the action of $\psi$ on the charged
$(d-2)$-dimensional object $O$ can be pictured as follows: a pair of
fermions is produced, one encircles $O$ getting an AB phase of $-1$
and it is then annihilated with its partner, leaving overall a new
state $-O$. This can be thought of as a spin structure defect, that
changes the boundary conditions of fermions by a sign, and this is
also the effect of a $\pi$-flux on a charged fermion. The
fractionalization choice $C=c_1(B_m)|_2$ is the link between the two
pictures, dressing the operator $O$ with a magnetic flux and making it
a suitable ANO vortex.

\subsubsection{Charge $q=2n$ superconductors}

The discussion of the Higgs phase can be
generalized straightforwardly to the $\U(1)_f\to \Z_{2n}$ case, with
$\Z_2^f \subset \Z_{2n}$, where the Higgs field has
charge $q=2n$ and represents a bound state of $2n$
electrons. The case $q=4$ is a realistic scenario
\cite{BergFradkinKivelson2009SC4e,Vishwanath2026TSC4e,GaoWangZhangYang2026SC4ePrimary,SC2e4eQPT2026,SC3D4e6e2026,GaoWangYangWu2025TSC2ne}.

We assume as before that the IR theory is
characterized only by the $\Z_{2n}$ remnant of the spin$_c$ connection
$a$. This is a cochain $a\in C^1(X;\R/\Z)$ that
takes the discrete set of values
$\oint a \in \{0,1/2n,...,(2n-1)/2n\}$ and obeys $\d a = 1/2\,w_2\mod 1$. By
suitably rescaling to integer valued periods, $a \to 2n a$, we obtain
$a\in C^1(X;\Z_{2n})$ with $\d a = nw_2 \mod 2n$. The resulting
topological action, with the Lagrange multiplier $b$ and magnetic
coupling analogous to \eqref{SC TQFT with w2 and Bm}, is
therefore\footnote{The magnetic coupling can be deduced as before, see
  footnote above \eqref{flat spinc theory}.}
\begin{equation}
\label{SC2ne TQFT}
S[B_m] =\frac{2\pi i}{2n} \int_X b \cup (\d a+ n w_2(TX)) +
c_1(B_m) \cup a \,.
\end{equation}  
As \eqref{SC TQFT with w2 and Bm}, this is a level $2n$ BF theory with
fractionalization choices $B = nw_2$ and $C= c_1(B_m)\mod 2n$. The BF
mixed anomaly gives \eqref{anom for gauge spinc} correctly. There are
no surprises at this point: there is a $\Z_{2n}$ topological order and
each vortex traps a $2\pi /2n =\pi/n$-flux, leading to fractional
anyonic-like braiding between a particle and a vortex.

According to our general observation above, the theory \eqref{SC2ne
  TQFT} has a $\Z_{2n}^{(d-2)}$ symmetry whose $\Z_2$ subgroup has the
anomaly \eqref{standard bos anom} (the fractionalization $B=nw_2$
makes the Wilson line of $a$ a fermion). We now discuss the
uniqueness of \eqref{SC2ne TQFT} using a dual DW formulation,
as done for \eqref{SC TQFT with w2} in the previous section.
We shall find that more theories are compatible
with the anomaly \eqref{standard bos anom}, in principle.

\paragraph{$3+1d $ theory.} The action \eqref{SC2ne TQFT}, with $B_m= 0$,
is equivalent to (as explained for \eqref{SC TQFT bb})
\begin{equation}
\label{SC2ne TQFT bb}
S =i\pi \int_X b|_2 \cup w_2= i\pi \int_X b \cup b=
\frac{2\pi i}{4n} p\int_X  \mathcal{P}(b), \qquad\quad p =2n \in \Z_{4n},
\end{equation} 
where $b\in Z^2(X;\Z_{2n})$ and
$\mathcal{P}: H^2(X;\Z_{2n})\to H^4(X;\Z_{4n})$ as \eqref{Pontr square
  def text}. $p\in \Z_{4n}$ labels the inequivalent $\Z_{2n}^{(1)}$
gauge theories, given that $H^4(B^2\Z_{2n};\R/\Z)\simeq \Z_{4n}$ (and
$H^4(B\Z_{2n};\R/\Z)=0$). The action \eqref{SC2ne TQFT bb} is a
twisted BF theory \eqref{twisted BF continuum} in the continuum with
$q= 2n =p$.

Let us consider the action \eqref{SC2ne TQFT bb} for general $p\in \Z_{4n}$.
The situation is more involved than for $n=1$ ($q=2$). Generically, the
theory \eqref{SC2ne TQFT bb} has a $\Z_L^{(1)}\times \Z_L^{(2)}$
symmetry, with $L \coloneqq \gcd(2n,p)$
\cite{gaiotto2015generalized,kapustinseiberg2014}. To match the
anomaly \eqref{anom for gauge spinc} we need $L$ even, so $p$ must be
even too. For generic $2n = rL$ and $p = sL$, the anomaly of
\eqref{SC2ne TQFT bb} is
\begin{equation}
    \int_Y \frac{2\pi}{L} B \cup C + i\pi \,rs\, C|_2\cup_1 C|_2,
\end{equation}
in terms of the $\Z_L$ backgrounds $B$ and $C$ 
(details in appendix \ref{app twisted BF}). So, for any choice of $p$
with $rs$ odd, the $\Z_L^{(2)}$ symmetry has correctly the self
anomaly \eqref{standard bos anom} that reproduces \eqref{anom for
  gauge spinc} for $C = c_1(B_m) \mod L$ (and $B=0$). The case
$rs \in 2\Z+1$ is precisely when the Wilson line of the dual variable
in \eqref{SC2ne TQFT bb} (like $\wh a$ in \eqref{SC TQFT bb}) is a
fermionic line.

We conclude that anomaly matching alone would allow
more general actions than \eqref{SC2ne TQFT}. However, for realistic
charge $q=4$ SCs, we still have only one choice compatible with
$rs$ odd, namely $p= q=4$, which is indeed \eqref{SC2ne TQFT}. For
$q=6$, which could also be realistic, there
are in principle three options: $p=2,6,10$, $p=6=q$ being \eqref{SC2ne
  TQFT}. The other two cases, $p=2$ and $p=10$, have a $\Z_2$
topological order like ordinary SCs, contrary to the expected
$\Z_6$ topological order.

Therefore, the topological theory \eqref{SC2ne TQFT} is the only one that
ensures both anomaly matching and $\Z_{2n}$ topological order. Note
that its action is derived by assuming that the spin$_c$ connection
is the relevant degree of freedom in the IR, which also guarantees
anomaly matching \eqref{anom for gauge spinc}.  We also remark that
the action \eqref{SC2ne TQFT bb}, with $p$ ($p\neq 2n$), is not
obviously related with the original formulation with $a$ (the dual
variable $\wh a$ is not $\Z_{2n}$ valued for generic twists).

\paragraph{$2+1d$ theory.}
The theory \eqref{SC2ne TQFT} is again equivalent to plain level $2n$
BF theory, since integrating out $a$ yields $b|_2\cup w_2 =0$. Indeed,
level $2n$ BF theory has already fermionic quasiparticles in its
spectrum, as for the toric code. Denoting $e$ and $m$ the bosonic
Wilson lines of $a$ and $b$, with braiding phases $e^{i\pi/n}$, the
composite $f=e^km^l$ has spin $h= kl/2n$. We conclude that $e^nm$,
$em^n$ are always fermionic quasiparticles, while $e^n m^n$ is a
fermion for $n$ odd. They all generate self-anomalous symmetries
(subgroups of $\Z_{2n}^{(1)}\times \Z_{2n}^{(1)}$) with anomaly
\eqref{standard bos anom}, which can be used to match \eqref{anom for
  gauge spinc}, as explained in the introduction to this
section. Among these choices, the fractionalization of the symmetry
generated by $em^n$ with respect to the UV magnetic symmetry is the
one that always assigns a fractional $\pi/n$-flux to the vortex line
of $b$.

As before, the above description coming from \eqref{SC2ne TQFT} is
motivated physically by the requirement to have a suitable higgsed
spin$_c$ connection $a$ in the low energy theory. We can be more
agnostic on the relevant IR degrees of freedom and just assume that
the Higgs phase with charge $2n$ condensate is a $\Z_{2n}$ gauge
theory. This allows us to consider more general DW actions, which are
classified by $H^3(B\Z_{2n};\R/\Z)\simeq \Z_{2n}$. Their continuum
action is \cite{seibergwitten2016gappedTI}
\begin{equation}
\label{twist DW Z2n 3d}
\int \frac{2n}{2\pi} b\d a + \frac{2p}{4\pi} a\d a,
\qquad\qquad p \in \Z_{2n}\,,
\end{equation}
($p=0$ is \eqref{SC2ne TQFT}). A quick way to see if any such theory
could match the anomaly \eqref{anom for gauge spinc} is to check if
there is a fermionic line in the spectrum: if there is one, then its
anomalous symmetry reproduces \eqref{anom for gauge spinc} by choosing
$C= c_1(B_m)$ (and setting to zero all backgrounds for other
symmetries with possible mixed anomalies with it). The spin of a
generic dyon\footnote{Computing
  $\braket{e^x m^y}=\exp(2\pi i(xy/q - p' y^2/2q^2))=\exp(2\pi ih)$
  for $\link(\g,\g)=1$ a framed loop, $q=2n$ and $p'=2p$.} $e^xm^y$ is
\cite{seibergwitten2016gappedTI}
\begin{equation}
    h(x,y)= \frac{xy}{2n}- \frac{py^2}{4n^2}\mod 1, \qquad\quad x, y \in\Z.
\end{equation}
We want to look for $h(x,y)=1/2\mod 1$. If $p$ is odd, there are no
solutions, while for even $p$ there could be fermionic lines. For
$n=1$, $p=1$ is the only option and this is not admissible for anomaly
matching, as we already commented above. For $n=2$ (charge $q=4$ SCs),
there is also the other possible choice $p=2$. In this latter case the
topological order is still $\Z_4$ as for $p=0$ (global symmetry
$\Z_4^{(1)}\times \Z_4^{(1)}$).\footnote{Generically, the global
  symmetry of \eqref{twist DW Z2n 3d} for $2n\to q$ and $2p=p'$ even
  is $\Z_{q^2/\gcd(q,p')}^{(1)}\times \Z_{\gcd(q,p')}^{(1)}$
  \cite{seibergwitten2016gappedTI}. The topological lines $e^xm^y$
  labeled by $(x,y)\in \Z \times \Z$ are subject to the identification
  $(x,y)\sim (x+q,y)$ and $(x,y)\sim (x+p',y+q)$ by looking at their
  correlators and spins. After this quotient we obtain the total group
  of the inequivalent topological line defects of \eqref{twist DW Z2n
    3d}.}


\subsubsection{Topological superconductors}
 
Topological superconductors (TSC) are usually treated as invertible fermionic
phases, for example in the ten-fold way classification
\cite{ryu2016classification}.\footnote{There is a different
  terminology in hep-th and cond-mat communities, in that SPT phases
  for a gravitational anomaly
  \cite{kapustinBordism,kapustinFermionicBordism} are called
  invertible topologically ordered phases from a condensed matter
  perspective
  \cite{wen2017colloquium,wen2013cohomologyBos,Wen2012cohomologyFer}
  (since phases protected by gravitational symmetries are robust
  against any local perturbation, gravitational symmetries cannot be
  broken). This mismatch of terminology applies to some TSCs, e.g. the
  $2d$ Kitaev chain (protected by fermion parity) and the chiral
  $p+ip$ superconductor in $3d$. It is also the case for the familiar
  integer quantum Hall effect. Here we use the term invertible to
  indicate that there is no bulk ground state degeneracy and no
  anyons.} However, being  gapped Higgs phases of a spin$_c$
 field, they should be bosonic systems with topological
 order to match the anomaly \eqref{anom for gauge spinc},
 as ordinary s-wave SCs.
 Notice that the fact the TSCs are actually topologically ordered when
 the electromagnetic field is properly taken into account is already
 known in the literature, see
 e.g. \cite{BernevigNeupert2015LectureTSCcat,ReadGreen1999}.\footnote{See also the answer
  by X.-G. Wen at
  \url{https://physics.stackexchange.com/questions/71351/do-topological-superconductors-exhibit-symmetry-enriched-topological-order}.} Moreover,
 chiral $p+ip$ TSCs in three-dimensions have non-Abelian anyons, which
 is the characteristic feature that makes them interesting for quantum
 computation.

As for ordinary s-wave superconductors, some properties of TSCs can also be
understood without dynamical electromagnetic field
\cite{BernevigNeupert2015LectureTSCcat}. It seems 
that breaking explicitly  particle number symmetry
in mean-field approximation and then gauging the
remaining $\Z_2^f$ symmetry looks really similar to
higgsing a $\U(1)$ gauge symmetry down to $\Z_2^f$.
Thus, it is understandable why the same properties of these
systems can be understood using the two different pictures,
being related just by a discrete gauging \cite{gaiottokulp2020orbifolds}.
Nonetheless, disregarding $\U(1)$ gauging  misses all consequences of anomaly we have been discussing in this paper. 

An extensive analysis of the consequences of the bosonization anomaly \eqref{anom for gauge spinc} in topological superconductors is 
clearly beyond the scope of this work. In particular,
in 2+1d TSC time reversal invariance 
is dynamically broken in the IR by the $p$-wave condensate
and anomaly matching requires a careful study.
Hereafter, we shall limit ourselves to some remarks, mostly on the
consequence of $\Z_2^f$ gauging.

The IR theory can be guessed as follows. For ordinary SCs, we saw that
 our reasoning leads us to the TQFT \eqref{SC TQFT with w2} uniquely,
 which can also be understood as the result of gauging $\Z_2^f$
 of the trivial spin TQFT
 $\mathbf{1}_\eta$. The spin structure is emergent at low energy from
 the higgsing of the spin$_c$ structure and the bosonization procedure
 comes from the microscopic theory, where $\Z_2^f$ is part of the
 electromagnetic gauge group. The s-wave SC is the trivial
 phase in the space of the topological superconductors and it is then
 natural to generalize the discussion to TSCs by simply taking as 
spin TQFT not the trivial one, but the invertible TQFT
 describing the defining fermionic anomaly
 \cite{witten2016fermion,witten2016parityanomalyUnoriented}, which we
 call $\T_\eta^{\rm Inv}$.\footnote{Generically, $\T_\eta^{\rm Inv}$
   should be some proper APS $\eta$ invariant ($\eta$ is not the spin
   structure here) for a Dirac operator in $d$-dimension
   \cite{wittenyonekura2019}.} Then, the TQFT for the TSCs in the IR
 is given by its bosonization,
\begin{equation}
    \label{TQFT for TSCs}
    \T_{\rm TSC} \simeq \T_\eta^{\rm Inv}/\Z_2^f = \sum_\eta \T_\eta^{\rm Inv}.
\end{equation}
Notice that any such theory could match the anomaly \eqref{anom for
  gauge spinc}: it has a $\Z_2^{(d-2)}$ symmetry with anomaly
\eqref{standard bos anom} that reduces to \eqref{anom for gauge spinc}
by fractionalization $C= c_1(B_m)|_2$. Since this symmetry is
generated by the emergent fermionic quasiparticle, it acts on the flux lines
 and the fractionalization condition ensures that they have flux
$\pi$.

\paragraph{$p+ip$ TSC in $2+1d$.} 
Consider the chiral $p_x+ip_y$ TSC 
\cite{ReadGreen1999}, whose invertible spin TQFT description is
$\SO(1)_1$ Chern-Simons theory, $\T_\eta^{\rm Inv}=\SO(1)_1$, called
Ising spin TQFT \cite{seibergwitten2016gappedTI}. Its
bosonization version, following \eqref{TQFT for TSCs}, is
Spin$(1)_1$ \cite{seibergwitten2016gappedTI,
  CordovaHsinSeiberg2018SONnCS}, the Ising TQFT.\footnote{It can
  alternatively be written as $\U(2)_{2,-4}$, where the two levels
  denote $\SU(2)_2 \times \U(1)_{-4}$ and then the $\Z_2$ quotient is
  obtained by gauging the $\Z_2$ one-form symmetry.} This should be
the correct topological description of a $p+ip$ superconductor in
the non-trivial phase. Spin$(1)_1$ has three lines, $1$, $\psi$ with
spin $1/2$ and $\sigma$ with spin $1/16$, with fusion rules
\cite{HsinShao2020}
\begin{equation}
\psi \times \psi = 1, \qquad \psi \times \sigma = \sigma,
\qquad \sigma \times \sigma = 1 + \psi,
\end{equation}
which is indeed the same as those of the Ising
chiral algebra with $c= 1/2$. These are also the non-Abelian fusion
rules for the topological defects of $p+ip$ TSC
\cite{BernevigNeupert2015LectureTSCcat}: $\psi$ represents the neutral
fermion (charged under the surviving $\Z_2^f$ part of $\U(1)$) and
generates the anomalous $\Z_2^{(1)}$ symmetry dual to $\Z_2^f$; $\s$
is the $\pi$-flux vortex that traps a Majorana zero-mode.\footnote{The
  important result is its vanishing energy. Localized modes are
 also present in vortices of s-wave SCs, but they are gapped.}
The natural boundary condition for $\spin(1)_1$ is the Ising CFT
(comparing to the Majorana CFT for $\SO(1)_1$).
This analysis suggests that the Majorana modes in TSC are not easily
accessible, but should be searched in the twisted sectors, as the
fermion state in the Ising CFT.

Despite being non-invertible phases, we can still argue that stacking
two TSCs gives $\spin(1)_1\boxtimes \spin(1)_1= \spin(2)_1$. This can be
seen from the fact that the electromagnetic field coupled to the two
systems must be the same, so that the product should be thought of as
\begin{equation}
  \spin(1)_1\boxtimes \spin(1)_1\coloneqq \sum_\eta \SO(1)_1\ \SO(1)_1 =
\sum_\eta \SO(2)_1=  \spin(2)_1\,,
\end{equation}
(remember that $\eta$ is the remaining part of the electromagnetic
field). The notion of stacking is thus the combination of the product
of the matter theories together with gauging the diagonal $\U(1)$
symmetry. This reproduces the 16-fold Kitaev way for fermions coupled
to $\Z_2$ gauge field \cite{Kitaev2006Anyons}, $\Z_2^f$ for us:
Spin$_1(16)$ represents the trivial phase since it is the toric code,
i.e. standard $\Z_2$ gauge theory, the topological order of the
'topologically-trivial' superconductor.

The existence of a topological order in the $p+ip$
TSC was already noted in the seminal work \cite{ReadGreen1999}. The
$2^{2g}$ ground-state degeneracy on a Riemann surface of genus $g$ (both
in the strong pairing (topologically trivial) and weak pairing
(topologically non-trivial) phases) in the gauged theory is a
consequence of the fact that there is one zero-mode in the ungauged
theory for each choice of spin structure, which have that multiplicity.

The standard effective field theory description of the phase
transition between  SC and TSC
is a free Majorana fermion changing mass sign at the transition
\cite{ReadGreen1999,witten2016fermion}. There is a conjectured duality
for the Majorana fermion with a scalar field coupled to $\SO(n)_1$ CS
theory
\cite{AharonyBeniniHsinSeiberg2017CSSODualities,metlitski2017MajoranaDuality},
analogous to the Dirac-boson duality
\cite{tongwebofduality,wittenwebofduality}. This implies a purely
bosonic description of the bosonized Majorana fermion after gauging
$\Z_2^f$, which is Eq. (5.20) in \cite{CappelliVillaBosDual2025}. This
theory indeed interpolates, for the two signs of the mass, between a
toric code phase and a $\spin(1)_1$ TQFT, being therefore a suitable
description of the SC-TSC gauged phase transition.
In conclusion, all these considerations support the bosonic nature
\eqref{TQFT for TSCs} of these superconductors.


\subsection{Phases of QED$_3$}

The anomaly \eqref{anom for gauge spinc} is also present in QED, namely $N_f$  Dirac fermions coupled to a spin$_c$
connection $a$. In four dimensions, the theory is weakly coupled, thus the anomaly is not particularly useful. In three dimensions,
QED$_3$ is strongly coupled in the IR, and anomaly matching
is extremely important. However, it requires time reversal symmetry
in $2+1d$, as discussed before, which is only present for even $N_f$.
Therefore, there are no implications for odd values. Nonetheless,
we shall see what happens for $N_f=1$.

\paragraph{$N_f=2$ theory.}
In the work \cite{DumitrescuNiroThorngren2024QED3}, the authors have shown
that the low-energy phase is characterized by spontaneous symmetry breaking of
the mixed flavor-magnetic symmetry
\be
\label{u2}
\U(2)=\frac{\SU(2)_f \times \U(1)_m}{\Z_2} \ \to \U(1)_{\rm unbroken}\,,
\ee
due to monopole condensation. Clearly, anomaly matching was a major
input. Note the the magnetic symmetry  $\U(1)_m^{(d-3)} $ is of ordinary
type and can mix with flavor symmetry. 
The $\spinc$ symmetry is obtained from the larger group
\be
\label{spinc3}
\frac{\SU(2)_f \times \U(1)_m\times \spin(3) \times \U(1)}{\Z_2}\,,
\ee
by the quotient of the diagonal $\Z_2$ elements of all four factors.
As a consequence, the anomaly has more terms than those considered before.
Nonetheless, the bosonization anomaly \eqref{anom for gauge spinc}
is part of the result in Ref.\cite{DumitrescuNiroThorngren2024QED3}:
in their Eq. (2.19), which we rewrite here,
\be
{\cal A}_{N_f=2}= i\pi\int_Y \frac{\d B_m}{2\pi} \wedge \frac{\d B_m}{2\pi}
- \frac{\d A_f}{2\pi} \wedge \frac{\d A_f}{2\pi}, 
\ee
the first term is identified with \eqref{anom for gauge spinc}, where
$B_m$ is the $\U(1)_m$ background, while the $A_f$ term pertains to the
Cartan subalgebra $\U(1)_f\subset \SU(2)_f$.
Note that in their analysis, the larger anomaly is crucial for imposing
the massless IR phase. It would be interesting to study this matching
for all even $N_f$ values in all detail.

\paragraph{$N_f=1$ theory.}
A single $2+1d$ fermion breaks time-reversal symmetry, explicitly by its
mass or by anomaly if massless, the breaking term being
the Chern-Simons action $\CS(a)$ with coupling $k={\rm sign}(m)/2$.
As discussed in section 3.4, in this case the
bosonization anomaly \eqref{anom for gauge spinc} reduces
to an allowed $2+1d$ counterterm, which is the Chern-Simons action
$\CS(B_m)$ with $k=1$. Therefore, the earlier analysis does not apply
in this case (and other odd $N_f$ values).
Nevertheless, let us show how our approach recovers 
the known results of $2+1d$ boson-fermion dualities
\cite{wittenwebofduality,CappelliVillaBosDual2025}.

The IR fixed point of QED$_3$ is conjectured in
\cite{wittenwebofduality}, by gauging the familiar three-dimensional
fermion-boson duality
\begin{equation}
\label{dirac boson duality}
\Bar{\psi} i \slashed{D}_{A} \psi \qquad
\longleftrightarrow \qquad |D_{b}\phi|^2 + V(|\phi|)+ \frac{i}{4\pi} b\d b +
\frac{i}{2\pi} b\d A,
\end{equation} 
where the complex scalar field on the right is suitably tuned to the
Wilson-Fisher fixed point and $A$ is a spin$_c$ connection ($b$ is
instead a standard $\U(1)$ gauge field). To make connection with our
discussion in section \ref{Grp ext sec}, we can integrate over $A$ by
splitting it as $(A_b,\eta)$, summing first over $\eta$ and then
$A_b$. Summing over $\eta$, i.e. gauging $\Z_2^f$, is already done in
\cite{CappelliVillaBosDual2025} and gives the bosonic dual of the
fermionic duality \eqref{dirac boson duality}:
\begin{equation}
\label{3d bosonization of dirac}
\begin{split}
    \sum_{\eta} \int_X \Bar{\psi} i \slashed{D}_{A_b,\eta} \psi + \frac{i}{\pi}\eta\wedge C\qquad &
\longleftrightarrow \\
\int_X|D_{2b}\phi|^2 + V(|\phi|)+\int_Y & \frac{4i}{4\pi} (\d b-C)\wedge(\d b-C) +\frac{i}{2\pi} (\d b-C) \wedge \d A_b +\d f(C).
\end{split}
\end{equation} 
$C$ is the $\Z_2$ 2-form background gauge field for
$\Z_2^{(1)}$. $\eta$ and $C$ are intended as constrained continuous
$\U(1)$ fields \cite{kapustinseiberg2014} and BF-like couplings are
written by extending $X$ to $Y$, with $\p Y =X$.\footnote{Note that
  the dynamical terms do not depend on the extension to $Y$. The
  non-dynamical terms, written solely in terms of the backgrounds $C$
  and $A_b$, give the anomaly \eqref{bos spinc anomaly}.} The
counterterm $f(C)$ is needed to ensure that the anomaly \eqref{bos
  spinc anomaly} takes the form $w_2\cup C$ also on the RHS:
\begin{equation}
    \d f(C) = \frac{i}{\pi} (\pi w_2) \wedge C - \frac{i}{\pi} C\wedge C.
\end{equation}
By the Wu formula, this is exact and thus local on $X$.

Now we integrate over $A_b$ on both sides of \eqref{3d bosonization of
  dirac}, together with a coupling to $B_m^b$, background for the dual
magnetic symmetry $\U(1)_{m}^b$ (a zero-form symmetry in $d=3$). On
the left side, we recover QED$_3$ with the magnetic coupling
\begin{equation}
\begin{split}
  &\int_Y \frac{i}{\pi} \d \eta\wedge C+\frac{i}{2\pi}\d a_b \wedge \d B_m^b  =\int_Y i\pi\, w_2 \wedge\frac C \pi + 2\pi i\, \frac{\d a_b}{2\pi}\wedge\left( \frac{\d B_m^b}{2\pi }+\frac{C}{2\pi}\right)= \\ &=\int_Y i\pi\, w_2 \wedge\frac C \pi + 2\pi i\, \frac{\d a}{2\pi}\wedge\left(2 \frac{\d B_m^b}{2\pi }+\frac{C}{\pi}\right) =\int_Y 2\pi i\,\left( \frac{\d a}{2\pi}+ \frac{w_2}2 \right)\wedge \frac{\d B_m}{2\pi },
\end{split}
\end{equation}
using the identification \eqref{Bm = Bmtilde + C}. On the right hand
side, on top of the complex scalar action coupled to $b$, we get,
integrating out $a_b$, $\d b = C-\d B_m^b$, such that
\begin{equation}
\begin{split}
       &\int_Y\frac{4i}{4\pi} (\d b-C)\wedge(\d b-C) +\frac{i}{2\pi} (\d b-C) \wedge \d a_b +\frac{i}{2\pi} \d a_b \wedge \d B_m^b+\d f(C)=
       \\ &= \int_Y\frac{4i}{4\pi} \d B_m^b \wedge\d B_m^b+\d f (C)=\int_Y\frac{i}{4\pi} \d B_m \wedge\d B_m+i\pi\, w_2\wedge \frac{\d B_m}{2\pi}.
\end{split}
\end{equation}
In this expression, we used \eqref{Bm = Bmtilde + C} globally, in the
notation here $\d B_m= 2\d B_m^b-2C$, and locally,
i.e. $2b=-2B_m^b=-B_m$. We also neglected a term
  $4\pi \,c_1(B_m^b)\wedge (C/\pi)=0 $ mod $2\pi \Z$. All
together, we obtain the following IR duality for three-dimensional
QED:
\begin{equation}
\label{IR FP QED3}
    \Bar{\psi} i \slashed{D}_{a} \psi +\frac{i}{2\pi}a\d B_m-\frac{i}{4\pi} B_m\d B_m\qquad 
    \longleftrightarrow    \qquad
    |D_{-B_m}\phi|^2 + V(|\phi|).
\end{equation} 
So strongly coupled $N_f=1$ QED$_3$ flows to the Wilson-Fisher fixed point in the IR.

This is the same result already obtained in \cite{wittenwebofduality} by gauging $A$ directly \eqref{dirac boson duality}. The RHS is manifestly bosonic and anomaly free, consistent with the fact that the anomaly \eqref{anom for gauge spinc} is not there if time reversal is broken, as $N_f=1$ QED$_3$ does. The counterterm on the LHS of \eqref{IR FP QED3} is exactly the one discussed in \eqref{CS ct B}.

\section{Conclusion and outlook}

In this work we discussed the global properties of superconductors
and the consequences of gauging the $\spinc$ symmetry.
A characteristic anomaly was analyzed which is present when the
Higgs phenomenon is due to pairing of electrons.
It follows from bosonization, the gauging of fermion
parity symmetry $\Z_2^f$, in dimensions three and four. Earlier studies
of bosonization were shown to be
relevant not only for superconductivity but also
for general gauged electronic systems.
For example, in identifying the minimal topological theory which describes
their massive non-trivial phases.

We also briefly discussed s-wave superconductors with
$q=2n>4$ and $p$-wave pairings, and verified the presence of
the bosonization anomaly in recently analyzed
$N_f=2$ electrodynamics in $2+1$ dimensions.
Our analysis of these systems is admittedly rather brief and will be continued in future work. In particular, it will be interesting 
to develop the consequences of gauging the $\spinc$ symmetry in
topological superconductors and other SPT phases.

Among the several directions that could be pursued, a very important one is the study of boundary conditions and
boundary degrees of freedom in superconductors, both s-wave and p-wave.
First of all, there exists many types of them, all physically relevant,
such as SC-metal, SC-vacuum, SC-TSC, SC-TI, SC-insulator-SC, etc..
The main question is how much the topological theory
can explain them, and thus predict robust physical consequences.
In particular, the characteristic anomaly described in this paper
might manifest itself in one of these settings. These very interesting
questions also need specific analyses.

Another theme which was only briefly touched upon is the effect of
gauging (i.e. bosonization) on the massless degrees of freedom that
are present on the boundary of topological superconductors.
It is not clear whether gauging
forbids boundary states of a single fermion, such as the Majorana.
This question was already asked in Ref. \cite{Moroz_gauging}, where it was
suggested that gauging at the boundary is not mandatory but could be a 
choice. A related, broader study is on the consequences of
gauging on our understanding of all other topological
superconductors which enter the tenfold classification.

\paragraph{Acknowledgments}

We would like to thank J. H. Bardarson, L. Chirolli, T. H. Hansson,
S. Moroz, K. Zarembo
for interesting discussions on superconductors and gauging.
MB thanks the G.Galilei Institute, Arcetri, Florence for hospitality.
AC thanks the Simons Center for Geometry and Physics, Stony Brook
for hospitality.

\appendix


\section{$\Z_2^f$ gauge theory}

\label{app Z2f gauge theory}

\paragraph{Calculus with $\Z_2$ cochains.}
Throughout the paper we often use the discrete formulation for
topological theories, involving cochains
and higher cup products, which may not have a corresponding
continuous representation in terms of U(1) differential forms. In particular,
the cup products obey modified graded commutativity and
Leibniz rule for the derivative (coboundary) as explained in
\cite{kapustinseiberg2014}. Let us give the simpler formulas in the
case of $\Z_2$ cochains, which are mostly used,
\begin{align}
\label{cup-alg}
& f \cup_k g +g \cup_k f= \d (f\cup_{k+1} g) + \d f \cup_{k+1} +
  f \cup_{k+1} \d g \,,
  \\
&  \d\left( f \cup_k g\right) = \d f \cup_k g + f \cup_k \d g +
f \cup_{k-1} g +g \cup_{k-1} f \,.
\end{align}
These relations are valid mod 2, so signs are not important, and for all
$\Z_2$ cochains $f,g$. Note that flat cochains obey ordinary rules of
differential calculus;
also $\cup_0\equiv \cup$ and $\cup_k$ can be discarded for $k<0$.\\
                                
\noindent A $\Z_n$ bosonic gauge theory in $d$ dimensions is a TQFT whose gauge
field is an element $[a]\in H^1(X;\Z_n)\simeq [X,B\Z_n]$ and the
action is given by a Dijkgraaf-Witten (DW) term
$[\w]\in H^d(B\Z_n;\R/\Z)$ as $\int_X a^*\w$. A `minimal' version is
obtained by considering the case with `zero action'. By introducing a
Lagrange multiplier $b$ to implement the constraint
$a\in Z^1(X;\Z_n)$, one arrives at (a discretized version of) level
$n$ BF theory \eqref{BF theory with backgrounds}, i.e.
\begin{equation}
\label{app Z_n gauge theory}
   S =\frac{2\pi i}{n} \int_X b \cup \d a,
\end{equation}
in cochain notation, $a\in C^1(X;\Z_n)$, $b\in C^{d-2}(X;\Z_n)$. This has a $\Z_n^{(1)}\times \Z_n^{(d-2)}$ symmetry \cite{gaiotto2015generalized} that can be probed by turning on its background fields $B\in Z^2(X;\Z_n)$, $C\in Z^2(X;\Z_n)$,
\begin{equation}
       S =\frac{2\pi i}{n} \int_X b \cup \d a + b\cup B +C \cup a.
\end{equation}
The $\Z_n^{(d-2)}$ symmetry is the dual symmetry of the zero-form $\Z_2$ gauge symmetry.\footnote{Think of \eqref{app Z_n gauge theory} as the result of gauging a trivially acting $\Z_n$ symmetry in the trivial theory $\1$.} The symmetry has a 't Hooft anomaly given by the SPT phase on a $d+1$ dimensional $Y$ \eqref{BF mixed anomaly}
\begin{equation}
\label{app BF mixed anomaly}
   \frac{2\pi i}{n} \int_Y C \cup B.
\end{equation}
For $n=2$, this is a $\Z_2$ gauge theory (e.g. in three dimensions it represents the low-energy limit of the toric code \cite{kitaev2003toricCode,mcgreevy2023generalized}). Notice that integrating out $a$ instead of $b$ tells us that a $\Z_n$ gauge theory is equivalent to a $\Z_n^{(d-3)}$ gauge theory for $b$.

We now apply a similar reasoning to the fermion parity symmetry $\Z_2^f$. A $\Z_2^f$ gauge theory is a $\Z_2$ gauge theory for the spin structures: it is topological field theory for an element $a\in C^1(X;\Z_2)$ such that $\d a =w_2(TX)$,\footnote{Here we call $a$ the spin structure, instead of $\eta$.} that we could take with zero action. Imposing this with a Lagrange multiplier yields 
\begin{equation}
\label{app Z_2^f gauge theory}
     S= i\pi \int_X b \cup (\d a+w_2(TX)).
\end{equation}
In this case the dual $\Z_2^{(d-2)}$ symmetry is broken on a non-spin manifold: the variation after $b \to b+\g$ is $\int \g w_2 $, which is indeed the bosonization anomaly \eqref{standard bos anom}. 

The $\Z_2^f$ gauge theory \eqref{app Z_2^f gauge theory} can be thought of as a $\Z_2$ gauge theory \eqref{app Z_n gauge theory} with a particular choice of symmetry fractionalization \cite{barkeshli2014symfrac,KomargodskiHsinSymFrac2022,DumitrescuCordovaBrennanSymFrac2022,brennanJacobsonRoumpedakis2025SymFrac,HsinShao2020}: $B = w_2$, with $B$ the background of the $\Z_2^{(1)}$ symmetry. While in the BF theory \eqref{app Z_n gauge theory} the Wilson line of $a$ represents a bosonic particle, this choice of symmetry fractionalization makes $e^{i\oint a}$ fermionic in \eqref{app Z_2^f gauge theory} (it is the Wilson line of the spin structure) \cite{thorngren2014framed,Villa2026Extensions}. Since this Wilson line is the generator of $\Z_2^{(d-2)}$, its fractionalization with respect to spacetime $\SO(d)$ implies the mixed gravitational anomaly \eqref{standard bos anom} \cite{Villa2026Extensions,brennanJacobsonRoumpedakis2025SymFrac}. Alternatively, the anomaly \eqref{standard bos anom} follows from the BF mixed anomaly \eqref{app BF mixed anomaly} choosing $B= w_2(TX)$; notice however that this discussion involving the $\Z_2^{(1)}$ symmetry is not intrinsic from a general bosonization point of view, which, on general grounds, implies the existence of just the fractionalized $\Z_2^{(d-2)}$ symmetry.

As already discussed, $\Z_2$ gauge theory \eqref{app Z_n gauge theory}
is equivalent to $\Z_2^{(d-3)}$ gauge theory by integrating out $a$.
The resulting theory is a $\Z_2^{(d-3)}$ gauge theory for $b$ with
action (on closed orientable $X$)
\begin{equation}
\label{app twisted Z_2 b action}
   i\pi \int_X b \cup w_2 = i\pi \int_X Sq^2 b= i\pi \int_X b\cup_{d-4}b,
\end{equation} 
by making use of the Wu formula and the definition of the Steenrod square \cite{kapustinthorngren2017,Steenrod1947,MilnorStasheff}. This corresponds to a twisted $\Z_2^{(d-3)}$ gauge theory, that, thanks to the Wu formula, can be described just in terms of $b$: the action \eqref{app twisted Z_2 b action} should be described by a DW term $[\w]\in H^d(B^{d-2}\Z_2;\R/\Z)$, with $B^{d-2}\Z_2\simeq K(\Z_2,d-2)$. This can be done case by case, see below. So, a minimal $\Z_2^f$ gauge theory corresponds to a twisted $\Z_2^{(d-3)}$ DW theory with action \eqref{app twisted Z_2 b action}.

One can add again a Lagrange multiplier $\wh a$ to enforce $\d b=0$, yielding\footnote{This is similar to the discussion in \cite{ThetaTRtemperature2017}, appendix A. There, $d=2$ BF theory $i\pi \int b \d a$ is generalized to non-orientable manifolds as $i\pi \int b \d a + b^2$. With the same trick, it is rewritten as $i\pi\int b\d a'+bw_1$. Here $a\neq a'$ (since $a'\to a'+f$ when $w_1\to w_1+\d f$) and the $\Z_2^{(1)}$ symmetry $b\to b+\g$ is broken when $[w_1]\neq 0$.}
\begin{equation}
     i\pi \int_X b \cup \d \wh a+ b \cup_{d-4} b + g(b), \qquad g(b)=0 \;\text{ if }\; \d b=0.
\end{equation}
 In the process, we could allow undetermined terms $g(b)$ such that $g(b)=0$ when $\d b=0$ (since integrating over $\wh a$ still yields $\int Sq^2b$). This extra term can also be understood as a consequence of lifting the cohomological operation $Sq^2b$ to non closed $b$: while the action \eqref{app twisted Z_2 b action} is gauge invariant for $b\to b+\d \l$, this is not the case for $b$ not closed. Requiring gauge invariance fixes $g(b)=b\cup_{d-3}\d b$, so that the final action is
\begin{equation}
\label{app Z_2^f gauge theory after Wu}
     i\pi \int_X b \cup \d \wh a+ b \cup_{d-4} b + b\cup_{d-3}\d b.
\end{equation}
It is reasonable to guess that the anomaly of $\Z_2^{(d-2)}$ in \eqref{app Z_2^f gauge theory after Wu} has the expression in terms of $Sq^2C$ in \eqref{standard bos anom}. This is indeed the case, as it can be simply computed by coupling the action to $C$ as
\begin{equation}
\label{app Z_2^f gauge theory after Wu with C}
i\pi \int_X b \cup \d \wh a+ b \cup_{d-4} b + b\cup_{d-3}\d b +
C\cup \wh a + C\cup_{d-2}\d b.
\end{equation}
The coupling to $b$ is needed to cancel variations depending on the dynamical field. Then, under $C\to C+\d \g$, $b\to b+\g$, the anomaly is given by inflow by
\begin{equation}
    i\pi \int_Y Sq^2 C = i\pi \int_Y C\cup_{d-3} C.
\end{equation}
In the following discussion of the action
\eqref{app Z_2^f gauge theory after Wu} we drop the hat on $a$.

\subsection{$1+1$ dimensional case}

In $d=2$ every manifold is spin, therefore $w_2=\d \eta$. The action \eqref{app Z_2^f gauge theory} can be rewritten as a standard $\Z_2$ gauge theory in terms of $a' = a+\eta$ (which is a standard $\Z_2$ gauge field, $\d a'=0$ and it does not transform under $w_2\to w_2+\d f$). This is expected, because $d=2$ is a limiting case, since the anomaly \eqref{standard bos anom} vanishes and therefore the dual $\Z_2$ symmetry is generic, since there is no much difference between bosons and fermions in $2d$. It follows that any theory with a $\Z_2$ symmetry can be suitably fermionized \cite{gaiottokapustinspinTQFT1,shaoTASI2023,thorngren2020anomalies}, which is just the old familiar bosonization method \cite{coleman1975Bos,ginspargCFT}. See also the discussion in the appendix of \cite{CappelliVillaBosDual2025}.

Consistently, the dual description in terms of $b$ has no twist, given that $Sq^2b=0$. Here $b$ is a zero-cochain, so it is just a function from $X$ to $\Z_2$. If we think of this as a zero-form gauge field for a $(-1)$-form $\Z_2$ symmetry,\footnote{Like a periodic scalar can be thought of as a $0$-form gauge field for a $(-1)$-form $\U(1)$ symmetry.} then we have a notion of classifying space $K(\Z_2,0)$ which is indeed isomorphic to $\Z_2$, so that $b: X \to \Z_2$. Being a set of points, its cohomology groups are trivial, so in particular $H^2(K(\Z_2,0);\R/\Z)\simeq 0$ and no DW twist is allowed.

\subsection{$2+1$ dimensional case}

In $d=3$ every manifold is spin and a discussion similar to the previous case applies. In particular, \eqref{app Z_2^f gauge theory} is equivalent to a familiar $\Z_2$ gauge theory \cite{HsinShao2020}, AKA the toric code, and indeed in the dual formulation \eqref{app twisted Z_2 b action} the action for $b$ is again zero, since $Sq^2b=0$. However, there can be added some comments.

In three dimensions there is a clear distinction between bosons and fermions. This is reflected in the bosonization anomaly \eqref{standard bos anom} that it is not trivial. Even if every three-manifold is spin, this is not the case for four-manifolds, therefore the theory could depend on the extension if the bulk is not chosen appropriately. The equivalence between \eqref{app Z_2^f gauge theory} and \eqref{app Z_n gauge theory} is because $\Z_2$ gauge theory has already a fermionic particle in its spectrum, namely 
\begin{equation}
\label{app BF level 2 fermion}
    \psi  = e^{i\pi\oint a+b}
\end{equation}
(as a consequence of the linking interaction between the Wilson lines of $a$ and $b$), which is the generator of the diagonal symmetry of $\Z_2^{(1)}\times \Z_2^{(1)}$. This anyon cannot condense unless a spin structure is chosen, reflected in the self-anomaly \cite{gaiottokapustinspinTQFT1},
\begin{equation}
\label{app bos anom 3d}
    \mathcal{A}_b = i\pi \int_Y Sq^2 B = i\pi \int_Y B\cup B,
\end{equation}
which is \eqref{standard bos anom}. This can be computed from \eqref{app Z_n gauge theory} turning on backgrounds $B=C$. The fact that the toric code has the correct properties to be fermionized is used in \cite{gaiottokapustinspinTQFT2} to classify three-dimensional fermionic SPTs, by first enriching the toric code with a global symmetry and then gauging the diagonal $\Z_2^{(1)}$ symmetry with a choice of spin structure. 

In the dual formulation in terms of $b$, since $Sq^2b=0$, there is no need to introduce the extra term $b\cup \d b$ in \eqref{app Z_2^f gauge theory after Wu} in the first place. However, this is in principle non-zero in three dimensions and can be considered, but it is trivial: the action $b\cup \d a +b \cup \d b$ is equivalent to \eqref{app Z_n gauge theory} by a simple change of variables $c= a+b$.

From a DW theory, the possible twists for $b$ are classified by elements in $H^3(B\Z_2;\R/\Z)$. To see what this implies, let us consider the more general case $\Z_n$ first. The group cohomology of $\Z_n$ is given by $H^{2k}(B\Z_n;A)=A/nA$, $H^{2k+1}(B\Z_n;A)=\tor(A,n)$ for an Abelian group $A$. In particular, $H^3(B\Z_n;\R/\Z) \simeq \Z_n \simeq H^4(B\Z_n;\Z)$, where the isomorphism is given by the Bockstein of the sequence $\Z\to \R \to \R/\Z$ given that $H^*(B\Z_n;\R)\simeq 0$. An element $k[\w]\in H^3(B\Z_n; \R/\Z)\simeq \Z_n$, $k\in \Z_n$, can thus be written in terms of $x_4= x\cup x= \b(\w)\in H^4(B\Z_n;\Z)$, with $x$ the generator of $H^2(B\Z_n;\Z)$, which can in turn be written, being $n$-torsion $n[x]=0$, as $x =\b'(y)$, with $[y]\in H^1(B\Z_n;\Z_n)$, the identity in $[B\Z_n,B\Z_n]$, and $\b'$ the Bockstein for $\Z \to \Z\to \Z_n$. Therefore, 
\begin{equation}
    b^*x_4 = b^* \b'(y)\cup \b'(y)=\b'(b)\cup \b'(b)= \frac{1}{n^2}\d \wt b\cup \d \wt b = b^* \d \w,
\end{equation}
for $\wt b\in C^1(X;\Z)$ a lift of $b$. The possible DW actions for a three-dimensional $\Z_n$ gauge theory are then 
\begin{equation}
    2\pi i\int_X kb^*\w=\frac{2\pi i}{n} \int_X  \frac{k}{n} \wt b\cup \d \wt b = \frac{2\pi ik}{n} \int_X  b\cup \b'( b).
\end{equation}
These can be thought of as a discrete version of the familiar Chern-Simons action (the expression in terms of $x_4$ by an extension to a bulk $Y$ is similar to the definition of $A\d A$ in terms of $F\wedge F$ \cite{seibergwitten2016gappedTI,wittenwebofduality}). 

For $n=2$, the only non trivial action is given by $1/2 b\cup \d b$. We see that this is not the $Sq^2b$, which is consistently zero, and the other twisted term appearing in \eqref{app Z_2^f gauge theory after Wu} has $k=2 \sim 0$, so it is indeed trivial.

\subsection{$3+1$ dimensional case}

In four dimensions not every manifold is spin, therefore the $\Z_2^f$ theory \eqref{app Z_2^f gauge theory} is different from the level 2 BF theory. It is also the first case where the torsion term is present, $b\cup b = Sq^2 b$. It is thus equivalent to a twisted $\Z_2^{(1)}$ gauge theory with non-trivial action $Sq^2b$.

The dual formulation in terms of $b$, written as in \eqref{app Z_2^f gauge theory after Wu}, is
\begin{equation}
\label{app twisted BF level 2}
    i\pi \int_X b\cup \d a + b\cup b + b \cup_1 \d b. 
\end{equation}
This has appeared in various forms in the literature \cite{ye2015vortex,YeGu2014BTIs,kapustinseiberg2014,kapustinthorngren2017}. Written in this way, it is the seed theory considered in \cite{kapustinthorngren2017}, where it is used to produce four-dimensional fermionic SPTs by fermionization using the same strategy mentioned above for the toric code. It is the discretized version of a twisted BF theory \cite{gaiotto2015generalized,kapustinseiberg2014}; see also \cite{hsinlam2020discretetheta,HsinLamSeiberg20181formSym3d4d} for this perspective and appendix \ref{app twisted BF}. It is sometimes called (four-dimensional) `fermionic' toric code, where the Wilson line of $a$ represents a fermionic particle, already mentioned for the low-energy theory of a Walker-Wang model \cite{walkerwang2011} and a superconductor in \cite{vonKeyserlingk2014WWModelBF}. It is argued in \cite{vonKeyserlingk2025finiteTTO} that \eqref{app twisted BF level 2} could imply a finite temperature topological order due to the presence of an anomalous $\Z_2^{(2)}$ symmetry. Equivalently to \eqref{app Z_2^f gauge theory}, the $\Z_2^{(2)}$ symmetry of \eqref{app twisted BF level 2} is broken on non-spin manifolds: under $b\to b+\g$, with $\g \in Z^2(X;\Z_2)$, the action has a variation $\int_X \g \cup \g = \int_X w_2 \cup \g$. When $X$ is spin, the $\Z_2^{(2)}$ symmetry is restored and the theory can be coupled to $C \in Z^3(X;\Z_2)$ as \eqref{app Z_2^f gauge theory after Wu with C}, 
\begin{equation}
\label{app twisted BF lev 2 with C}
        i\pi\int_X b\cup \d a + b\cup b +b \cup_1 \d b + C \cup a + C\cup_2 \d b.
\end{equation}
It has the anomaly \eqref{standard bos anom}, here $\int_Y C \cup_1 C$.

The action for $b$, namely $Sq^2b$, can again be understood in a (generalized) DW theory, as explained in \cite{CordovaBeniniHsin2group,kapustinthorngren2013TQFTlattice}. For a $\Z_n^{(1)}$ gauge theory, the possible actions are classified by $H^4(B^2\Z_n;\R/\Z)$, with $B^2\Z_n\simeq K(\Z_n,2)$. For a general coefficient Abelian group $A$, there is the following isomorphism \cite{EilenbergMacLaneOriginal}
\begin{equation}
    H^4(K(\Z_n,2);A)\simeq \hom(\Z_{n\gcd(n,2)},A).
\end{equation}
So, DW actions are given by $q \in H^4(K(\Z_n,2);\R/\Z)\simeq \hom(\Z_{n\gcd(n,2)},\R/\Z)\simeq \wh \Z_{n\gcd(n,2)}$. Moreover, taking $A  \simeq \Z_{n\gcd(n,2)}$, we see that there is a preferred element in the cohomology group $H^4(K(\Z_n,2);\Z_{n\gcd(n,2)}) \simeq \hom(\Z_{n\gcd(n,2)},\Z_{n\gcd(n,2)})$, namely the identity morphism, which corresponds to a natural cohomological operation \cite{HatcherAT,FomenkoFuchs2016} $\mathcal{P}: H^2(-;\Z_n)\to H^4(-;\Z_{n\gcd(n,2)})$. This operation is called the Pontryagin square \cite{whitehead1949,whitehead1950} and it follows that DW actions are given by 
\begin{equation}
\label{app Pontryagin square DW action}
    2\pi i\int_Xq(b^*\mathcal{P}) = 2\pi i\int_X q(\mathcal{P}(b)) = \frac{2\pi i p}{n\gcd(n,2)}\int_X \mathcal{P}(b),\qquad p\in \Z_{n\gcd(n,2)}.
\end{equation}
This operation can be explicitly defined as\footnote{The explicit definition works for a general map $H^k(X;\Z_n)\to H^{2k}(X;\Z_{n\gcd(n,2)})$.}
\begin{equation}
\label{app pontryagin square def}
    \mathcal{P}(b)=\begin{cases}
        \wt b \cup \wt b - \wt b \cup_1 \d \wt b, \qquad &n \in 2\Z,\\
        b\cup b, \qquad &n\in 2\Z+1.
    \end{cases} 
\end{equation}
for a lift $\wt b\in C^2(X;\Z)$.  Indeed, it is easy to check\footnote{Using
\[ a \cup_i b - (-1)^{pq-i} b \cup_i a = (-1)^{p+q-1-i}(\d (a \cup_{i+1} b) -\d a \cup_{i+1} b - (-1)^pa \cup_{i+1} \d b),\quad \cup_0 \coloneqq \cup,\]
with $a\in C^p(X;\Z_k)$ and $b\in C^q(X;\Z_k)$ \cite{CordovaBeniniHsin2group,Steenrod1947}.} that \eqref{app pontryagin square def} is closed mod $n\gcd(n,2)$ and well-defined (i.e. by considering $\mathcal{P}(b+\d \l)$ for a different representative of $[b]$ and by considering an equivalent lift $\wt b' = \wt b+n f$, $f\in C^2(X;\Z)$, $\mathcal{P}(b)$ defines an element mod $n\gcd(n,2)$ in cohomology, i.e. up to exact terms). Notice that $\mathcal{P}(b) = b\cup b \mod n$.\footnote{Since $\d \wt b =n w$, $w\in Z^3(X;\Z)$. If $b$ has a lift to $\wt b \in Z^2(X;\Z)$, then $\mathcal{P}(b)=\wt b \cup \wt b \mod 2n$.} 

Consider the case $n=2$, where the possible DW actions are given by 
\begin{equation}
     \pi i \frac{p}{2}\int_X \mathcal{P}(b),\qquad p\in \Z_4.
\end{equation}
For $p=2$, only $\mathcal{P}(b) \mod 2$ matters, so $\mathcal{P}(b)=b\cup b = Sq^2(b) \mod 2$, which is indeed the action \eqref{app twisted Z_2 b action}. Therefore, the $\Z_2^f$ gauge theory \eqref{app Z_2^f gauge theory} is equivalent to a (generalized) $\Z_2^{(1)}$ DW theory labeled by $p=2 \in \Z_4\simeq H^4(K(\Z_2,2);\R/\Z)$.

\subsection{$4+1$ dimensional case}
As in four dimensions, the $\Z_2^f$ gauge theory \eqref{app Z_2^f gauge theory} is different from the level 2 BF theory and dual to a $\Z_2^{(2)}$ gauge theory with twisted action \eqref{app twisted Z_2 b action}, i.e. $Sq^2b$. This applies similarly to every higher dimension.

The action for $b$ is still described by a generalized DW theory. Here the possible actions are classified by $H^5(B^3\Z_n;\R/\Z)$, with $B^3\Z_n \simeq K(\Z_n,3)$. For a generic Abelian group $A$ \cite{EilenbergMacLaneOriginal}
\begin{equation}
    H^5(K(\Z_n,3);A)\simeq \hom(\Z_n/2\Z_n, A) \simeq \hom(\Z_{\gcd(n,2)},A).
\end{equation}
This is trivial for $n$ odd, so also no DW twists are allowed. For $n=2m$ even, $H^5(K(\Z_n,3);A)\simeq  \hom(\Z_2,A)$. Taking $A \simeq \R/\Z$, we see that there is only one possible non trivial term $\phi \in \hom(\Z_2,\R/\Z)\simeq\wh \Z_2$. To understand it, take $A\simeq \Z_2$, which gives a preferred cohomological operation $H^3(-;\Z_{2m})\to H^5(-;\Z_2)$ corresponding to the identity in $\hom(\Z_2,\Z_2)$. If $n=2m=2$ the only option is indeed $Sq^2$; for general $n=2m$, we can take it as the combination of the mod $2$ reduction of $b$ and $Sq^2$. We thus obtain the only possible non-trivial DW action for a five dimensional $\Z_{2m}^{(2)}$ gauge theory,
\begin{equation}
\label{app Zn 5d dw theory}
    2\pi i p\int_X \phi(Sq^2(\wt b))= \pi ip\int_X Sq^2 \wt b, \qquad \wt b \coloneqq b\mod 2, \qquad p\in \Z_2.
\end{equation}
This is indeed the action for $b$ in \eqref{app twisted Z_2 b action} for $p=1$.\\

\noindent Notice that the DW actions \eqref{app Pontryagin square DW action} and \eqref{app Zn 5d dw theory}, considering $b$ as a background field, are the SPT phases that classify the anomaly for a $\Z_n$ symmetry generated by lines in dimensions three and four, respectively. In three dimensions there can be anyons and these richer statistics is reflected in the possible anomalies \eqref{app Pontryagin square DW action} (where $p/n\gcd(n,2)$ is the spin of the generating anyon). In dimensions four, particles are either bosons or fermions, thus the anomaly is a mod 2 index \eqref{app Zn 5d dw theory} (either representing an anomaly free boson, or a fermion with the bosonization anomaly).


\section{Twisted BF theory}
\label{app twisted BF}

In this appendix we collect some facts regarding the twisted BF theory in four dimensions. See also \cite{kapustinseiberg2014,gaiotto2015generalized,HsinLamSeiberg20181formSym3d4d,hsinlam2020discretetheta,ThetaTRtemperature2017,vonKeyserlingk2014WWModelBF,vonKeyserlingk2025finiteTTO}.

\subsection{Cochain formulation}

Consider the action \eqref{app Pontryagin square DW action}. Rename $n \to k$ and, when $k$ is odd, we rescale $p\to 2p$, so that $p\in \Z_{2k}$ in both cases but when $k$ is odd only even values of $p$ are considered. This leads to the following action, with $b\in Z^2(X;\Z_k)$ and $p\in \Z_{2k}$:\footnote{Notice that for $p$ even only $\mathcal{P}(b)$ mod $k$ matters, so $\mathcal{P}(b) = b\cup b $ also for even $k$.}
\begin{equation}
\label{pontryagin square action}
    S = \frac{2\pi i p}{2k} \int_X   \mathcal{P}(b) =\begin{cases}
        \frac{2\pi i p}{2k} \int_X   b \cup b, \qquad &p\in 2\Z,\; k \in \Z\\
        \frac{2\pi i p}{2k} \int_X   \wt b\cup \wt b- \wt b\cup_1 \d \wt b,\qquad &p\in 2\Z +1, \;k \in 2\Z.
    \end{cases}
\end{equation}
It has always $pk \in2\Z$. One can extend the definition \eqref{pontryagin square action} to both $p$ and $k$ odd but only on spin manifolds, so that the resulting theory is a spin TQFT. Here we are mostly interested in the bosonic case.

For $p=0$, \eqref{pontryagin square action} is a zero-action level $k$ BF theory in four dimensions, with symmetry $\Z_k^{(1)}\times \Z_k^{(2)}$ \cite{gaiotto2015generalized}. For $p\neq 0$, the global symmetry is reduced to $\Z_L^{(1)}\times \Z_L^{(2)}$, with $L\coloneqq \gcd(k,p)$. It is easy to see the fate of the electric two-form symmetry from \eqref{pontryagin square action}. Consider the naive shift $b\to b+\b'$, with $\b' \in Z^2(X;\Z_k)$. The variation of the action \eqref{pontryagin square action} is\footnote{This follows from $\mathcal{P}(x+y) = \mathcal{P}(x)+\mathcal{P}(y)+2x\cup y$, mod $2k$ and up to exact pieces, as can be easily computed from the definition \eqref{app pontryagin square def}, i.e. $\mathcal{P}$ is the quadratic refinement of twice the cup product \cite{kapustinthorngren2013TQFTlattice}.}
\begin{equation}
    \D S= \frac{2\pi i p}{2k} \int_X  2b \cup \b' + \mathcal{P}(\b').
\end{equation}
The first term vanishes mod $2\pi\Z$ if $\b' = k/L \b$, with $\b\in Z^2(X;\Z_L)$. The second term thus becomes
\begin{equation}
\label{variation pontr action two form sym}
    \D S= \frac{2\pi i pk}{2L^2} \int_X \b\cup \b =2\pi i \frac{rs}{2} \int_X \b\cup \b, \qquad k = rL, \; p = sL.
\end{equation}
For $rs\in 2\Z$, we see that the $p$-twisted theory \eqref{pontryagin square action} has a $\Z_L^{(2)}$ symmetry. This exhausts the cases for $L$ odd, when $p$ and $k$ have different parity. For even $L$, the symmetry is there for $rs\in 2\Z$ and broken otherwise. In this latter case, the variation is just a $\Z_2$ sign given by
\begin{equation}
\label{ZL^2 anomaly Pontr action discrete}
    \D S= \pi i rs \int_X \b\cup \b= \pi i rs \int_X w_2(TX) \cup \b.
\end{equation}
This variation is trivial on spin manifolds, therefore we could say that (a $\Z_2$ subgroup of) the $\Z_L^{(2)}$ symmetry has a gravitational anomaly controlled by $w_2$. This is exactly the bosonization anomaly \eqref{standard bos anom}, which is captured by inflow by the SPT phase $w_2\cup C = Sq^2C$, where $C$ is the (reduction mod $2$) of the $\Z_2^{(2)}$ background. Notice that all this discussion is coherent with \eqref{app Zn 5d dw theory}: a $\Z_L^{(2)}$ symmetry in four dimensions can be anomalous only for $L$ even with the bosonization anomaly. This also says that the topological line generating $\Z_L^{(2)}$, namely the 't Hooft line $V(\g)$ of $b$, is a fermionic particle for $rs$ odd (and bosonic otherwise).

The reduction of the two-form symmetry to $\Z_L$ implies that also the dual magnetic one-form symmetry is $\Z_L$. This can be understood as follows. The $\Z_L^{(2)}$ symmetry is generated by 't Hooft line operators $V(\g)$, dual to $b$, which are defined by the locus $\g$ such that $\d b = - k/L\,\PD(\g)$, for $\PD(\g)\in Z^3(X;\Z_L)$. This twists the boundary condition of $b$ in the path integral and gives 
\begin{equation}
\label{VU correlator discrete}
    \braket{V(\g)U(\S)}=e^{-\frac{2\pi i}{L}\link(\g,\S)}(-1)^{rs\, \rm Int(\S',\S')}, \qquad U(\S) \coloneqq \exp{\left(\frac{2\pi i}{k}\int_\S b\right)},
\end{equation}
where $\p\S' = \g$ (so $b = -k/L\,\PD(\S')$).\footnote{One can check that the action \eqref{pontryagin square action} is still well-defined, i.e. invariant under $b\to b+\d \l+kf$, also for $b= -k/L\,\PD(\S)$.} Neglecting the extra sign for a moment, the linking term gives the action of $\Z_L^{(2)}$ on $U(\S)$. However, the interpretation of \eqref{VU correlator discrete} can be reversed: since $U(\S)$ is topological, it is the generator of a one-form symmetry whose charged objects are the 't Hooft lines $V(\g)$: this is manifestly a $\Z_L$ action. The correlator also implies the mixed anomaly \eqref{app BF mixed anomaly}, as can be seen by adding a coupling $b \cup B$ to the action \eqref{pontryagin square action}, with $B$ the $\Z_L^{(1)}$ background, which breaks the two-form symmetry $b\to b+\b$.

Now we consider the extra sign in \eqref{VU correlator discrete}, which comes from the non-trivial action \eqref{pontryagin square action}. It gives the one-point function
\begin{equation}
\label{V discrete 1pt}
    \braket{V(\g)} = (-1)^{rs \int_X \PD(\S)\cup\PD(\S)+ \PD(\S)\cup_1\PD(\g)}, \qquad \p\S = \g.
\end{equation}
The extra mysterious term involving $\cup_1$ is there only for $rs \in 2\Z$ in principle, that is why we neglected it in \eqref{VU correlator discrete}. However, we can formally consider it for every value of $k$, $p$ in \eqref{pontryagin square action}, which is convenient in the following. The one-point function \eqref{V discrete 1pt} says that, when $rs$ is odd, the definition of $V(\g)$ requires a choice of framing. This implies a subtle dependence of $V$ on a surface bounding $\g$. Indeed, comparing $V(\g)$ defined using either $\S$ or $\S'$ such that $\p \S =\p\S'=\g$,  we obtain
\begin{equation}
\label{V discrete is fermion}
    \frac{\braket{V(\g,\S)}}{\braket{V(\g,\S')}} = (-1)^{rs \int_X \PD(\W)\cup \PD(\W)} =  (-1)^{rs \int_X \PD(\W)\cup w_2} = (-1)^{rs \int_\W w_2}.
\end{equation}
Here $\W =\S-\S'$ is the closed surface obtained by gluing $\S$ and $\S'$ along their common boundary $\g$ (reversing the orientation of $\S'$) and to arrive at \eqref{V discrete is fermion} it is necessary to take into account the $\cup_1$ term in \eqref{V discrete 1pt} (which removes the one coming from $\PD(\S)\cup\PD(\S)$ after expanding $\PD(\S)=\PD(\S')+\PD(\W)$). Notice that, while the framing can depend on choices, \eqref{V discrete is fermion} is independent of any choice and it says that the 't Hooft line $V(\g)$ carries a topological surface dependence due to a non trivial class $w_2(TX)$, only for $rs$ odd. So, for $rs$ odd, $V(\g)$ is a fermionic particle \cite{thorngren2014framed}, coherently with the gravitational anomaly \eqref{ZL^2 anomaly Pontr action discrete}.

The discussion for the symmetries is expected for $k=p=2$, since \eqref{pontryagin square action} is equivalent to \eqref{app Z_2^f gauge theory}, see appendix \ref{app Z2f gauge theory}. More generally, when $p=k$ even, the action \eqref{pontryagin square action} becomes 
\begin{equation}
\label{bb action as w2 b}
    S =  i\pi \int_X  b\cup b = i\pi \int_X w_2(TX)\cup b 
\end{equation}
for all $k$. So only $b$ mod $2$ is relevant. This is thus equivalent to $\Z_k$ gauge theory with zero-action \eqref{app Z_n gauge theory} with a background $B$ of the one-form symmetry $\Z_k^{(1)}$ activated by $B= k/2\, w_2(TX)$. This shows clearly that the Wilson lines charged under it, i.e. the 't Hooft lines $V(\g)$, are fermionic particles that carry a $w_2$ surface dependence \cite{thorngren2014framed} as discussed above. This is an instance of symmetry fractionalization \cite{KomargodskiHsinSymFrac2022,DumitrescuCordovaBrennanSymFrac2022,brennanJacobsonRoumpedakis2025SymFrac}. So, for $p=k$, the twisted BF theory \eqref{pontryagin square action} is equivalent to a $\Z_k$ gauge theory with a specific choice of symmetry fractionalization that makes the Wilson line a fermion. See the discussion for $k=p=2$ in appendix \ref{app Z2f gauge theory}.

Everything said so far can also be shown by adding a Lagrangian multiplier dual field $a$ in the cohomological action \eqref{pontryagin square action}, which becomes
\begin{equation}
\label{twisted BF action cochain}
    S =  \int_X \frac{2\pi i}{k} b \cup \d a + \frac{2\pi ip}{2k}\left( b\cup b -b\cup_1 \d b\right).
\end{equation}
For $p=k=2$, we obtain the twisted BF theory already considered for bosonization, namely \eqref{app twisted BF level 2}. This is a particularly convenient formalism that treats democratically both higher-form symmetries. The 't Hooft lines $V$ of $b$ are the Wilson lines of $a$ (up to other terms, since $\d a \neq0$  by itself). To achieve gauge invariance, $a$ has a twisted gauge transformation  
\begin{equation}
\label{twisted BF gauge trans discrete}
    b \to b+\d \l, \qquad a \to a-p\l +\d \e.
\end{equation}
This will be discussed also in the continuous approach below. Notice that the extra term using $\cup_1$ is required to ensure gauge invariance also for $k$ odd. However, the action \eqref{twisted BF action cochain}, which should be thought to be written in terms of lifts of $a$ and $b$ to integral valued cochains $\wt a$, $\wt b$, is correctly invariant under $\wt a \to \wt a+kg$, $\wt b\to \wt b+kf$, with $g \in C^1(X;\Z)$, $f\in C^2(X;\Z)$ only for $p$ even. When $p$ is odd it is invariant only on-shell (for $\d \wt b= kw$).\footnote{The variation is $\D S =-\int_X i\pi p\,\d f \cup_2\d \wt b$.} Despite this subtlety, it is easy to analyze the higher form symmetries in this language.

The one-form symmetry of \eqref{twisted BF action cochain} is $a\to a+\a$, $\a \in Z^1(X;\Z_k)$. Because of the gauge transformations \eqref{twisted BF gauge trans discrete}, $p\a$ is actually a gauge transformation, so that the true symmetry variation is given by $\a$ valued in $\Z_k$ mod $p\Z_k$. Therefore, the one-form symmetry is $\Z_L^{(1)}$, using $\Z_L\simeq \Z_k/p\Z_k$.\footnote{Consider the projection $\pi:\Z_k\to\Z_L$ given by $x\mapsto\pi(x)= rx$, with $k=rL$. This is surjective and $\ker(\pi)= \{ Lx \}\simeq \Z_r$. It follows that $\Z_L\simeq \Z_k/\Z_r$. $p\Z_k$ is a finite Abelian group generated by $px$, so it has order $r$, given that $rp\,x = np/L\, x =0$. Therefore $p\Z_k\simeq \Z_r$ and $\Z_L \simeq \Z_k/p\Z_k$. } For the two-form symmetry, the discussion is completely analogous to \eqref{variation pontr action two form sym}. Namely, the variation $b\to b+k/L\,\b$ is a symmetry for $\b \in Z^2(X;\Z_L)$ modulo the gravitational anomaly \eqref{ZL^2 anomaly Pontr action discrete}.

The discussion can be completed by turning on the backgrounds $B$ and $C$ for the one-form and two-form symmetries, respectively. Adding the coupling $b\cup B$, i.e. the minimal coupling
\begin{equation}
    S \supset \frac{2\pi i}{k}\int_X b\cup (\d a - B'),\qquad B'=\frac{k}{L}B,
\end{equation}
shows that the two symmetries have a mixed anomaly
\begin{equation}
    \frac{2\pi i}{L} \int_Y C \cup B,
\end{equation}
as a $\Z_L$ gauge theory \eqref{app BF mixed anomaly}. This is a consequence of \eqref{VU correlator discrete}, as mentioned. The coupling to $C$ can be written as
\begin{equation}
\label{generic twisted BF with C cochain}
    S =  \int_X \frac{2\pi i}{k} (C'-\d b) \cup a + \frac{2\pi ip}{2k}\left( b\cup b -b\cup_1 \d b\right) + \frac{2\pi i p}{2k}C' \cup_2 \d b, \quad C' = \frac{k}{L}C.
\end{equation}
The non minimal term $C\cup_2 \d b$ is needed to remove variations depending on $b$, as already observed for $k=p=2$ in \eqref{app twisted BF level 2}. Under $b\to b+k/L\, \b$, $C \to C+\d \b$, the variation is 
\begin{equation}
    \D S= \pi i rs \int_X C\cup_2 \d \b + \b \cup \b-\b \cup_1 \d \b, \qquad C, \b .
\end{equation}
This is again \eqref{ZL^2 anomaly Pontr action discrete}, namely the bosonization anomaly in the dual formulation after Wu formula given by the SPT phase $Sq^2C=C\cup_1 C$. Again, there is anomaly only for $rs$ odd.

\subsection{Continuum version}

In this section we review the continuous field formulation of \eqref{pontryagin square action} \cite{gaiotto2015generalized,kapustinseiberg2014}. It is mostly a rephrasing of the results above in a different language, to understand what the continuum picture can or cannot describe.

The action \eqref{pontryagin square action} can be written as
\begin{equation}
    S = \int_X \frac{kp}{4\pi} b\wedge b, \qquad \oint b \in \frac{2\pi}{k}\Z,
\end{equation}
sending $b \to k/2\pi \,\wt b$, with $\wt b $ a two-form $\U(1)$ flat gauge field with quantized periods, and further renaming $\wt b$ as $b$. Adding a Lagrange multiplier one-form $\U(1)$ gauge field $a$ yields the twisted BF action
\begin{equation}
    \label{twisted BF generic}
    \begin{split}
    S &= \int_X \frac{k}{2\pi} b\wedge \d a+ \frac{kp}{4\pi}b \wedge b =\\
      &= \int_X \frac{k}{4\pi p} (\d a +pb)\wedge (\d a+pb)- \frac{k}{4\pi p}\d a \wedge \d a,
    \qquad k,p\in \Z.
    \end{split}
\end{equation}
This is the analogue of \eqref{twisted BF action cochain}. The gauge transformations are
\begin{equation}
\label{gauge tranf twist BF}
    b \to b+\d \l, \qquad a \to a +\d\e-p\l.
\end{equation}
The action \eqref{twisted BF generic} is invariant under \eqref{gauge tranf twist BF} provided that $k$ and $p$ are integers with $kp\in 2\Z$. If $X$ is spin, then $k$, $p$ are generic integers with no extra condition. Notice that actually $p \in \Z_{2k}$, since shifting $p\to p+2k$ yields an extra term
\begin{equation}
    \int_X \frac{k^2}{2\pi} b \wedge b
\end{equation}
which is trivial (namely $2\pi \Z$) after integrating out $a$ (since $\d b=0$ and $\oint b \in (2\pi/k)\Z$).

Consider now the global symmetries of \eqref{twisted BF generic}.
\begin{itemize}
    \item Under the shift $a \to a +\a$, with $\d \a =0$, the action changes as 
    \begin{equation}
        \D S = \int_X  \frac{k}{2\pi} \a \d b.
    \end{equation}
    This is an integer multiple of $2\pi$ provided that $\oint \a = 2 \pi / k$, i.e. $\a$ is a representative of an element of $Z^1(X;\Z_k)$. This suggests that the theory \eqref{twisted BF generic} has a $\Z_k^{(1)}$ symmetry. However, there is also the gauge transformation \eqref{gauge tranf twist BF} to take into account. When $\d \l = 0$, $b$ is invariant, but $a \to a-p\l$ and\footnote{Notice that $\oint \l= 2\pi/p$ is just a standard gauge transformation for $a$, since $\oint p \l = 2\pi$.}
    \begin{equation}
        \D S= \int_X -\frac{kp}{2\pi} \l \d b \in 2\pi \Z \qquad \Leftrightarrow \qquad \oint \l = \frac{2\pi}{k}.
    \end{equation}
    This says that, if $\a$ parameterizes an element of $Z^1(X;\Z_k)$, $a\to a+\a$ is a symmetry of \eqref{twisted BF generic}, but $a\to a-p\a$ is a gauge transformation. This is equivalent to say that a subgroup $p\Z_k$ of the naive $\Z_k^{(1)}$ symmetry is gauged, so that the total one-form symmetry is
    \begin{equation}
        \frac{\Z_k}{p\Z_k} \simeq \Z_L, \qquad L \coloneqq \gcd(k,p).
    \end{equation}
    \item The two-form symmetry is the main difference with the discrete formalism. Because $b$ is not integer valued generically, the two-form symmetry should involve also $a$, as for the gauge transformation, i.e.
    \begin{equation}
        b \to b+\b, \qquad \d a \to \d a-p \b, \qquad \d \b=0.
    \end{equation}
    The variation of the action is 
    \begin{equation}
        \D S= \int_X \frac{k}{2\pi} \b \d a- \frac{kp}{4\pi}\b\b.
    \end{equation}
    This again requires some quantization on the periods of $\b$, so it is convenient to define 
    \begin{equation}
        \b \coloneqq \frac{1}{N} \b', \qquad \oint b' \in 2\pi \Z,
    \end{equation}
    i.e. $b'$ is the familiar curvature of a standard $\U(1)$ gauge field. The variation then reads
    \begin{equation}
        \D S = \begin{cases}
            2\pi \frac{k}{N} \Z - 2\pi\frac{kp}{N^2}\Z, \qquad &\text{$X$ spin};\\
            2\pi \frac{k}{N} \Z - \pi\frac{kp}{N^2}\Z, \qquad &\text{otherwise}.
        \end{cases}        
    \end{equation}
    On $X$ spin, $N$ must be a divisor of both $k$ and $p$, therefore $N=L$ and there is a $\Z_L^{(2)}$ symmetry. When $kp\in 2\Z$, the theory can be considered also on non spin $X$ and the two-form symmetry is more delicate. If either $k$ or $p$ is odd, then $L$ is odd and $pk / L^2 \in 2\Z$ when $kp \in 2\Z$. There is still a $\Z_L^{(2)}$ symmetry even on non spin $X$. If both $k$ and $p$ are even, then $L$ is even and there can be subcases. To understand this situation, it is convenient to write
    \begin{equation}
        k= 2^n (2x+1), \qquad p =2^m(2y+1), \qquad n,m,x,y \in \Z,
    \end{equation}  
    so that $L = \gcd(k,p)=2^{\min(n,m)}\gcd(2x+1,2y+1)$. The $\Z_L^{(2)}$ symmetry is broken for $n=m$, otherwise it is still there. When $n=m$, even if $\Z_L^{(2)}$ is broken, there is still a symmetry $\Z_{L/2}^{(2)}$ by taking $N=L/2$ (since $L$ is even). Thus, in this latter case, the theory considered on $X$ spin has a $\Z_L^{(2)}$ symmetry that it is broken to $\Z_{L/2}^{(2)}$ when $X$ is not spin: a subgroup $\Z_2\subset \Z_L$ must be broken by a gravitational anomaly depending on $[w_2(TX)]$.

    All in all, we arrive at the same results as in the discrete formalism.
    \begin{equation}
    \nonumber
        \begin{split}
            & kp \notin 2\Z, \text{ $L$ odd} \; \Rightarrow \; \text{$X$ must be spin, } \, \Z_L^{(2)} \text{ symmetry};\\
            & kp \in 2\Z, \; k \in 2\Z+1 \vee p\in 2\Z+1, \text{ $L$ odd} \; \Rightarrow \; \text{$X$ orientable, } \, \Z_L^{(2)} \text{ symmetry}; \\
            & k, p \in 2\Z,\; k= rL, \; p = sL, \; r,s \in \Z, \text{ $L$ even }\; \Rightarrow  \;\text{$X$ orientable:}  \\
            &\begin{cases}
                r \in 2\Z \vee s\in 2\Z   \quad  &\Z_L^{(2)} \text{ symmetry always};\\
                r \in 2\Z+1 \wedge s\in 2\Z+1   \quad &\Z_L^{(2)} \text{ if $X$ spin,  $\Z_{L/2}^{(2)}$ on $X$ not spin},\\
                &\text{$\Z_2^{(2)}\subset \Z_{L}^{(2)}$ has a gravitational anomaly.}
            \end{cases}
        \end{split}
    \end{equation}
\end{itemize}

The higher-form symmetries can also be understood looking at the observables of \eqref{twisted BF generic}. The one-form symmetry is generated by the Wilson surfaces 
\begin{equation}
\label{twisted BF U}
    U(\S)= e^{i\oint_\S b}, \qquad \p \S=0.
\end{equation}
Integrating $a$ out, $b$ is a $\Z_k$ gauge field, so $U^k=1$. Moreover, since $p\Z_k$ inside $\Z_k$ is part of the gauge transformation, we have also $U^p=1$. As a result, $U^L=1$ and therefore the one-form symmetry is $\Z_L^{(1)}$. The two-form symmetry is generated by the Wilson lines of $a$, but gauge invariance under \eqref{gauge tranf twist BF} requires that these lines must be dressed with $b$-surfaces:
\begin{equation}
    \wt V(\g,\S) = e^{i\oint_\g a} e^{ip \int_\S b}, \qquad \p \S =\g.
\end{equation}
This is again a consequence of the fact that $p\Z_k^{(1)}$ is gauged, since the Wilson lines of $a$ are charged under it.  On the other hand, the equations of motion of \eqref{twisted BF generic} imply $k(\d a +pb) =0$, so $\d a \neq 0$ in general and its Wilson lines are not topological by themselves. $\wt V$ depends on the surface, so it is not a line operator, but the exponentiated version
\begin{equation}
\label{twisted BF V}
    V(\g) = \wt V^{k/L}(\g,\S)
\end{equation}
does not depend on $\S$ anymore, since $b$ is $\Z_k$ valued, and it is thus a genuine line operator. This is the generator of the two-form symmetry and since $V^L=1$, it follows that this is a $\Z_L^{(2)}$ symmetry.

The correlators of $U$ and $\wt V$ can be computed straightforwardly from the action \eqref{twisted BF generic} \cite{putrovwang2017LinkNum}:
\begin{equation}
\label{twisted BF correlators}
    \begin{split}
        &\braket{U^n(\S)\wt V^m(\g, \S')}=e^{-\frac{2\pi inm}{k}\link(\g,\S)}e^{-\frac{i\pi p m^2}{k} \rm Int(\S',\S')},\\
        &\braket{V^n(\g,\S)\wt V^m(\g', \S')}=e^{-\frac{2\pi i p nm}{k}\rm Int(\S,\S')}e^{-\frac{i\pi p}{k} (n^2\rm Int(\S,\S)+m^2\rm Int(\S',\S'))}.
    \end{split}
\end{equation}
Inserting in these correlators the genuine line operator $V$, i.e. $\wt V^{k/L}$, yields
\begin{equation}
    \begin{split}
        &\braket{U(\S)V(\g)}=e^{-\frac{2\pi i}{L}\link(\g,\S)}e^{-\frac{i\pi p k}{L^2} \rm Int(\S',\S')},\\
        &\braket{V(\g) V(\g')}=e^{-\frac{2\pi i p k}{L^2}\rm Int(\S,\S')}e^{-\frac{i\pi p k}{L^2} (\rm Int(\S,\S)+\rm Int(\S',\S'))},
    \end{split}
\end{equation}
which finally reduce to, by writing $k=rL$, $p=sL$, 
\begin{equation}
\label{twisted BF correlators for V U}
    \begin{split}
        &\braket{U(\S)V(\g)}=e^{-\frac{2\pi i}{L}\link(\g,\S)}e^{-i\pi rs \rm Int(\S',\S')},\\
        &\braket{V(\g) V(\g')}=e^{-i\pi rs (\rm Int(\S,\S)+\rm Int(\S',\S'))}.
    \end{split}
\end{equation}
Neglecting the self-intersection terms for a moment, the correlator $\braket{UV}$ is exactly the one for a $\Z_L$ gauge theory, with a $\Z_L^{(1)}\times \Z_L^{(2)}$ symmetry, coherent with the discussion above. However, there do exist self-intersection terms for $V$, which are non-trivial when both $r$ and $s$ are odd. They come from 
\begin{equation}
\label{V depends on Sigma}
    \braket{V(\g)} = \braket{\wt V^{k/L}(\g,\S)}= e^{-i\pi rs\rm Int(\S,\S)}=\begin{cases}
        (-1)^{\rm Int(\S,\S)}, \quad &r,s\; \rm odd;\\
        1, &\rm otherwise.
    \end{cases}
\end{equation}
This is irrelevant in the two point functions \eqref{twisted BF correlators for V U}, since we can normalize them with $\braket{V}$. However, it shows that, when $r$, $s$ are both odd, $V$ is not actually a genuine line operator as expected, but it is slightly sensitive to the surface to which is attached. 

The result of \eqref{V depends on Sigma} is not intrinsic, since $\S$ is open and therefore it depends on a choice of framing for $\g$ \cite{putrovwang2017LinkNum}, but it is possible to derive from it an intrinsic canonical effect. Indeed, even if naively \eqref{twisted BF V} does not depend on $\S$, the correlator \eqref{V depends on Sigma} yields
\begin{equation}
    \frac{\braket{\wt V^{k/L}(\g,\S)}}{\braket{\wt V^{k/L}(\g,\S')}} = \frac{e^{-i\pi rs\rm Int(\S,\S)}}{e^{-i\pi rs\rm Int(\S',\S')}} = e^{i\pi rs(\rm Int(\S,\W)+\rm Int(\W,\S))} e^{i\pi rs\rm Int(\W,\W)},
\end{equation}
for $\S' = \S + \W$, with $\p \W =0$. Since $\W$ is closed, its self-intersection number is well-defined and, since only its mod 2 value is relevant in the exponent above, it follows that
\begin{equation}
    \inters(\W,\W)=\int_X \PD(\W)\cup \PD(\W)=\int_X w_2(TX) \cup \PD(\W)=\int_\W w_2(TX) \mod 2,
\end{equation}
for PD$(\W)$ the Poincar\'e dual of $\W$. Therefore, there is always a surface-dependent term independently from the choice of framing, 
\begin{equation}
\label{V depends on w2}
        \frac{\braket{V(\g,\S)}}{\braket{V(\g,\S')}} = (-1)^{rs \int_{\S-\S'}w_2(TX)}.
\end{equation}
When $r$ and $s$ are both odd, it implies that $V$ is really the Wilson line of a fermionic quasiparticle. On a non-spin manifold, it must be attached to a $w_2$ surface and this is the origin of the gravitational anomaly explained above for the $\Z_L^{(2)}$ symmetry. Clearly, $V^2$ is independent of $w_2$, so a $\Z_{L/2}^{(2)}$ subgroup survives even on non-spin manifolds.

The twisted BF theory \eqref{twisted BF generic} has also the same mixed $\Z_L^{(1)}\times \Z_L^{(2)}$ anomaly of a standard level $L$ BF theory with $p=0$. This can be seen by turning on a background $B$ for the one-form symmetry as follows
\begin{equation}
    S = \int_X \frac{k}{2\pi} b\wedge (\d a-B')+ \frac{kp}{4\pi}b \wedge b.
\end{equation}
Gauge invariance under $b\to b+\d\l$ requires $B'$ to be a $\Z_k$ gauge field in general, so the actual $\Z_L$ gauge field for $\Z_L^{(1)}$ should be $B = k/L B'$ (consistent with the fact that the general one-form variation of $a$ could be $a \to a+\a$, with $\a$ valued in $\Z_k$, but then a subgroup $p\Z_k$ is actually a gauge transformation). Under the two-form symmetry $b \to b+\b$, the action changes as
\begin{equation}
\label{twisted BF same mixed anomaly}
    \D S= \int_X \frac{k}{2\pi}\b \wedge B' = \int_X \frac{L}{2\pi}\b \wedge B \qquad \Rightarrow \qquad \A = \frac{L}{2\pi}\int_Y C \wedge B,
\end{equation}
where we used inflow to write the anomaly on $Y$, $\p Y =X$, and $C$ is the background for $\Z_L^{(2)}$. This is exactly \eqref{app BF mixed anomaly}.

\section{Number of spin$_c$ structures}

A spin$_c$ structure is a spin structure on $TX \oplus L$, with $L$ a complex line bundle on $X$ defined by its first Chern class $c\in H^2(X;\Z)$. In our discussion, we identify a spin$_c$ structure as an element $\eta^c \in C^1(X;\Z_2)$ that satisfies \eqref{spinc structure} ($c=c_1(A_b)$ and $\eta =\eta^c$ here). As for standard spin structures, one can shift $\eta^c \to \eta^c+s$, with $[s]\in H^1(X;\Z_2)$, to obtain another spin$_c$ structure. For spin structures on $TX$, this is the final result after modding out by all the gauge redundancies. Instead, in this case, we still need to consider gauge transformations of the $\U(1)_f$ bundle, which correspond to sending $c_1(A_f) \to c_1(A_f) +\d n$, with $n \in C^1(X;\Z)$. After lifting the relation \eqref{spinc structure} to integral values (denoting the lifted quantities with tildes), the shift $\wt \eta^c \to \wt \eta^c + \wt s$ is equivalent to consider
\begin{equation}
    \d \wt \eta^c = \wt w_2 + 2 c_1(A_f)- \d \wt s= \wt w_2 + 2 (c_1(A_f)- \b(s)),
\end{equation}
where $\b(s) = \d \wt s/2$ is the Bockstein map \cite{HatcherAT}
\begin{equation} 
\label{bockstein ZZZ_2} 
... \to H^1(X;\Z_2) \xrightarrow[]{\beta} H^2(X;\Z) \xrightarrow[]{\times 2} H^2(X;\Z) \to ... 
\end{equation}
associated to the exact sequence 
\begin{equation}
\label{ZZZ_2 sequence}
    \Z \xrightarrow[]{\times 2} \Z \xrightarrow[]{\mod 2} \Z_2.
\end{equation}
It follows that, if $\beta(s)$ is zero in cohomology, i.e. $\b(s)=\d n$, then the shift $\eta^c \to \eta^c+s$ is actually a $\U(1)$ gauge transformation (indeed, in this case $s$ is the reduction mod 2 of an integer class in $H^1(X;\Z)$, which represents a familiar large gauge transformation).\footnote{This can be seen also working with \v{C}ech cocycles and considering the lifting problem \eqref{spinc group ext}. Consider the principal bundle with transition functions $g_{ij}u_{ij}$, $g \in \SO(d)$, $u\in \U(1)_b$. The lift \eqref{spinc group ext} is obtained by considering the transition functions $\l_{ij}\wt u_{ij}$, with $\l \in \spin(d)$ and $\wt u \in \U(1)_f$, $\pi (\l)=g$ and $\wt u^2 =u$. The cocycle condition for the lifted transition functions is what gives \eqref{spinc structure}. A choice of lift for $g \to \l$ fixes the representative of $w_2$ and then inequivalent lifts are given by $\l_{ij}\to (-1)^{s_{ij}}\l_{ij}$, $[s]\in H^1(X;\Z_2)$. However, given the identification $(\l_{ij},\wt u_{ij})\sim (-\l_{ij},-\wt u_{ij})$, this is equivalent to twist $\wt u$ by $s$, obtaining
\begin{equation}
    (-1)^{\wt s_{ij}}\wt u_{ij} (-1)^{\wt s_{jk}}\wt u_{jk} (-1)^{\wt s_{ki}} \wt u_{ki} = e^{2\pi i (c_1(A_f)+\b(s))},
\end{equation} 
where $\wt s $ is a lift of $s$. It follows that if $\b(s) = \d n$, it can be removed by a gauge transformation of $c_1(A_f)$. The inequivalent lifts are therefore $|H^1(X;\Z_2)|/|\ker(\beta)| = |\Im(\beta)|$, with $\Im(\beta)\simeq$ 2-Torsion of $H^2(X;\Z)$: they are the torsion part of $H^2(X;\Z)$ that is not seen by $c_1(A_b) = 2c_1(A_f)$.} So the total number of spin structures on $TX\oplus L$ is given by $|H^1(X;\Z_2)|/|\ker(\beta)|$ and gauging $\Z_2^f$ means to sum over all possible such $\eta^c$ satisfying \eqref{spinc structure}.

This discussion fits naturally in our physics context where the $\U(1)$ principal bundle is part of the data of the system and represents a fixed background. However, if a manifold $X$ is spin$_c$ or not is a question regarding $X$ by itself: $X$ admits a spin$_c$ structure if $w_2(TX)$ is the mod $2$ reduction of an integral class in $H^2(X;\Z)$. This can be formalized by saying that the obstruction to define a spin$_c$ structure on a manifold $X$ is given by the third integral Stiefel-Whitney class $[W_3(X)] \in H^3(X;\Z)$ \cite{avisisham1980spinG}, $W_3(X) \coloneqq \beta (w_2(X))$, where $\beta$ is the Bockstein homomorphism associated to the sequence \eqref{ZZZ_2 sequence}.\footnote{This time $\b: H^2(X;\Z_2)\to H^3(X;\Z)$.} Indeed, when the class of $W_3$ is trivial, $[\beta(w_2)]=0$, then $w_2$ is the reduction mod 2 of an integer class, $[w_2]=[c]\mod 2$. The total number of inequivalent spin$_c$ structures on a manifold $X$ requires also to sum over possible choices of the integer class $c \in H^2(X;\Z)$.\footnote{Which is not required for bosonization, where we leave the background $\U(1)$ bundle fixed. In the bosonization approach, we are considering the extension problem \eqref{spinc group ext}, with a specific $\U(1)$ bundle. Neglecting this datum, it is like considering the inequivalent lifts given by the extension $\U(1)\to \spin_c(d)\to\SO(d)$, which is another presentation of $\spin_c(d)$. In this latter case only $X$ is kept fixed. This is what we do when we gauge $\U(1)_f$ in \eqref{spinc group ext} and it could give yet another perspective on the anomaly \eqref{anom for gauge spinc}.} As said above, there are $|H^1(X;\Z_2)|/|\ker(\beta)|$ inequivalent choices for $\eta^c$ at fixed $c$. To any such $c$ that solves \eqref{spinc structure}, we can add other first Chern classes whose mod 2 reduction is trivial, which are therefore in the image of $\times 2$ in \eqref{bockstein ZZZ_2} (these are indeed the bundles that we sum over when we gauge also $\U(1)_b$ after $\Z_2^f$). So total number of inequivalent spin$_c$ structures is $|H^1(X;\Z_2)||\Im (\times 2)|/|\ker(\beta)|$, which can also be rewritten, using \eqref{bockstein ZZZ_2}, as $|H^2(X;\Z)|$.

This result can also be obtained by considering the trivialization $W_3= \d \l$, whose inequivalent solutions are given by $\l \to \l + c$, with $[c]\in H^2(X;\Z)$. $\l\in C^2(X;\Z)$ here can be considered as another way to represent a spin$_c$ structure, albeit it is not the same object as $\eta^c$ of \eqref{spinc structure}. We see that the two approaches give the same result, refining the computation of \cite{avisisham1980spinG}.\footnote{The same result can be found in section 4 of the lectures \url{https://webhomes.maths.ed.ac.uk/~v1ranick/papers/mellor.pdf}.}

\newpage

\bibliographystyle{ieeetr}
\bibliography{bibliography.bib}

\end{document}